\RequirePackage{fix-cm}
\documentclass[natbib]{svjour3}                     %
\smartqed  
\usepackage[pdftex]{graphicx}
\usepackage{latexsym}
\usepackage{epsf}
\usepackage{amssymb}
\usepackage{amsmath}
\usepackage{placeins}
\usepackage{color}
\usepackage{hyperref}
\usepackage{physics}
\usepackage{threeparttable}
\usepackage{booktabs}
\usepackage{dcolumn}
\usepackage{longtable}
\usepackage{threeparttablex}
 \journalname{SSRv}
\newcommand{\be}{\begin{eqnarray}}
\newcommand{\ee}{\end{eqnarray}}

\def\ssr{Space~Sci.~Rev.}%
\def\apj{\rmfamily{ApJ~}}         
\def\apjl{\rmfamily{ApJ~}}        
\def\apjs{\rmfamily{ApJS~}}       

\def\aap{\rmfamily{A\&A~}}        

\def\araa{\rmfamily{ARA\&A~}}     
\def\aapr{\rmfamily{A\&A~Rev.~}}  
\def\mnras{\rmfamily{MNRAS~}}     
\def\pasp{\rmfamily{PASP~}}       
\def\pasj{\rmfamily{PASJ~}}       
\def\prd{\rmfamily{Phys.~Rev.~D~}}
\def\prl{\rmfamily{Phys.~Rev.~Lett.~}}
\def\nat{\rmfamily{Nature~}}      

\def\ssr{\rmfamily{Space~Sci.~Rev.~}}

\begin{document}

\title{Towards precision measurements of accreting black holes\\using X-ray reflection spectroscopy}


\author{Cosimo~Bambi$^{1}$, Laura~W.~Brenneman$^{2}$, Thomas~Dauser$^{3}$, Javier~A.~Garc\'ia$^{4,3}$, Victoria~Grinberg$^{5}$, Adam~Ingram$^{6}$, Jiachen~Jiang$^{7}$, Honghui~Liu$^{1}$, Anne~M.~Lohfink$^{8}$, Andrea~Marinucci$^{9}$, Guglielmo~Mastroserio$^{4}$, Riccardo~Middei$^{10,11}$, Sourabh~Nampalliwar$^{5}$, Andrzej~Nied\'zwiecki$^{12}$, James~F.~Steiner$^{2}$, Ashutosh~Tripathi$^{1}$, Andrzej~A.~Zdziarski$^{13}$}

\institute{
$^1$ Center for Field Theory and Particle Physics and Department of Physics, Fudan University, 200438 Shanghai, China\\
$^2$ Center for Astrophysics, Harvard \& Smithsonian, Cambridge, MA 02138, USA\\
$^3$ Dr. Karl Remeis Observatory and Erlangen Centre for Astroparticle Physics, D-96049 Bamberg, Germany\\
$^4$ Cahill Center for Astronomy and Astrophysics, California Institute of Technology, Pasadena, CA 91125, USA\\
$^5$ Institut f\"ur Astronomie und Astrophysik (IAAT), Eberhard-Karls Universit\"at T\"ubingen, D-72076 T\"ubingen, Germany\\
$^6$ Department of Physics, University of Oxford, OX1~3PU Oxford, UK\\
$^7$ Department of Astronomy, Tsinghua University, 100084 Beijing, China\\
$^8$ Department of Physics, Montana State University, Bozeman, MT 59717, USA\\
$^{9}$ ASI - Unit\`a di Ricerca Scientifica, I-00133, Rome, Italy\\
$^{10}$ INAF -- Osservatorio Astronomico di Roma, I-00078 Monte Porzio Catone (Rome), Italy\\
$^{11}$ Space Science Data Center -- ASI, I-00133 Rome, Italy\\
$^{12}$ Faculty of Physics and Applied Informatics, \L\'od\'z University, PL-90-236 \L\'od\'z, Poland\\
$^{13}$ Nicolaus Copernicus Astronomical Center, Polish Academy of Sciences, PL-00-716 Warsaw, Poland\\
\email{bambi@fudan.edu.cn}
}

\date{\today}

\maketitle

\abstract
Relativistic reflection features are commonly observed in the X-ray spectra of accreting black holes. In the presence of high quality data and with the correct astrophysical model, X-ray reflection spectroscopy can be quite a powerful tool to probe the strong gravity region, study the morphology of the accreting matter, measure black hole spins, and possibly test Einstein's theory of general relativity in the strong field regime. In the last decade, there has been significant progress in the development of the analysis of these features, thanks to more sophisticated astrophysical models and new observational facilities. Here we review the state-of-the-art in relativistic reflection modeling, listing assumptions and simplifications that may affect, at some level, the final measurements and may be investigated better in the future. We review black hole spin measurements and the most recent efforts to use X-ray reflection spectroscopy for testing fundamental physics.

\keywords{Black Holes -- X-ray Astronomy -- Reflection Spectrum -- Iron Line -- Black Hole Spins}


\section{Introduction \label{s-intro}}

Black holes (BHs) are among the most intriguing predictions of general relativity (GR). The first BH solution was discovered by Karl Schwarzschild in 1916 \citep{Schwarzschild:1916uq}, less than one year after the announcement by Albert Einstein of his theory. However, the actual properties of these solutions were not understood immediately. David Finkelstein was the first, in 1958, to realize that these solutions had an event horizon acting as a one-way membrane \citep{Finkelstein:1958zz}: whatever crosses the horizon, it cannot influence the exterior region any longer. For a long time, the astronomy community was very skeptical about the possibility of the existence of BHs in the Universe. The situation changed in the early 1970s, when Thomas Bolton and, independently, Louise Webster and Paul Murdin identified the X-ray source Cygnus X-1 as the first stellar-mass BH candidate \citep{1972Natur.235..271B,1972Natur.235...37W}. Since then, an increasing number of astronomical observations have pointed out the existence of stellar-mass and supermassive BHs \citep{Kormendy:1995er,Remillard:2006fc,2008ApJ...689.1044G,2009ApJ...692.1075G,Abbott:2016blz,Akiyama:2019cqa}; for pedagogical reviews, see, for instance, \citet{Narayan:2005ie} and \citet{Bambi:2017iyh}. Thanks to technological progress and new observational facilities, in the past 10--20~years there have been tremendous advancements in the understanding of these objects and of their environments.

\begin{figure}[t]
\vspace{0.3cm}
\begin{center}
\includegraphics[type=pdf,ext=.pdf,read=.pdf,width=8cm]{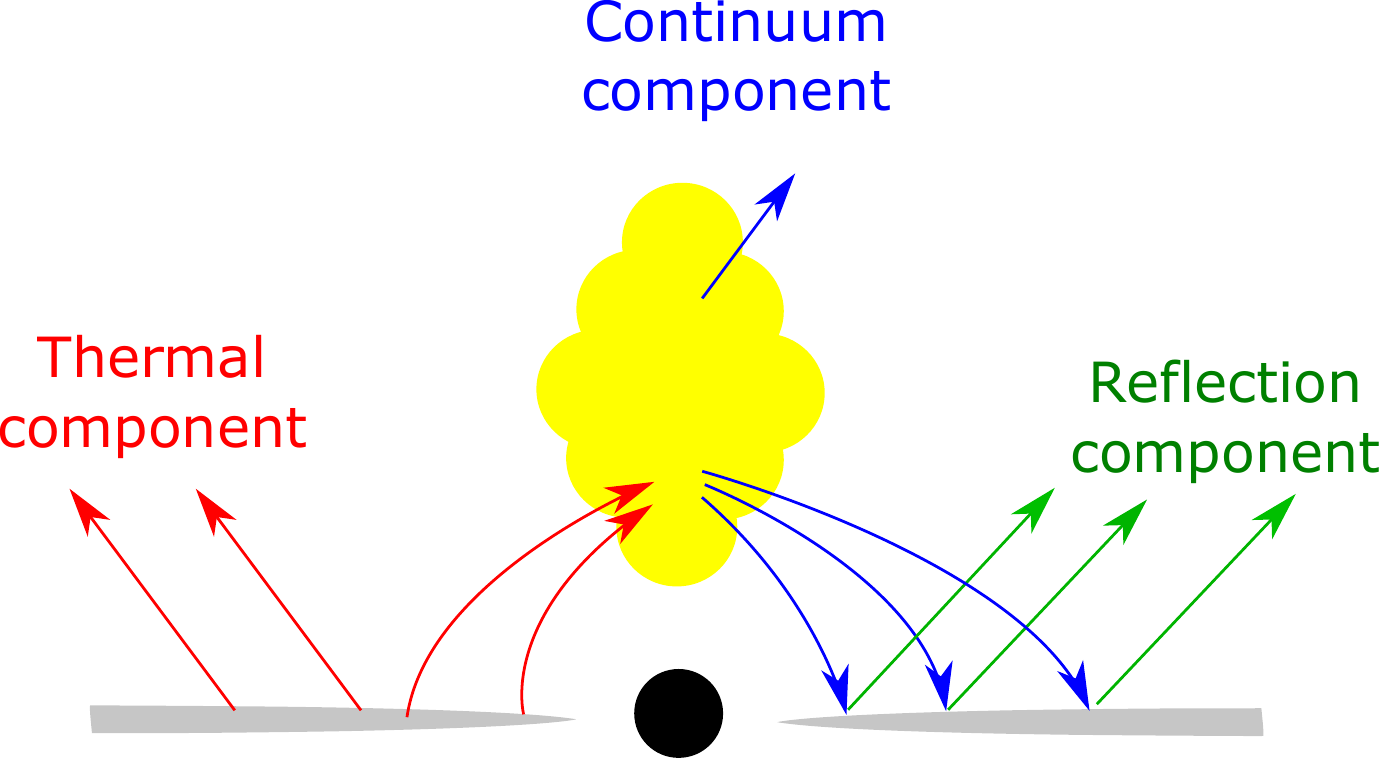}
\end{center}
\vspace{0.3cm}
\caption{Cartoon of the disk-corona model. A BH (black circle) accretes from a geometrically thin and optically thick disk (gray layers). Thermal photons (red arrows) from the disk can inverse Compton scatter off free electrons in the corona (yellow cloud) generating the continuum component (blue arrows). A fraction of the Comptonized photons illuminate the disk, producing the reflection component (green arrows). 
\label{f-diskcorona1}}
\end{figure}

Relativistic reflection features are commonly observed in the X-ray spectra of accreting BHs, both BH binaries \citep{Fabian:1989ej,Blum:2009ez,Fabian:2012kv,Miller:2013rca,Tomsick:2013nua, Basak17, Xu:2017yrm} and AGN \citep{Tanaka:1995en,Nandra:1996vv,Nandra:2007rp,Walton:2012aw, 2013mams.book.....B,Reynolds:2019uxi}, and are thought to be generated by illumination of the inner part of the accretion disk by a hot corona \citep{Fabian:1995qz,Zoghbi:2009wd,Risaliti:2013cga}. This physical process is illustrated in Fig.~\ref{f-diskcorona1}.

The standard set-up has a BH (black circle in Fig.~\ref{f-diskcorona1}) accreting from a geometrically thin and optically thick disk. The gas in the disk is close to a local thermal equilibrium: every point on the surface of the disk has a blackbody-like spectrum and the whole disk radiates like a sum of blackbodies, leading to a multi-temperature blackbody spectrum ({\it thermal component} in Fig.~\ref{f-diskcorona1}). Thermal photons from the disk inverse Compton scatter off free electrons in the ``corona'' (yellow cloud in Fig.~\ref{f-diskcorona1}), which is some energetic plasma near the BH but its morphology remains not well understood \citep{1979ApJ...229..318G, Haardt:1991tp, PVZ18}. The Comptonized photons have a power-law spectrum with an exponential high energy cut-off ({\it continuum component} in Fig.~\ref{f-diskcorona1}). A fraction of the Comptonized photons illuminate the accretion disk. Compton scattering and absorption followed by re-emission produce a reflection component ({\it reflection component} in Fig.~\ref{f-diskcorona1}) \citep{Lightman88, George:1991jj, MZ95, Ross:2005dm, Garcia:2010iz}.

\begin{figure}[t]
\vspace{0.3cm}
\begin{center}
\includegraphics[trim={0.5cm 0.0cm 3.5cm 17.0cm},clip,width=1\columnwidth]{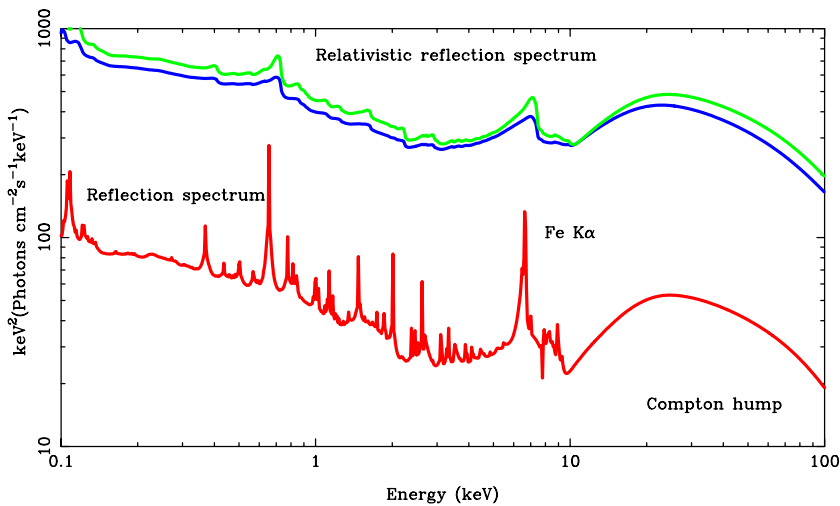}
\end{center}
\vspace{-0.3cm}
\caption{Reflection spectrum in the rest-frame of the gas in the disk (red curve) and relativistic reflection spectra of the whole disk as seen by a distant observer (green and blue curves). Narrow fluorescent emission lines in the reflection spectrum become broadened and skewed for the distant observer. In the figure, we employed a different normalization between non-relativistic and relativistic spectra in order to avoid overlapping. The reflection spectrum is generated with {\tt xillver}~\citep{Garcia:2010iz,Garcia:2013oma} assuming that the incident radiation has photon index $\Gamma = 2$ and high energy cut-off $E_{\rm cut} = 300$~keV, the reflector has solar iron abundance ($A_{\rm Fe} = 1$), ionization $\log\xi = 3$ ($\xi$ in units erg~cm~s$^{-1}$), electron density $n_{\rm e}=10^{15}$ cm$^{-3}$, and the local viewing angle is $\vartheta_{\rm e} = 45^\circ$. The relativistic reflection spectra are generated with {\tt relxill}~\citep{Dauser:2010ne,Dauser:2013xv} and we also assume that the BH spin parameter is $a_* = 0$ (green curve) and $a_* = 0.998$ (blue curve), the inclination angle of the disk is $i = 45^\circ$, and the emissivity profile of the disk scales as $1/r^3$. 
\label{f-sp}}
\end{figure}

The most prominent features in the X-ray reflection spectrum are fluorescent emission lines below 10~keV, the Fe K-edge at 7--10~keV, and the Compton hump peaked at 20--30~keV, see Fig.~\ref{f-sp}. Among the fluorescent emission lines, the most prominent one is often the iron K$\alpha$ complex, which is at 6.4~keV for neutral or weakly ionized iron atoms and shifts up to 6.97~keV in the case of H-like iron ions. The Compton hump is a bump in the reflection spectrum resulting from the combination of photo-electric absorption of low-energy photons and multiple electron down-scattering of high energy photons.
The reflection spectrum of the whole accretion disk seen far from the source is the result of the reflection emission at every point of the disk and of relativistic effects (Doppler boosting, gravitational redshift, light bending, and photon capture by the BH) occurring in the strong gravity region around the BH \citep{Fabian:1989ej,Laor:1991nc, Dauser16}. Spectral features are thus broadened and skewed for a distant observer, as illustrated in Fig.~\ref{f-sp}.
The reflection spectrum of the disk may also account for the so-called ``soft excess'', which is an excess of counts below 2~keV. It has been ubiquitously observed in Seyfert galaxies~\citep{2009A&A...495..421B,2020MNRAS.491..532G} and cannot be explained by the Comptonized photons from the hot corona\footnote{The origine of the soft excess is still matter of debate and two scenarios have been proposed and tested, namely blurred ionized reflection and warm Comptonization. In the first case, the soft excess would be due the relativistic reflection and it has been successfully tested on a number of different sources \citep[see, e.g.,][]{walton13}. On the other hand, the warm Comptonization scenario has proved to be a viable explanation from the ultraviolet to soft X-rays of a number of nearby Seyfert galaxies for which simultaneous broadband data were used \citep[see, e.g.,][]{2018A&A...609A..42P,2019A&A...622A.116U,2020A&A...640A..99M,2020MNRAS.497.2352M}.}.

X-ray reflection spectroscopy refers to the analysis of reflection features in the spectra of accreting BHs. The technique has been historically called the iron line method because the iron K$\alpha$ line was the first feature used for X-ray reflection spectroscopy measurements. This is an intrinsically narrow feature, so it is particularly suitable to measure the effects of relativistic blurring. Moreover, it is normally the most prominent fluorescent line for four reasons: $i)$ the probability of fluorescent line emission is higher than other heavy elements (for neutral matter, the fluorescence yield is proportional to $Z^4$, where $Z$ is the atomic number), $ii)$ iron is more abundant than other heavy elements (iron-56 nucleus is more tightly bound than lighter and heavier elements, so it is the final product of astrophysical nuclear reactions); around 6~keV there are indeed other ions (Cr, Mg, Ni, etc.) emitting or absorbing radiation, but their abundances are much lower than iron and their lines are thus much weaker, $iii)$ for typical continua, the flux in the power-law component per unit energy decreases with increasing energy, so the iron line at 6--7~keV is going to be stronger relative to the continuum than lines at a few keV, and $iv)$ around 6~keV the Galactic absorption is negligible.

Since the reflection spectrum is mainly emitted from the very inner part of the accretion disk, the analysis of reflection features in X-ray spectra of accreting BHs can be a powerful tool to study the region around these compact objects. We can learn about the structure of the inner part of accretion disks and the properties of other material near BHs, such as the corona. We can measure BH spins or, more in general, test fundamental physics in the strong gravity regime.

In the past $\sim10$~years, there have been tremendous advancements in the analysis of relativistic reflection features. Today sophisticated theoretical models are available for the analysis of these features as well as suitable observational facilities -- e.g., \textsl{Chandra}~\citep{2002PASP..114....1W}, \textsl{XMM-Newton}~\citep{2001A&A...365L...1J}, \textsl{NuSTAR}~\citep{2013ApJ...770..103H}, \textsl{NICER}~\citep{2012SPIE.8443E..13G}, \textsl{Insight-HXTM}~\citep{2020SCPMA..63x9502Z} -- for their detection. Spin measurements of $\sim50$~sources among BH binaries and AGN have been reported (see Section~\ref{s-spin}) and provide important information about the birth and the evolution of these systems and their environment.

However, all relativistic reflection models necessarily have a number of simplifications, and therefore care is needed to understand the effect of these simplifications on precision BH measurements. Instrumental effects in current and future instruments are another source of systematic uncertainties. The next generation of X-ray missions -- e.g., \textsl{XRISM}~\citep{2020SPIE11444E..22T}, \textsl{eXTP}~\citep{2016SPIE.9905E..1QZ}, \textsl{Athena}~\citep{2013arXiv1306.2307N}, \textsl{STROBE-X}~\citep{2018SPIE10699E..19R} -- promises to provide unprecedented high quality data, which will necessarily require more accurate synthetic reflection spectra than those available today and a better knowledge of the instrumental effects.

The aim of this article is to review the state-of-the-art of reflection spectroscopy, discuss current limitations, and outline plans for future improvements. The content of the paper is as follows. In Section~\ref{s-d-c}, we briefly review the disk-corona models. In Section~\ref{s-ref}, we describe the available relativistic reflection models and how synthetic relativistic reflection spectra are calculated. In Section~\ref{s-mod}, we list the simplifications in the available relativistic reflection models that might introduce modeling bias in the final measurements of the properties of accreting BHs. In Section~\ref{s-instr}, we discuss instrumental effects. In Section~\ref{s-spin}, we review BH spin measurements. In Section~\ref{s-gr}, we briefly discuss current efforts of using X-ray reflection spectroscopy for testing Einstein's theory of GR in the strong field regime. 
Our conclusions are reported in Section~\ref{s-c}.


\section{Disk-corona models \label{s-d-c}}

\subsection{Accretion disks \label{ss-ad-3}}

The theoretical framework describing a BH surrounded by a geometrically thin and optically thick accretion disk was developed in the early 1970s \citep{Shakura73,Novikov:1973kta,Page:1974he}. In BH binaries, stable optically thick accretion disks are seen when the disk luminosity, $L_{\rm d}$, is more than a few per cent of the Eddington luminosity, $L_{\rm E}$, say $L_{\rm d}\gtrsim 0.01$--0.04~$L_{\rm E}$ \citep{Maccarone03}, while an inner disk is absent at lower luminosities, i.e., in quiescence and in early stages of outbursts \citep[see, e.g.,][]{DHL01,Bernardini16,2001ApJ...555..483E}. The disk can remain geometrically thin up to $\sim0.3$~$L_{\rm E}$, and becomes puffed up at higher luminosities \citep{Abramowicz88}.

The gas in optically thick accretion disks is close to a local thermal equilibrium. The local effective temperature of the disk scales approximately as $M^{-1/4}R^{-3/4}$, where $M$ is the BH mass and $R$ is the disk radius. This scaling is valid regardless of the disk physics, as long as the disk is optically thick. Neglecting the inner boundary condition \citep[as in the {\tt diskbb} spectral model,][]{Mitsuda84}, the peak of the $EF_E$ spectrum can be found to be at $E_{\rm peak}\approx {2.36}$~$k_{\rm B}T_{\rm in}$, where $T_{\rm in}$ is the color temperature at the disk inner edge, $R_{\rm in}$. The inner temperature is given by\footnote{We note that since we include no zero-stress inner boundary condition, we do not have the factor of 3 enhancement of the dissipation at large radii, which would be present with that condition due to the resulting redistribution of the gravitationally released power. Moreover, the expression is valid for a pure hydrogen plasma. In the general case, $L_{\rm E}$ is $2/(1+X)$ times larger, where $X$ is the hydrogen abundance.}
\begin{equation}
T_{\rm in}=\kappa\left[\frac{(L_{\rm d}/L_{\rm E})m_{\rm p}c^5}{2 r_{\rm in}^3 \eta \sigma\sigma_{\rm T} G_{\rm N} M}\right]^{1/4},
\label{Tin}
\end{equation}
where $r_{\rm in}\equiv R_{\rm in}/r_{\rm g}$, $r_{\rm g}\equiv G_{\rm N}M/c^2$ is the gravitational radius, $\eta \sim 0.1$ is the accretion efficiency, $m_{\rm p}$ is the proton mass, $\sigma$ is the Stefan-Boltzmann constant, $\sigma_{\rm T}$ is the Thomson cross section, and $\kappa\approx 1.5$--2 \citep{Davis05} is the color correction. Numerically, we have
\begin{equation}
E_{\rm peak}\approx {2.9}\,{\rm keV} \frac{\kappa}{1.7}
\left(\frac{L_{\rm d}/L_{\rm E}}{0.01}\right)^{1/4} \left(\frac{\eta}{0.1}\right)^{-1/4}
\left(\frac{M}{10~M_\odot}\right)^{-1/4}
\left(\frac{r_{\rm in}}{2}\right)^{-3/4}.
\label{Epeak}
\end{equation}
Thus, blackbody spectra of disks extending near the innermost stable circular orbit (ISCO) peak in the X-ray band ($\sim 1$--10~keV) for stellar-mass BHs ($M \sim 10$~$M_\odot$) and in the UV band ($\sim 1$--100~eV) for the supermassive ones ($M \sim 10^5$--$10^9$~$M_\odot$).

However, the geometrically thin solution is not unique. \citet{SLE76} found the existence of another branch in which the disk is geometrically thick and the accreting gas is hot, $kT_{\rm e}\sim 100$~keV (see also \citealt{Thorne75,Eardley75}). That solution was found to be thermally unstable \citep{Pringle76}, but originally it did not include advection of hot ions onto the BH. The advection was then found to stabilize the disk \citep{NY94,NY95b,Abramowicz95}, establishing the Advection Dominated Accretion Flow (ADAF) paradigm, which exerted a strong influence on the astrophysics of accreting BHs for the following decade. In particular, it became the standard explanation for the X-ray emission in the hard state of BH binaries and of low and medium luminosity AGN. In this model, the outer geometrically thin disk is truncated at some radius and replaced by a hot flow, in which the ions are viscously heated and transfer a part of their energy to electrons via Coulomb collisions before being advected to the BH. The ADAF model was then developed to include compressive heating and direct heating of electrons by some magnetic processes. This allowed the maximum possible luminosity of this solution to be as high as $\sim 0.2$--0.3~$L_{\rm E}$ \citep{Yuan07}. See \citet{YN14} for a comprehensive review. Reflection features in this model are from the surrounding cold disk and/or cold clumps within the hot flow.

The range of the luminosities possible for the geometrically thin and thick solutions overlap. Either of them is possible in the $\sim 0.01$--0.3~$L_{\rm E}$ range and may be realized in the outbursts of BH transients (e.g., \citealt{Zdziarski04, Belloni10}), as a result of their hysteretic behavior. The initial outburst rise is along the hard state, which is followed by a hard-to-soft transition, and the outburst decline down to $\sim 0.01$~$L_{\rm E}$ is along the soft state, at which point the source goes back to the hard state. The outer cold disk/inner hot flow model has been used to explain many phenomena in accreting BH binaries; see \citet{2007A&ARv..15....1D} for a review.

\subsection{Coronae \label{ss-corona3}}

\begin{figure}[t]
\begin{center}
\includegraphics[trim={0cm 0cm 0cm 0cm},clip,width=1\columnwidth]{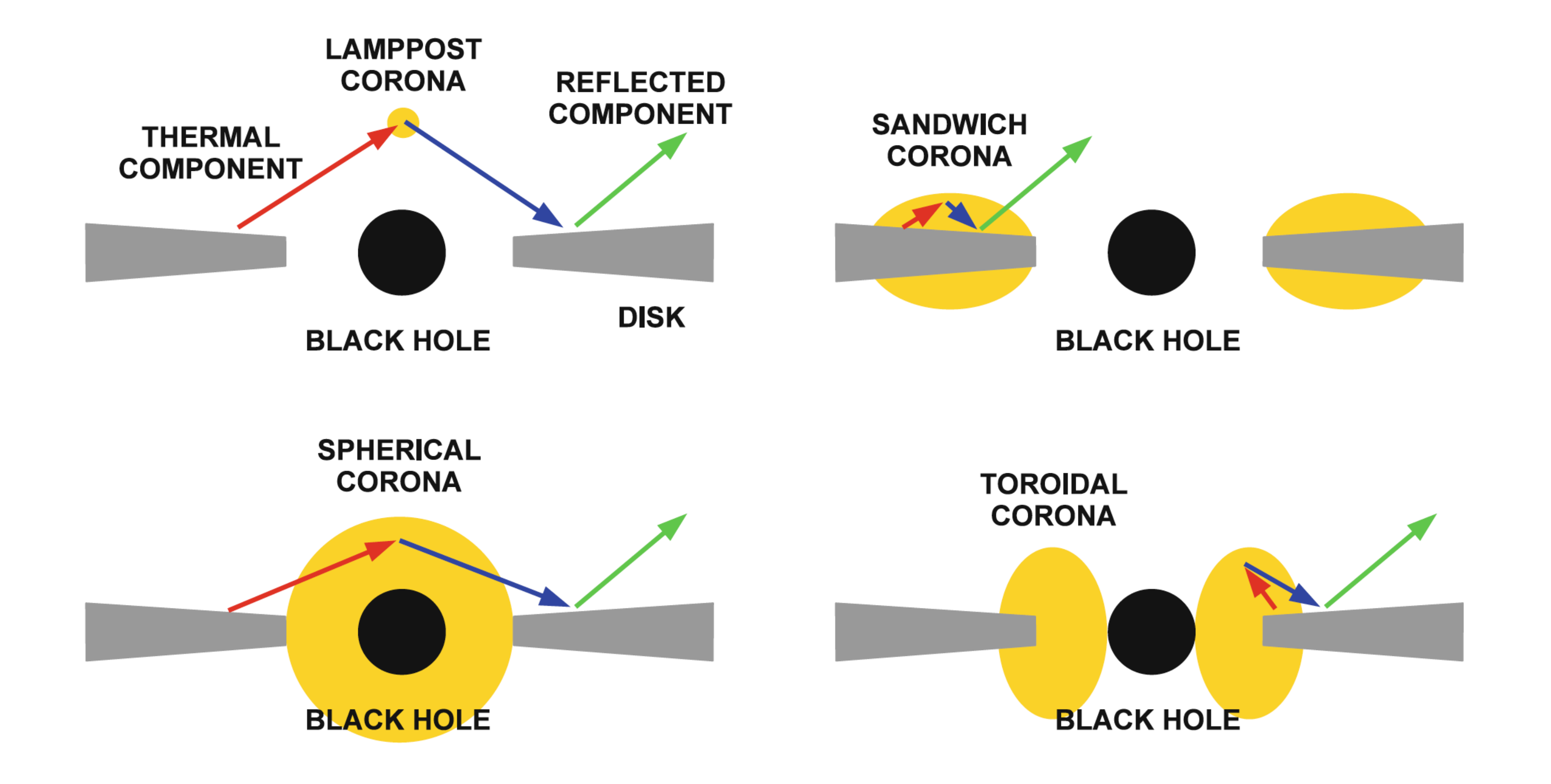}
\end{center}
\caption{Examples of possible corona geometries: lamppost geometry (top left panel), sandwich geometry (top right panel), spherical geometry (bottom left panel), and toroidal geometry (bottom
right panel). Figure from \citet{Bambi:2017khi}.
\label{f-coro}}
\end{figure}

The term corona is generically used to denote some energetic plasma around the BH~\citep{1979ApJ...229..318G, Haardt:1991tp, PVZ18}. For example, in the ``lamppost'' model the corona is a compact source on the rotation axis of the BH; such a configuration may, in principle, correspond to the base of a jet \citep{markoff05}. In the ADAF scenario, the corona would be the inner hot accretion flow between the inner edge of the outer cold disk and the BH. In the so-called ``sandwich'' model, the corona is the atmosphere above inner parts of the accretion disk. The corona may be in approximate hydrostatic equilibrium or be outflowing \citep{Beloborodov99}. Fig.~\ref{f-coro} shows some examples of possible coronal geometries.

The coronal geometry can evolve with time, and this is seen either in BH binaries~\citep[e.g.,][]{2006A&A...448.1125M,kara19} and AGN~\citep[e.g.,][]{Wilkins:2015nfa}. More than one corona may coexist at the same time \citep[see, e.g.,][]{Rozanska15, Fuerst:2015ska, Petrucci:2017niz,2020MNRAS.491.3553B}.

The coronal spectrum is likely to depend on its location, in particular on its distance from the BH. Frequency-resolved spectroscopy shows that X-ray spectra harden with increasing frequency, both in BH binaries \citep{RGC99,Axelsson18} and in AGN \citep{Markowitz05, Arevalo08}. An inhomogeneous corona with each region varying linearly is the simplest way to create harder spectra at higher frequencies \citep{Mahmoud18a,Mahmoud18b, Mahmoud19,Alston19,Zdziarski20b}. However, even a single compact corona \citep[see, e.g., ][]{2016AN....337..356C} with spectrum that varies in shape as well as normalization leads to have harder spectra at higher frequencies \citep[e.g.,][]{Mastroserio2018}.

The electrons in the corona can have either a purely thermal (Maxwellian) distribution or also contain a high energy non-thermal tail. The electrons in the corona Compton upscatter soft photons present in the source (e.g., \citealt{Sunyaev:1979nz}). The soft photons come either from the blackbody emission of the disk or from the synchrotron emission of the electrons. The latter process is especially efficient if there is a non-thermal tail beyond the Maxwellian distribution (e.g., \citealt{PV14}). There is strong evidence that such tails are present in the soft state of BH binaries; they are required to explain the power-law like spectra extending to the MeV range \citep{Gierlinski99, Gierlinski03,ZMC17}. A non-thermal tail can affect the reflection spectrum at high energies, where reflection is purely Compton scattering. On the other hand, the bulk of the X-ray spectra in the hard state are well explained by Comptonization of thermal electrons, though either a weak non-thermal tail can still be present \citep{McConnell02, PV14} or most of the Comptonization may be on the bulk motion of plasmoids in the accretion flow \citep{Beloborodov17}. In AGN, spectra tend to be too soft, and therefore lack sufficient flux at higher energies, to be able to detect a potential non-thermal tail.

If the coronal electrons are thermal, the spectrum of the Comptonized photons is determined by the coronal temperature, $T_{\rm e}$, and the optical depth to Thomson scattering $\tau_{\rm e}$ \citep{1957JETP....4..730K,PS96, Zdziarski20a}. At energies much lower than $k_{\rm B} T_{\rm e}$, multiple scatterings form a power-law spectrum, $F(E) \propto E^{1-\Gamma}$, with photon index $\Gamma$, which, for the case of $k_{\rm B} T_{\rm e}\ll m_{\rm c}c^2$ and $\tau_{\rm e}^2\gg 1$, is given by \citep{ST80}
\be
\Gamma = \sqrt{\frac{9}{4} + \frac{\pi^2 m_{\rm e} c^2}{3 k_{\rm B} T_{\rm e} 
\left( \tau_{\rm e} +2/3 \right)^2}} - \frac{1}{2},
\ee
where $m_{\rm e}$ is the electron mass. At higher energies, the spectrum cuts off exponentially since the electrons reach an approximate equipartition with the upscattered photons.


\section{Relativistic reflection models \label{s-ref}}

\subsection{Reflection spectroscopy}

Relativistic reflection models require the calculation of photon paths from the emission point in the accretion disk near the BH to the detection point in the flat faraway region. This is normally done by implementing the formalism of the transfer function~\citep{Cunningham:1975zz,1995CoPhC..88..109S,Bambi:2017khi}. The method splits the calculations in a few blocks and this turns out to be particularly convenient in the data analysis process.

The flux detected by a distant observer (measured, for instance, in erg~s$^{-1}$~cm$^{-2}$~keV$^{-1}$) can be written as
\be\label{eq-flux}
F_{\rm o} (E_{\rm o}) &=& \int I_{\rm o} (E_{\rm o}) \, d\Omega \nonumber\\
&=& \frac{1}{D^2} \int_{R_{\rm in}}^{R_{\rm out}} \int_0^1 \frac{\pi r_{\rm e} g^2}{\sqrt{g^* (1 - g^*)}}
\, f (g^*,r_{\rm e},i) \, I_{\rm e} (E_{\rm e},r_{\rm e},\vartheta_{\rm e}) \, dr_{\rm e} \, dg^* \, ,
\ee
where $E_{\rm o}$ and $E_{\rm e}$ are, respectively, the photon energy measured by the observer and the photon energy at the emission point in the rest-frame of the gas in the disk, $I_{\rm o}$ and $I_{\rm e}$ are, respectively, the specific intensity of the radiation measured by the observer and at the emission point in the rest-frame of the gas in the disk, $d\Omega$ is the element of the solid angle subtended by the image of the accretion disk in the observer's sky, $D$ is the distance of the observer from the source, $R_{\rm in}$ and $R_{\rm out}$ are, respectively, the inner and the outer edge of the accretion disk, $r_{\rm e}$ is the emission radius in the disk, $i$ is the viewing angle (i.e., the angle between the normal to the disk and the line of sight of the observer), $\vartheta_{\rm e}$ is the emission angle (i.e., the angle between the normal to the disk and the photon 4-momentum in the rest-frame of the gas), $g = E_{\rm o}/E_{\rm e}$ is the redshift factor, and $g^*$ is the relative redshift defined as 
\be
g^* = \frac{g - g_{\rm min}}{g_{\rm max} - g_{\rm min}} \, ,
\ee
where $g_{\rm max} = g_{\rm max} (r_{\rm e},i)$ and $g_{\rm min} = g_{\rm min} (r_{\rm e},i)$ are, respectively, the maximum and the minimum redshift factor $g$ for the photons emitted with radial coordinate $r_{\rm e}$ and detected by an observer with viewing angle $i$. $f$ is the {\it transfer function}~\citep{Cunningham:1975zz} 
\be
f (g^*,r_{\rm e},i) = \frac{g \sqrt{g^* (1 - g^*)}}{\pi r_{\rm e}} 
\left| \frac{\partial \left( X , Y \right)}{\partial \left( g^*,r_{\rm e}\right)} \right| \, ,
\ee
where $|\partial(X,Y)/\partial(g^*,r_{\rm e})|$ is the Jacobian between the Cartesian coordinates $(X,Y)$ of the image of the disk in the observer's sky and the integration variables $(g^*,r_{\rm e})$.

The transfer function $f$ acts as an integration kernel to calculate the reflection spectrum detected by the observer for a given local spectrum at every point of the accretion disk stored in $I_{\rm e}$. The transfer function $f$ depends on $1)$ the spacetime metric, $2)$ the accretion disk model (e.g., gas velocity, location of the accretion disk surface, etc.), and $3)$ the position of the distant observer via the viewing angle $i$. In practice, the transfer function is calculated with a ray-tracing code, starting from the image plane of the distant observer with Cartesian coordinates $(X,Y)$ and firing photons to the accretion disk to determine the emission point and, in turn, to calculate the redshift and the Jacobian.

In GR, the spacetime metric of astrophysical BHs should be approximated well by the Kerr solution~\citep{Kerr:1963ud}. Indeed, deviations from the Kerr geometry induced by the presence of the accretion disk, stars orbiting the BH, or the BH electric charge are normally very small~\citep{Bambi:2017khi} and their impact on the background metric can be ignored when the transfer function is calculated. The Kerr metric is completely specified by the BH mass $M$ and the dimensionless spin parameter $a_* = cJ/(G_{\rm N}M^2)$, where $J$ is the BH spin angular momentum, but the mass sets the length scale of the system and does not enter the calculation of the transfer function. In the end, the only parameter of the background metric is the BH spin parameter $a_*$.

The accretion disk is normally described the Novikov-Thorne model \citep{Novikov:1973kta,Page:1974he}. The disk lies on the equatorial plane, perpendicular to the BH spin, and is usually assumed to be infinitesimally thin. The gas in the disk moves on nearly geodesic, equatorial, circular orbits (Keplerian motion). The inner edge of the accretion disk is at the ISCO, $R_{\rm in} = R_{\rm ISCO}$, or, if the disk is truncated, at a larger radius, $R_{\rm in} > R_{\rm ISCO}$.

\begin{figure}[t]
\begin{center}
\includegraphics[trim={0cm 0cm 0cm 0cm},clip,width=0.49\columnwidth]{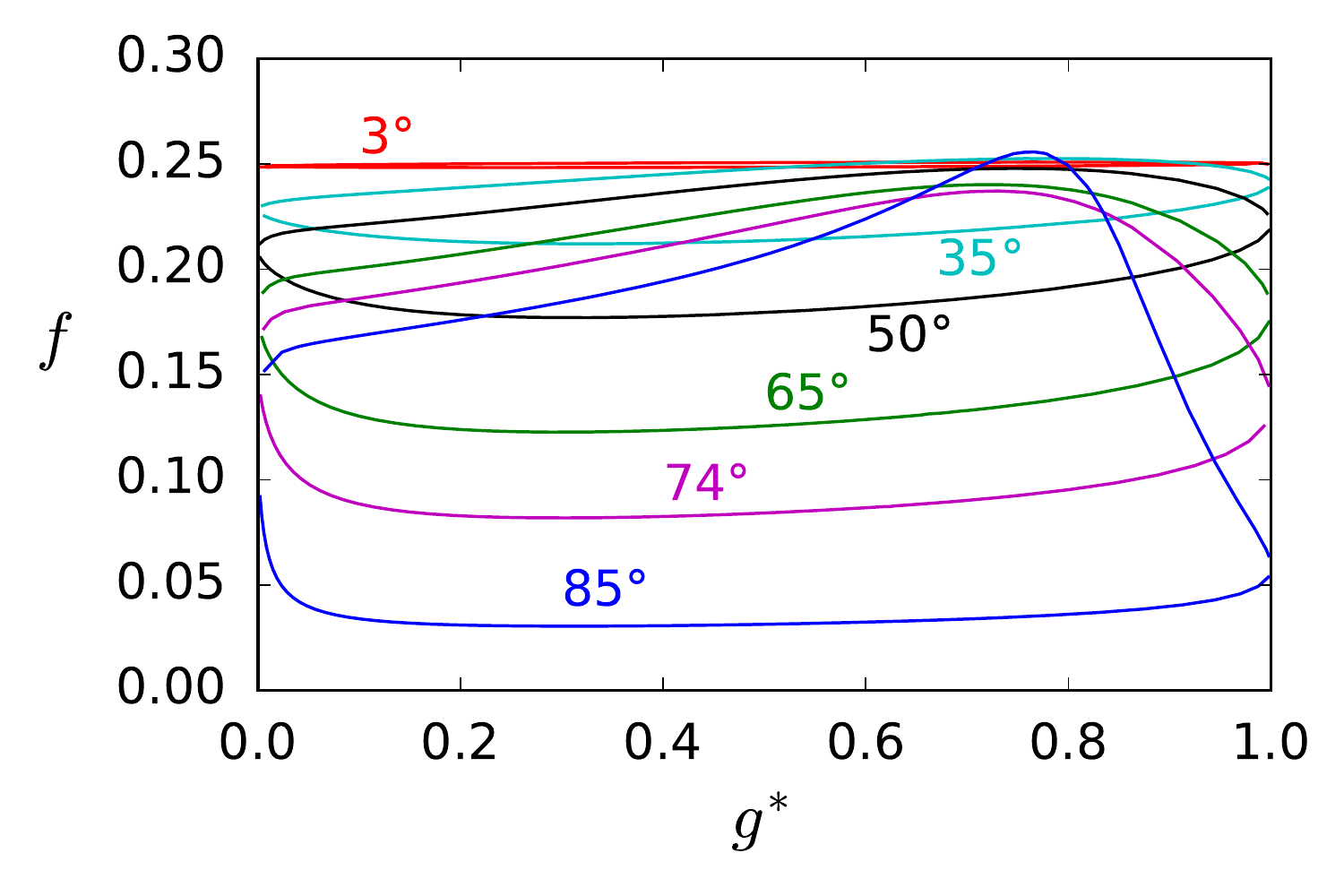}
\includegraphics[trim={0cm 0cm 0cm 0cm},clip,width=0.49\columnwidth]{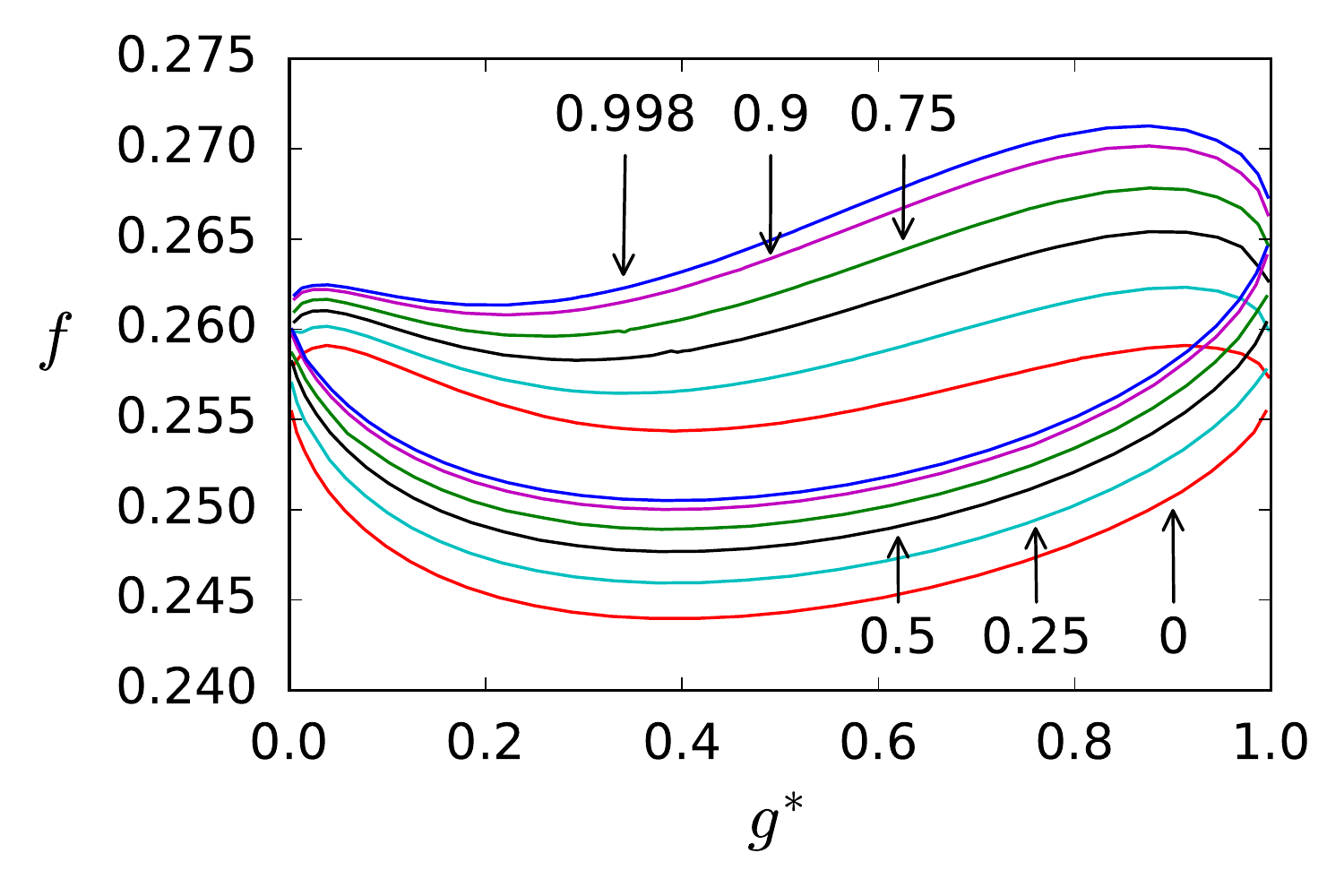}
\end{center}
\caption{Impact of the viewing angle $i$ and of the BH spin parameter $a_*$ on the transfer function $f$ in Kerr spacetime and for an infinitesimally thin accretion disk. In the left panel, $a_* = 0.998$, $r_{\rm e} = 4~r_{\rm g}$, and the values of the viewing angle $i$ are indicated. In the right panel, $i = 30^\circ$, $r_{\rm e} = 7~r_{\rm g}$, and the values of the BH spin parameter $a_*$ are indicated. From \citet{Bambi:2016sac}. \copyright AAS. Reproduced with permission.
\label{f-trf}}
\end{figure}

For an infinitesimally thin disk in the Kerr spacetime, for fixed emission radius $r_{\rm e}$ and viewing angle $i$, the transfer function is a closed curve parametrized by $g^*$ (except in the cases $i=0$ and $\pi/2$). There is only one point in the disk for which $g^* = 0$ and only one point for which $g^* = 1$. These two points are connected by two curves, corresponding to two branches of the transfer function, say $f^{(1)} (g^*,r_{\rm e},i)$ and $f^{(2)} (g^*,r_{\rm e},i)$. 
Eq.~(\ref{eq-flux}) thus becomes
\be
F_{\rm o} (E_{\rm o}) &=&
\frac{1}{D^2} \int_{R_{\rm in}}^{R_{\rm out}} \int_0^1 \frac{\pi r_{\rm e} g^2}{\sqrt{g^* (1 - g^*)}}
\, f^{(1)} (g^*,r_{\rm e},i) \, I_{\rm e} (E_{\rm e},r_{\rm e},\vartheta_{\rm e}^{(1)}) \, dr_{\rm e} \, dg^*
\nonumber\\ &&
+ \frac{1}{D^2} \int_{R_{\rm in}}^{R_{\rm out}} \int_0^1 \frac{\pi r_{\rm e} g^2}{\sqrt{g^* (1 - g^*)}}
\, f^{(2)} (g^*,r_{\rm e},i) \, I_{\rm e} (E_{\rm e},r_{\rm e},\vartheta_{\rm e}^{(2)}) \, dr_{\rm e} \, dg^* \, .
\nonumber\\
\ee
Fig.~\ref{f-trf} shows the impact of the inclination angle $i$ and of the BH spin parameter $a_*$ on the transfer function for a fixed emission radius $r_{\rm e}$.

If the disk has a finite thickness, a fraction of the inner part of the accretion disk may not be visible to the observer; for the values of $r_{\rm e}$ in which this happens, the transfer function is an open curve~\citep{Taylor:2017jep,2020arXiv200309663A}. In non-Kerr spacetimes, it is not guaranteed that the transfer function can be constructed as in the Kerr case, because we may not be able to use $g^*$ to parametrize the two curves connecting $g^* = 0$ with $g^* = 1$ (we may have more than two points with the same value of $g^*$ at some emission radii $r_{\rm e}$).

The function $I_{\rm e} (E_{\rm e},r_{\rm e},\vartheta_{\rm e})$ gives the shape and the normalization of the reflection spectrum at every point of the disk. The shape of the spectrum (in the rest-frame of the gas in the disk) is calculated by solving the problem of reprocessed X-ray radiation in an optically-thick medium illuminated by an external source.  The complexity of this problem involves the knowledge of the local radiation and the conditions of the gas at each point in the atmosphere, including its temperature, density, and ionization state. This requires (at least) the solution of: the energy equation, which provides the local temperature; the level population for all relevant ions, which determines the ionization state; and the radiative transfer equation, which describes the radiation field at every point in the atmosphere. These are coupled equations that need to be solved simultaneously, typically through an iteration procedure. The density of the gas is a fundamental parameter that affects all heating and cooling rates, and thus it must be also known. In general, the density of the accretion disk can be found through magneto-hydrodynamic (MHD) simulations. In practice, coupling the MHD equations with those listed above will make the computations unfeasible, and thus current models rely on simplifications. While some reflection calculations have considered hydrostatic atmospheres, most current models assume a constant density profile in the direction perpendicular to the disk surface. This approximation is likely to suffice in most cases because the effects of the reprocessed radiation take place in a relatively small portion of the upper layers of the disk, near the illuminated surface (most calculations are confined to a few Thomson depths). Moreover, inner accretion disks are expected to be magnetically supported, and thus density will remain nearly constant when magnetic pressure is dominant over gas pressure.

The normalization of the reflection spectrum is determined by the illumination pattern of the disk and, therefore, by the geometry of the corona. However, the geometry is highly uncertain and two  approximations are commonly used to cover this uncertainty. Namely, either the lamppost geometry is applied, or a phenomenological radial emissivity, e.g., a broken power-law (i.e., an emissivity $\varepsilon \propto 1/r^{q_{\rm in}}$ for $r < R_{\rm br}$ and $\varepsilon \propto 1/r^{q_{\rm out}}$ for $r > R_{\rm br}$, where $q_{\rm in}$ and $q_{\rm out}$ are the inner and the outer emissivity indices, respectively, and $R_{\rm br}$ is the breaking radius), is assumed, without any explicit assumption about the geometry and location of the primary X-ray source.

The simplest modeling  of relativistic reflection features uses the relativistic Fe K$\alpha$ line superimposed on a continuum (typically power-law) component. The first relativistic line models were  {\tt diskline} \citep{Fabian:1989ej}, valid for $a_*=0$, and {\tt laor} \citep{Laor:1991nc}, valid for $a_*=0.998$. Later models were {\tt kyrline} \citep{2004ApJS..153..205D}, {\tt kerrdisk} \citep{Brenneman:2006hw}, and {\tt relline} \citep{Dauser:2010ne}, and could compute the line profile for an arbitrary value of $a_*$. This approach to spectral modeling was used in the early reports of relativistically distorted spectra. It has two major drawbacks. First, the observed line profiles strongly depend on the assumed angular emissivity in the disk frame \citep{2004MNRAS.352..353B,Niedzwiecki:2007jy,2009A&A...507....1S}. The above models either intrinsically assume a specific angular law, e.g.\ $I_{\rm e} \propto 1 + 2.06 \cos \vartheta_{\rm e}$ used by \citet{Laor:1991nc} and adopted in most of the later models, or allow one to choose among some angular laws. All these angular prescriptions, however,  differ from the angular distribution of reflected radiation found in simulations using the most advanced reflection codes. \citet{2009A&A...507....1S} estimated that the uncertainty on the angular law leads to uncertainties of about 20\% on the estimate of the inner edge of the accretion disk (which, in turn, impacts on the BH spin parameter measurement). More importantly, the Fe K$\alpha$ line is only the most prominent line feature, but for the typical quality of current observations it is necessary to employ a model to fit even the other prominent (soft excess and Compton hump) and less prominent (other fluorescent emission lines in the soft X-ray band) features in the reflection spectrum.

The reflection spectra in the rest-frame of the accretion disk can be predicted by detailed radiative transfer calculations. The most advanced reflection models for the calculation of $I_{\rm e}$ are {\tt reflionx}~\citep{Ross:2005dm} and {\tt xillver}~\citep{Garcia:2010iz,Garcia:2013oma}. Another popular reflection model is {\tt ireflect}~\citep{MZ95}, which is a convolution model instead of a table model like {\tt reflionx} and {\tt xillver} and is currently the most accurate public model for the Compton hump. {\tt xillver} represents the state-of-the-art of reflection codes: it provides a superior treatment of the radiative transfer and an improved calculation of the ionization balance, by implementing the photoionization routines from the {\tt xstar} code~\citep{Kallman:2001zz}, which incorporates the most complete atomic database for modeling synthetic photoionized X-ray spectra. Tab.~\ref{t-r}  lists the typical parameters of reflection models like {\tt reflionx} or {\tt xillver}. We note that the ionization parameter of the accretion disk, normally indicated with the Greek letter $\xi$ and measured in units of erg~cm~s$^{-1}$, is defined as
\be
\xi = \frac{4 \pi F_X}{n_{\rm e}} \, ,
\label{xi}
\ee
where $F_X$ and $n_{\rm e}$ are, respectively, the X-ray flux from the corona and the electron density.

The full relativistic reflection spectra can then be calculated by applying relativistic convolution models to the rest-frame reflection spectra, as demonstrated in Eq.~(\ref{eq-flux}). The convolution kernels for the above line models are {\tt rdblur} (using {\tt diskline}), {\tt kdblur} ({\tt laor}), {\tt kyconv} ({\tt kyrline}), {\tt kerrconv} ({\tt kerrdisk}) and {\tt relconv} ({\tt relline}). These convolution models are normally applied to rest-frame reflection spectra calculated for $\vartheta_{\rm e} = i$. Tab.~\ref{t-c} lists the typical parameters of convolution models for an arbitrary corona geometry, like {\tt relconv}.

\begin{table}
{\renewcommand{\arraystretch}{1.3}
\begin{tabular}{cl}
\hline\hline
Parameter & Description \\
\hline
Photon index $\Gamma$ & Photon index of the continuum component \\ 
& illuminating the reflector and producing the reflection \\
& component \\
High energy cut-off $E_{\rm cut}$ & High energy cut-off of the continuum component \\ 
& illuminating the reflector and producing the reflection \\
& component. In some models it is replaced by the \\
& coronal temperature $T_{\rm e}$ \\
Ionization parameter $\xi$ & Ionization of the reflector \\
Iron abundance $A_{\rm Fe}$ & Iron abundance of the reflector \\
Emission angle $\vartheta_{\rm e}$ & Angle between the normal to the reflector and  \\
& the photon momentum \\
\hline\hline
\end{tabular}}
\vspace{0.0cm}
\caption{Typical parameters of reflection models like {\tt reflionx} or {\tt xillver}. Here we use the term ``reflector'' instead of ``disk'' because the model can be applied to any material illuminated by a continuum component. \label{t-r}}
\centering
\vspace{0.5cm}
\centering
{\renewcommand{\arraystretch}{1.3}
\begin{tabular}{cl}
\hline\hline
Parameter & Description \\
\hline
Inner emissivity index $q_{\rm in}$ & Inner emissivity index of the disk's emissivity \\
& profile: $\varepsilon \propto 1/r^{q_{\rm in}}_{\rm e}$ for $r_{\rm e} < R_{\rm br}$ \\ 
Outer emissivity index $q_{\rm out}$ & Outer emissivity index of the disk's emissivity \\
& profile: $\varepsilon \propto 1/r^{q_{\rm out}}_{\rm e}$ for $r_{\rm e} > R_{\rm br}$ \\
Breaking radius $R_{\rm br}$ & Breaking radius of the disk's emissivity profile \\
Spin $a_*$ & BH spin parameter \\
Viewing angle $i$ & Angle between the normal to the disk and the \\
& line of sight of the observer \\
Inner edge $R_{\rm in}$ & Inner edge of the disk $(R_{\rm in} \ge R_{\rm ISCO})$ \\
Outer edge $R_{\rm out}$ & Outer edge of the disk \\
\hline\hline
\end{tabular}}
\vspace{0.0cm}
\caption{Typical parameters of convolution models like {\tt relconv}. \label{t-c}}
\end{table}

Applying these relativistic kernels to a detailed rest-frame reflection spectrum is a large improvement to fitting a single relativistically broadened emission line. However, by design a convolution model is not able to account for the distribution of different emission angles $\vartheta_{\rm e}$ seen by an observer at a certain inclination $i$. Because of light bending in the strong gravity region, $\vartheta_{\rm e} \neq i$ and an observer will see a combination of reflection spectra emitted under different angles. \citet{Garcia:2013lxa} combined \texttt{relconv} and \texttt{xillver} to the angle-resolved model \texttt{relxill} and showed that, without angle-resolved calculations of the reflection spectrum, 
the estimate of some model parameters is affected by systematic uncertainties. The impact of the angle-averaged approach on the final measurements of the parameters of the system depends in a non-trivial way on the BH spin, disk ionization, and disk inclination angle, as well as on the quality of the data. The resulting bias in the final measurement can be negligible in some cases but not in others~\citep[see, e.g.,][]{Garcia:2013lxa,2014MNRAS.444L.100D,2016MNRAS.457.1568M,2019MNRAS.488..324I,2020arXiv200715914T}.

Therefore the most advanced relativistic reflection models directly combine reflection and relativistic smearing and also take the dependence on $\vartheta_{\rm e}$ correctly into account: {\tt relxill}\footnote{\url{www.sternwarte.uni-erlangen.de/~dauser/research/relxill/}} \citep{Dauser:2010ne,Dauser:2013xv}, {\tt kyn}\footnote{\url{https://projects.asu.cas.cz/stronggravity/kyn/}} \citep{2004ApJS..153..205D}, {\tt reflkerr}\footnote{\url{https://users.camk.edu.pl/mitsza/reflkerr}}~\citep{Niedzwiecki:2007jy,2019MNRAS.485.2942N}, and {\tt reltrans}\footnote{\url{https://adingram.bitbucket.io/reltrans.html}}~\citep{2019MNRAS.488..324I}. They all use the angle-resolved {\tt xillver} reflection spectra and are the most accurate models used in studies of relativistic reflection today. The first three are available in the phenomenological as well as the lamppost versions, while {\tt reltrans} works only with the lamppost set-up. Each of these families also has its own specific features addressing various aspects of spectral modeling. Importantly, while using different code architectures, the relativistic parts of these four model families are in good agreement\footnote{For the comparison of the line shape among {\tt relline}, {\tt kyn}, and {\tt reflkerr}, see Fig.~11 in \citet{2019MNRAS.485.2942N}. The agreement between the reflection spectra of {\tt relxill} and {\tt reltrans} is shown in Fig.~5 in \citet{2019MNRAS.488..324I}.}

{\tt relxill} is currently the most popular model among the four and {\tt xillver} is developed as a part of it, so updates in {\tt xillver} are first available in the  {\tt relxill}-family models. The models {\tt relxill}, {\tt kyn}, and {\tt reltrans} have the option to consider the ionization gradient in the disk, which can be determined self-consistently in the lamppost geometry. Both {\tt kyn} and {\tt reflkerr} models permit to take into account reflection from free-falling material in the plunging region within $R_{\rm ISCO}$. {\tt kyn} permits even to model specific portions of the disk or the obscuration of the disk by material crossing the line of sight~\citep{nprb18, mb20}.

{\tt reflkerr} takes into account several effects which may be important for the physical self-consistency of reflection models. The lamppost version, \texttt{reflkerr\_lp}, allows to include the X-ray emission of the source located on the opposite to observer side of the disk, which may be visible if the disk is truncated and whose contribution may be strongly enhanced by gravitational focusing  \citep{2018MNRAS.477.4269N}. The {\tt reflkerr\_elp} model \citep{2020A&A...641A..89S} extends \texttt{reflkerr\_lp} by taking into account the spatial extent of the X-ray source and then it allows to constrain the corona size as well as to study the effect of its rotation. The \texttt{reflkerr\_lpbb} model extends \texttt{reflkerr\_lp} by accounting for the quasi-thermal re-emission of the irradiating flux absorbed by the disk (see Section \ref{ss-rspec}). The {\tt reflkerr}-familiy models  implement the accurate model of  thermal Comptonization of \citet{PS96}, correctly describing spectra produced at relativistic electron temperatures, and then allowing for a proper modeling of gravitational redshift acting on the direct coronal radiation (in contrast, e.g., to the non-relativistic model, {\tt nthcomp}, used in the thermal Comptonization versions of {\tt relxill}). This is particularly important if the X-ray source is close to the horizon \citep{2020arXiv201008358S}, i.e., at the location needed to explain the very broadened reflection components. The {\tt reflkerr}-family models also  use a hybrid model of the rest-frame reflection, {\tt hreflect}, which improves the precision of reflection modeling above $\sim 10$~keV, by combining {\tt xillver} in the soft X-ray range with the exact model for Compton reflection of \citet{MZ95} in the hard X-ray range. The latest version of {\tt xillver}, which will be public shortly, also solves the problem at high energies with a new Comptonization routine~\citep{2020ApJ...897...67G}.

\subsection{Reverberation mapping}

The X-ray flux from X-ray binaries and AGN is routinely observed to vary across a wide range of timescales ($\sim [0.01-10](M/M_\odot)$ seconds) with fractional rms variability amplitude as high as tens of percent \citep[see, e.g.,][]{McHardy2006,McHardy2010}. Fluctuations in the coronal luminosity will first be seen in the direct continuum spectrum, and again after a light crossing delay in the reflection spectrum due to the longer path length travelled by the reflected radiation \citep[see, e.g.,][]{Uttley:2014zca}. This leads to variations of the flux in more reflection-dominated energy bands lagging behind those in more continuum dominated bands.

\begin{figure}[!h]
\centering
\includegraphics[width=8cm,trim=1cm 3cm 2cm 12cm,clip=true]{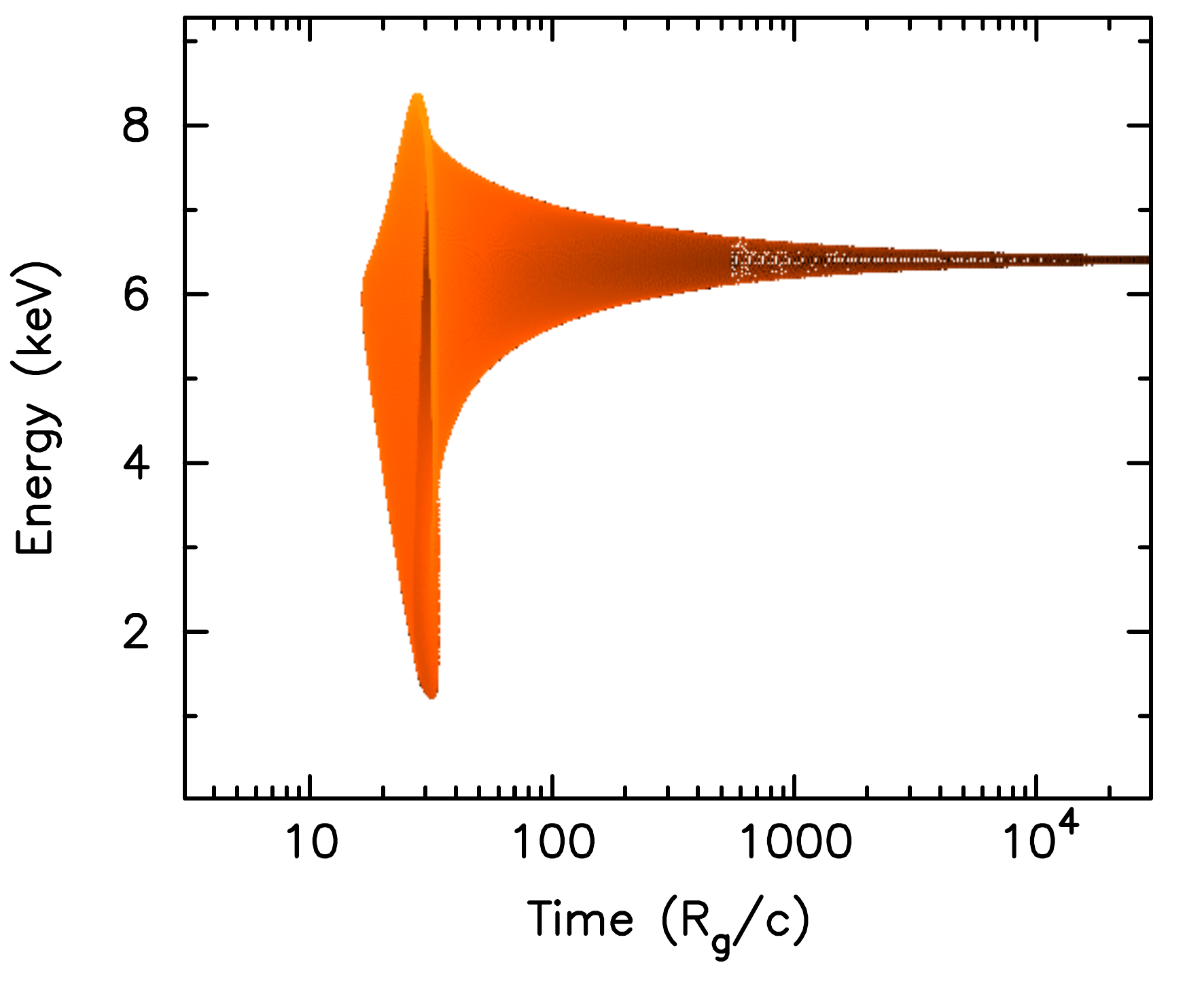}
\caption{Impulse-response function for a rest frame reflection spectrum consisting only of a narrow iron line at $6.4$ keV. The lamppost geometry has been assumed with a source height of $h=6~r_g$ and BH spin of $a_*=0.9$, the disk extends down to the ISCO and is viewed from an inclination angle of $i=70^\circ$.}
\label{fig:impresp}
\end{figure}

In the simplest case, the continuum spectrum is varying in normalization only: $F(E,t)=A(t) F(E)$. In this case, the time-dependent reflection spectrum can be represented as a convolution between the continuum normalization and an {\it impulse-response} function, $R(E,t)=A(t) \otimes w(E,t)$, where the impulse-response function is given by
\begin{eqnarray}
w(E,t) = \frac{1}{D^2} \int_{R_{\rm in}}^{R_{\rm out}} \int_0^1 \frac{\pi r_{\rm e} g^2}{\sqrt{g^*(1-g^*)}}~ f(g^*,r_{\rm e}) ~\delta( t - \tau(r_{\rm e},g^*) ) \nonumber \\
\times \mathcal{R}(E_{\rm e},r_{\rm e},\theta_{\rm e})~dr_{\rm e} dg^*.
\end{eqnarray}
Here, $\tau(r_{\rm e},g^*)$ is the extra time that photons reflecting from the disk patch with coordinates ($r_{\rm e}$, $g^*$) take to reach the observer compared with those that travel directly from the corona. $\mathcal{R}(E_{\rm e})$ is $I_{\rm e}(E_{\rm e})$ for $A=1$, and is therefore the normalized emergent reflection spectrum in the rest frame of the emission point. The impulse-response function describes the response of the reflection spectrum to a $\delta$-function flare in the luminosity of the corona, such that a $\delta$-function flare in the direct continuum spectrum would be observed at time $t=0$. Fig.~\ref{fig:impresp} shows an example impulse-response function calculated in the lamppost geometry considering only a narrow iron line instead of a full reflection spectrum. The first iron line photons we see after the continuum flash are those that reflected from the inner disk, and so the line is initially very broad and subsequently becomes narrower and dimmer as we see photons that reflected from progressively larger disk radii.

\begin{figure}[!h]
\centering
\includegraphics[width=12cm,trim=0cm 1.5cm 0cm 10cm,clip=true]{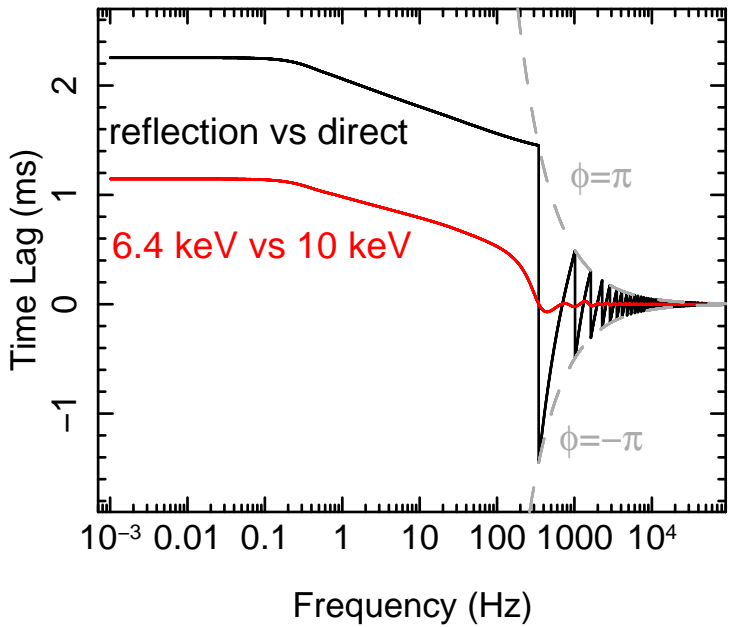}
\caption{Time lag as a function of Fourier frequency between reflected and directly observed photons (black line), and between 6.4~keV and 10~keV photons (red line), calculated using \texttt{reltrans}. The lamppost geometry is assumed with $h=6~r_{\rm g}$, $a_*=0.9$, $i=70^\circ$. The BH mass is $M=10~M_\odot$, the continuum spectrum has power-law index $\Gamma=2$ and high energy cut-off (in the observer's frame) $E_{\rm cut}=300$~keV and the disk ionization parameter is $\log\xi$ (independent of radius). The gray dashed lines mark time lags corresponding to a phase lag of $\pm \pi$ radians, for which the Fourier phase ambiguity causes phase wrapping.}
\label{fig:lagfreq}
\end{figure}

The Fourier transform of the time-dependent reflection spectrum can then be written very simply as $R(E,f) = A(f)W(E,f)$ due to the convolution theorem, where $f$ is Fourier frequency and the transfer function, $W(E,f)$, is the Fourier transform of $w(E,t)$. We can recover the Fourier frequency dependent time lag between the reflected and direct (continuum) flux from the transfer function. To do this, we first calculate the cross-spectrum between these two signals $R(f) F^*(f)$, where $R(f) = \int R(E,f) dE$ and $F(f) = \int F(E,f) dE$. The phase lag between reflected and direct components is given by the argument of this cross-spectrum, $\phi(f) = {\rm arg}[ R(f) F^*(f) ] = {\rm arg}[ W(f) ]$, where $W(f) = \int W(E,f) dE$. The time lag is then $t_{\rm lag}(f) = \phi(f) / (2\pi f)$.  Fig.~\ref{fig:lagfreq} (black line) shows this time lag calculated for the same parameters as were used for Fig.~\ref{fig:impresp}, and we have additionally assumed a $10~M_\odot$ BH. We see that at low frequencies, the lag is approximately constant at $\sim 2.2$~ms. This corresponds to $\sim 45~r_{\rm g}/c$, which is approximately the time it takes in Fig.~\ref{fig:impresp} for the bulk of the reflected flux to be observed after a sharp continuum flare. Above $f\sim 0.2$~Hz, the lag starts to drop off with frequency. This is because the total width of the impulse-response function is $\sim 5$~s $=1/ (0.2~{\rm Hz})$ (corresponding to $10^5~r_{\rm g}/c$ in Fig.~\ref{fig:impresp}). Therefore selecting variability timescales shorter than $\sim 5$ s will exclude the longest lags due to reflection from the very outer parts of the disk. The oscillatory structure above $\sim 300$~Hz is due to {\it phase wrapping}: the cyclical nature of the Fourier transform means that the measured time lag is equal to the true lag minus $n/f$, where $n \geq 0$ is an integer (see the gray dashed lines). This effect is similar to car wheels appearing to rotate backwards on a film because the frame rate of the camera is lower than the rotational frequency of the wheels.

To compare reverberation models to observational data, we must appreciate that we can never fully isolate the direct and reflected signals, we can only measure the time lags between photons in different energy channels. We can calculate this observable from the transfer function by first calculating the total specific flux, summed over direct and reflected components, which is $S(E,f) = A(f)[F(E) + W(E,f)]$. From this we can calculate the flux in some reference band, $F_r(t)$. The reference band could be defined as a narrow energy range, which is conceptually the simplest, or as a broad energy range, which results in a higher signal to noise ratio \citep{Uttley:2014zca,2019MNRAS.488..324I}. Defining the Fourier transform of the reference band flux as $F_r(f)$, the phase difference between fluctuations at energy $E$ and those in the reference band is $\phi(E,f)= {\rm arg}[ S(E,f) F_r^*(f) ]$, which can be written in terms of the transfer function as\footnote{Note that the $\arctan$ function must properly resolve the phase ambiguity associated with the $\tan$ function (e.g. typically the function \texttt{atan2(y,x)} in many programming languages).}
\begin{equation}
    \phi(E,f) = \arctan\left[ \frac{{\rm Im}W(E,f)}{F(E)+{\rm Re}W(E,f)} \right] - \phi_{\rm ref}(f).
    \label{eqn:tlag}
\end{equation}
Here, $\phi_{\rm ref}(f)$ is the phase lag between $F_r(t)$ and $A(t)$, which simply ensures that the phase lag between the reference band and itself is zero, and the time lag is again
$t_{\rm lag}(E,f) = \phi(E,f)/(2\pi f)$. The red line in Fig.~\ref{fig:lagfreq} shows the frequency dependent time lag between the 6.4~keV and 10~keV photons. The lag is positive at low frequencies because the ratio of reflected to direct flux is greater at 6.4~keV than it is at 10~keV. However, the amplitude of the lag is diluted compared with the black line because the observed flux at both energies contains a contribution from both direct and reflected emission. The red line will only converge to the black line in a hypothetical case whereby there is no continuum emission at 6.4~keV and  no reflected flux at 10~keV.

\begin{figure}[!h]
\centering
\includegraphics[width=10cm,trim=1cm 1.5cm 2cm 10cm,clip=true]{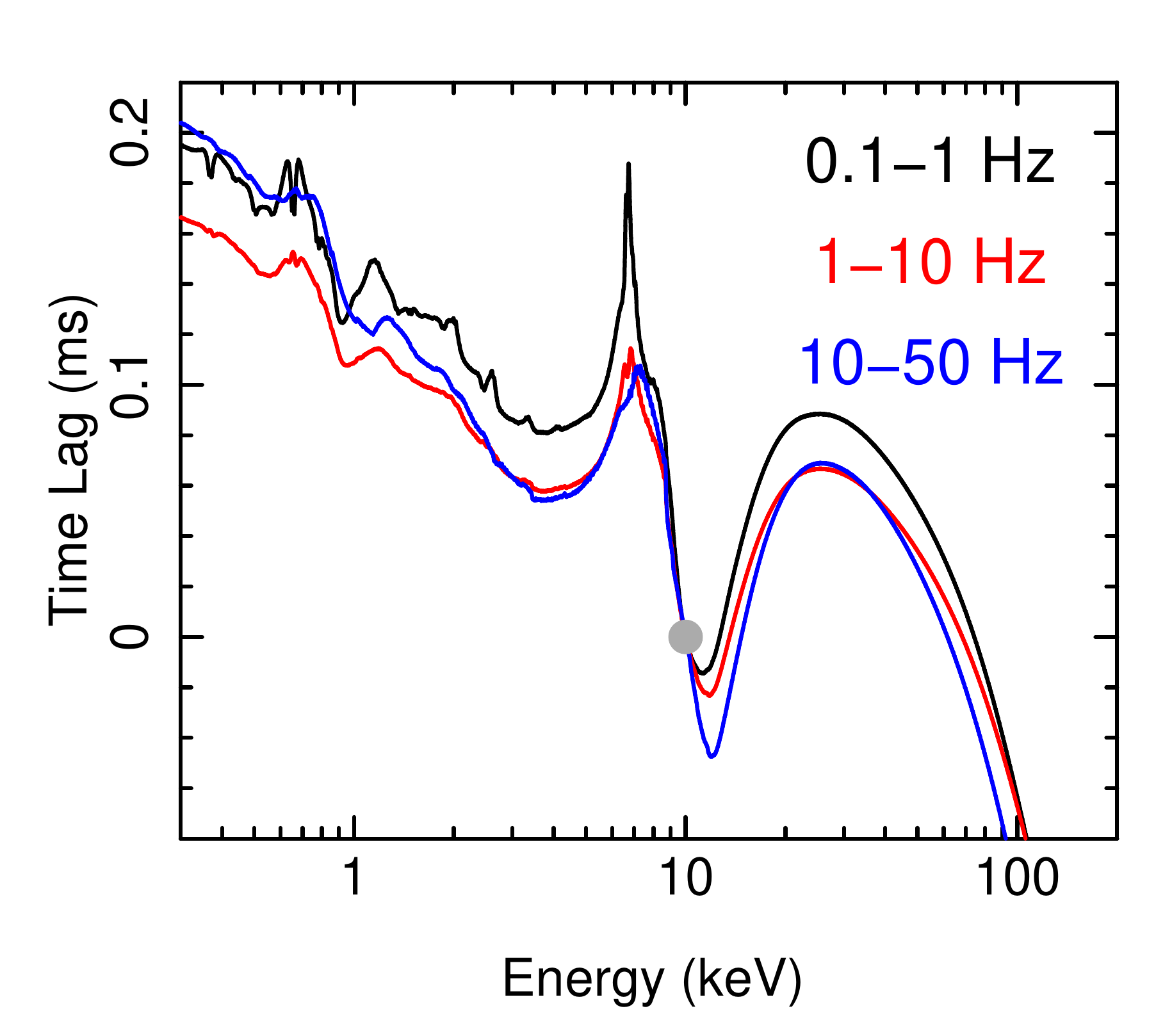}
\caption{Time lags relative to the 10~keV flux (lag-energy spectrum) calculated using \texttt{reltrans} for three different Fourier frequency ranges (as labelled). The model parameters are the same as for Fig.~\ref{fig:lagfreq}. The gray circle illustrates that the lag at 10~keV is trivially zero.}
\label{fig:lagmod}
\end{figure}

The time lag is dependent on the two energy bands chosen. This energy dependence can be explored by calculating the lags between many different energy channels and one common reference band, averaged over some frequency range. Fig.~\ref{fig:lagmod} shows an example of such a {\it lag-energy spectrum}, that has been calculated using \texttt{reltrans}~\citep{2019MNRAS.488..324I}. We see reflection features such as the iron line and Compton hump because the reverberation lag is less diluted in energy channels with a greater ratio of reflected to direct flux. We also see that the shape of the reflection features depends on Fourier frequency, which is because the fastest variability in the coronal luminosity is washed out in the reflection signal by light-crossing delays between rays that reflect from different parts of the disk. The amplitude of the reverberation lag increases linearly with BH mass (in the figure, we have assumed an $M=10~M_\odot$ BH). It is therefore possible to measure BH mass by fitting a reverberation model to the observed lag-energy spectrum.

\begin{figure}[!h]
\centering
\includegraphics[width=12cm,trim=1cm 0cm 0cm 0cm,clip=true]{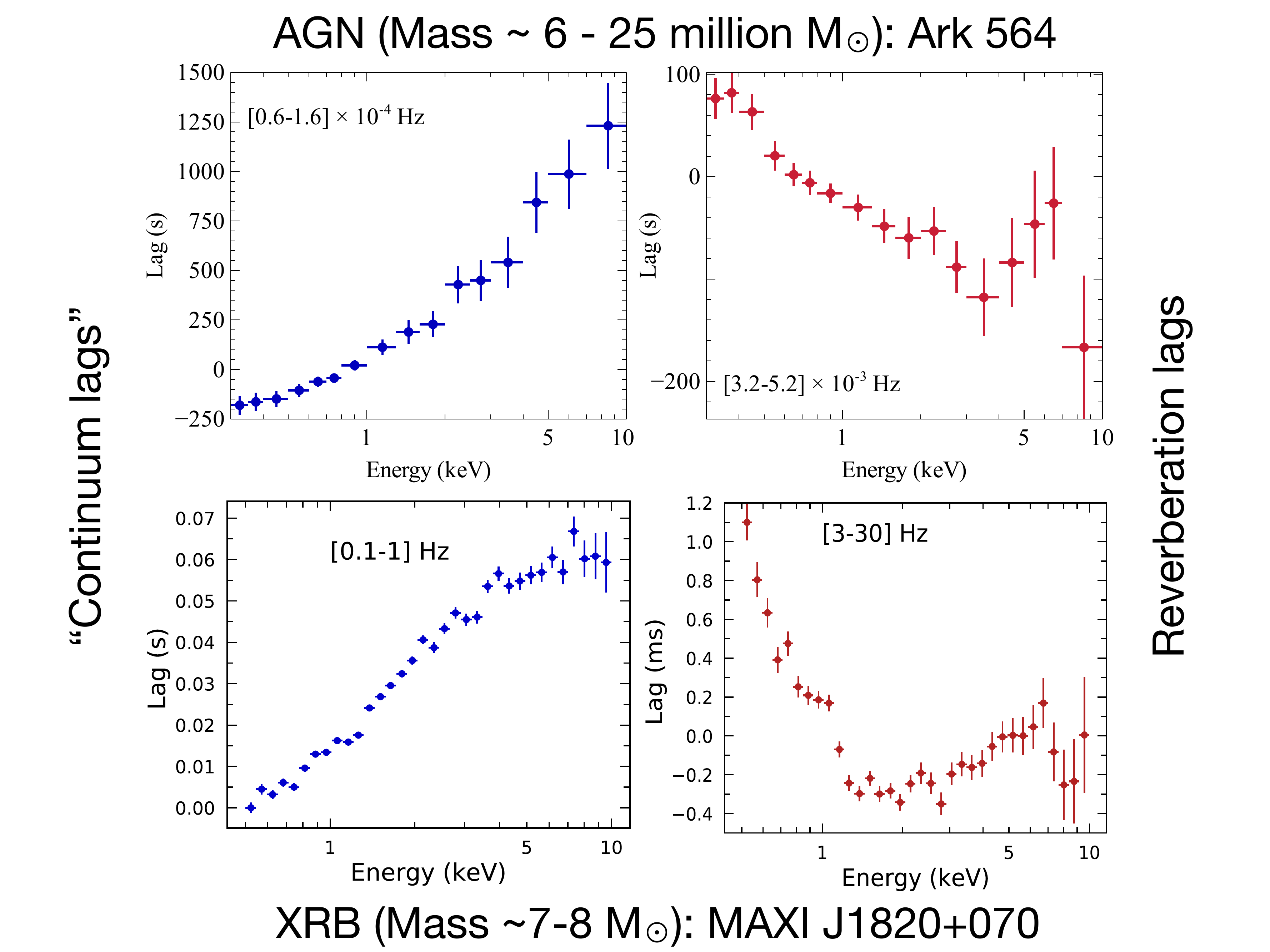}
\caption{Example observed lag-energy spectra of an AGN (top) and an X-ray binary (bottom) in a low (left) and high (right) Fourier frequency range for the system. Reverberation features are seen in the high frequency ranges (fastest variability timescales), whereas the low frequency ranges are dominated by much longer continuum lags.}
\label{fig:revobs}
\end{figure}

Fig.~\ref{fig:revobs} shows examples of lag-energy spectra observed for an AGN (top) and an X-ray binary (bottom) in a low (left) and high (right) Fourier frequency range. Note that, since all timescales of the system scale linearly with BH mass, what is considered a ``high'' or ``low'' Fourier frequency for a given system scales as $1/M$. We see an iron line feature in the high frequency range, whereas the low frequency range is dominated by featureless ``hard'' lags (meaning that hard photons arrive after soft photons; i.e., higher energies have a larger lag value). Since these lags are likely to be caused by spectral variability of the direct continuum rather than being associated with reverberation, they are often referred to as \textit{continuum lags}. Note that the continuum lags are much longer than the reverberation lags, which are very likely still present in the low frequency range but are completely dominated over by the much larger continuum lag signal. The amplitude of the continuum lags drops off with frequency \citep[$\propto \sim f^{-0.7}$; see, e.g.,][]{Nowak1999,2001ApJ...554L.133P}, and so the reverberation signal, which is roughly constant with frequency, can be detected in a high enough frequency range ($f \gtrsim (10~M_\odot/M)$~Hz). Iron line reverberation features have now been observed in many AGN \citep{kara16}, but only so far in one X-ray binary with high statistical confidence (\citealt{kara19}; but also see \citealt{DeMarco17}). This is because detection is more challenging in X-ray binaries due to the smaller lag amplitude and higher Fourier frequency range.

The continuum lags are thought to be caused by inward propagating fluctuations in mass accretion rate \citep{Lyubarskii1997, Kotov2001, Arevalo2006, Ingram2011, Ingram2013, Mushtukov2018}. In an extended corona, hard lags are caused by accretion rate fluctuations first propagating through the cooler outer region of the corona (with a softer spectrum) and then, after a viscous propagation time, through the hotter inner region (with a harder spectrum). Alternatively, hard lags can be generated in a compact corona by variable heating and cooling: fluctuations in the disk accretion rate first cool the corona via increased seed photon production, before propagating into the corona itself and heating it via increased gravitational dissipation (Uttley \& Malzac, in preparation). There have been attempts to model propagating fluctuations together with reverberation, but these models are either too computationally expensive to fit directly to observed data \citep{Wilkins2016}, or have employed a very simplified reverberation prescription \citep{Rapisarda2017}. It is more common to either only consider a reverberation-dominated frequency range (\citealt{Cackett2014,Chainakun2015,Chainakun2016,2019MNRAS.488..324I}) or incorporate a phenomenological model for the continuum lags (\citealt{Emmanoulopoulos2014,Chainakun2017,cg18,Mastroserio2018,al20}), typically assuming a lamppost corona for the calculation of reverberation lags. A number of authors have fit the time lags between two energy bands as a function of Fourier frequency by modeling the continuum lag amplitude as a power-law function of frequency and the reverberation signal with a transfer function (\citealt{Emmanoulopoulos2014,Epitropakis2016}; {\tt kynreverb}/{\tt kynrefrev}: \citealt{cg18}).

However, we note that the very presence of continuum lags contradicts the fundamental assumption of the transfer function formalism: that the continuum varies only in normalization and not in shape. \citet{Mastroserio2018} instead introduced fluctuations into the power-law index of the continuum spectrum; $F(E,t) \propto A(t) E^{1-\Gamma(t)}$, and linearized with a first order Taylor expansion (following \citealt{Kotov2001, Kording2004, Poutanen2002}). The \citet{Mastroserio2018} formalism accounts for the effect that fluctuations in the slope of the illuminating spectrum will have on the shape of the emergent reflection spectrum and has since been included in the {\tt reltrans} model \citep[][soon to be released publicly]{2019MNRAS.488..348M,Mastroserio2020}. The ``two blobs'' model of \citet{Chainakun2017} instead considers two lamppost sources, with fluctuations propagating from the cooler lamppost source to the hotter one (i.e., a very simplified propagating fluctuations model). Mathematically, this model is similar to {\tt reltrans}, in that continuum and reverberation lags are produced by the use of two transfer functions.

In the case of AGN, it is worth mentioning that a variable warm absorber can also produce lags between the light curve in the highly absorbed energy bands and the light curve in the continuum dominated energy bands~\citep[see, e.g.,][]{2019ApJ...884...26Z}. These lags are caused by the response of the warm absorber to changes in the ionizing flux, either by photoionization or radiative recombination \citep{2016A&A...596A..79S}. Simulations have shown that these lags can be of the order of hundreds of seconds at the relatively long timescales (low Fourier frequency) where the continuum lags dominate \citep{2016A&A...596A..79S}. They become negligible at shorter timescales, where the reverberation lags are detected. \citet{2015MNRAS.446..737K} found that lower energy light curves are lagging behind higher energy light curves at low Fourier frequency in NGC~1365. They argued that this unusual behavior of the low-frequency lags is due to a change in the column density of the absorber. It is important to model correctly these additional sources of lags in order to constrain the characteristic of the systems.

A number of X-ray reverberation studies have yielded BH mass measurements. Fitting the lag-energy spectrum for only one Fourier frequency range, or the frequency dependent lags between two energy bands, does not provide very tight mass constraints. This is because $M$ is degenerate (anti-correlated) with the source height in $r_g$: the same reverberation lag results from $h$ being many small $r_g$ or a few large $r_g$ \citep[e.g.][]{Cackett2014}. The degeneracy is further increased by phase-wrapping (e.g. \citealt{2019MNRAS.488..324I}). \citet{al20} broke this degeneracy by fitting the lags for many observations of the AGN IRAS~13224--3809 to find that the source height varied across observations, whereas the BH mass cannot. This yielded a BH mass measurement ($M=[1.9\pm 0.2]\times10^6~M_\odot$) of comparable accuracy to optical reverberation mapping, but is very expensive in terms of observing time. Alternatively, fitting the lag-energy spectrum for a number of frequency ranges simultaneously with the time-averaged spectrum breaks the $h/M$ degeneracy for a single observation. This is because the time-averaged iron line profile and the frequency dependence of the lag-energy spectrum (see Fig.~\ref{fig:lagmod}) both independently constrain $h$. Extra constraints are achieved by additionally fitting for the modulus of the cross-spectrum, which is related to the correlated variability amplitude, as \citet{2019MNRAS.488..348M} demonstrated by measuring the BH mass of Cygnus~X-1 with {\tt reltrans}. Their best-fitting mass of $M=26^{+9.6}_{-8.6}~M_\odot$ is higher than the earlier dynamical mass measurement of $M=14.8^{+1.0}_{-1.0}~M_\odot$ \citep{Orosz2011}. However, the dynamical mass measurement has recently been revised to $M=21.2^{+2.2}_{-2.2}~M_\odot$ after improvements to the radio parallax distance that agree with the \textit{Gaia} value were adopted \citep{Miller-Jones2021}, meaning that the reverberation mass now agrees remarkably well with the dynamical measurement. The \texttt{reltrans} mass measurement for the AGN Mrk~335 \citep[$1.1^{+2.0}_{-0.7} \times 10^6~M_\odot$][]{Mastroserio2020} is inconsistent with the existing optical reverberation measurements \citep[$M=14.2\pm 3.7 \times 10^6M_{\odot}$][]{pet04}, possibly indicating that the assumption of a lamppost geometry is incorrect.


\section{Current Status of Relativistic Reflection Modeling \label{s-mod}}

In order to understand the possible systematic uncertainties in the final measurements of the properties of accreting BHs, the assumptions and simplifications inherent in the theoretical models employed to fit the data need to be well understood. In the following, we will discuss the most recent developments in modeling, focusing on the assumptions and also limitations of the currently available reflection models.

\subsection{Reflection spectrum \label{ss-rspec}}

Theoretical models of X-ray reflection have been undergoing active development over the past three decades~\citep[see, for instance,][and reference therein]{2010SSRv..157..167F}. As of now, {\tt xillver} represents the state-of-the-art in reflection modeling~\citep{Garcia:2010iz,Garcia:2013oma}. Compared to earlier reflection codes, {\tt xillver} provides a superior treatment of the radiative transfer and an
improved calculation of the ionization balance by implementing the photoionization routines from {\tt xstar}~\citep{Kallman:2001zz}. The {\tt xstar} code includes the most complete atomic database for modeling synthetic photoionized X-ray spectra. The microphysics captured by {\tt xillver} is much more rigorous than any earlier code, mainly because of the detailed treatment of the K-shell atomic properties of the prominent ions~\citep{2004ApJS..155..675K,2005ApJS..158...68G,2009ApJS..185..477G}.

{\tt xillver} calculates reflection spectra assuming that the radiation from the corona illuminates the disk with an angle $\theta = 45^\circ$, while this angle depends on the coronal geometry and the position in the accretion disk~\citep[see, e.g., ][]{Dauser:2013xv}. As shown in the aforementioned reference, this incident angle strongly affects the reflection spectrum, which can be partly compensated by a change in (effective) ionization. Other reflection models, like \texttt{ireflect} or \texttt{reflionx}, assume that an isotropic X-ray source is located above a semi-infinite (i.e., subtending a $2\pi$ solid angle at the source) slab.

The {\tt xillver} model assumes Thomson scattering (with a first-order correction), which makes it quite
inaccurate at high energies and this causes a lack of energy conservation for hard incident spectra~\citep[but a new version fixing this problem should be released soon,][]{2020ApJ...897...67G}. Calculations also assume that the disk density is constant along the vertical direction of the disk, while we should expect that in a real disk the density increases as we move towards the disk center. The disk is assumed to be cold, in the sense that {\tt xillver} does not consider the thermal radiation from the disk when calculating the ionization equilibrium of the slab where ``reflection'' occurs. Such an approximation can be well justified for the X-ray spectral modeling of AGN, where the disks are found to have a temperature of up to $\sim 10$~eV. However, the ignorance of the thermal radiation from the disk can be important for $i)$ reflection-based spectral energy distribution modeling of AGN and $ii)$ X-ray spectral modeling of BHs in X-ray binaries where the disks are found to be hot. The impact of the additional thermal radiation has been calculated in the {\tt reflionx} model~\citep{ross07} and systematically studied for some sources; for example, for GX~339--4~\citep{Reis:2008ja}.

\begin{figure}
    \centering
    \includegraphics[width=11cm]{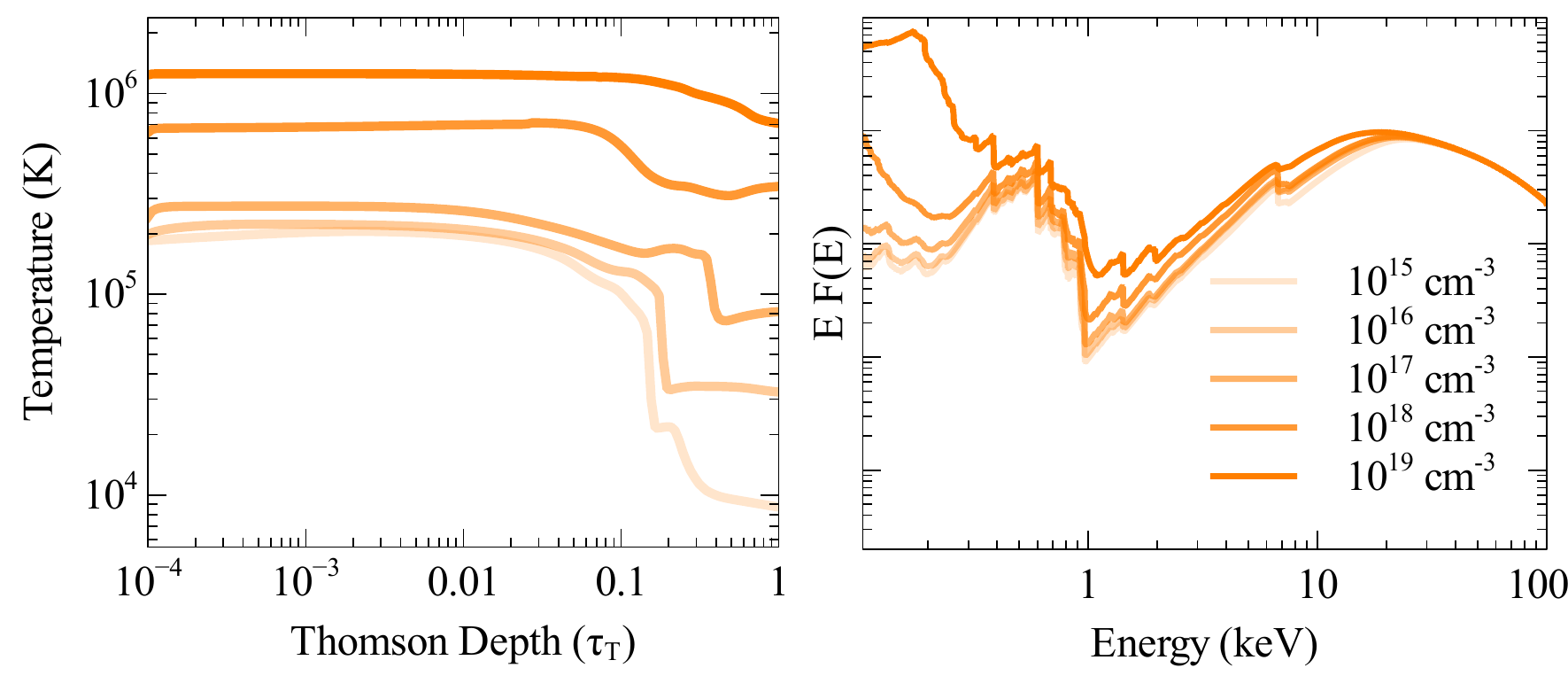}
    \caption{Temperature profiles of an illuminated slab under hydrostatic equilibrium (left panel) and relativistic disk reflection spectra (right panel) for different electron densities. The BH spin is $a_* = 0.998$, the disk inclination angle is $i = 30^\circ$, the ionization parameter is $\xi = 50$~erg~cm~s$^{-1}$, and a disk emissivity profile described by a power-law with emissivity index $q=3$ is assumed. From \citet{jiang19b}. }
    \label{f-s-density}
\end{figure}

Recent studies suggest that the number density within the optical depth of the disk has to be higher than the previous assumption of $n_{\rm e}=10^{15}$~cm$^{-3}$ in order to fit the broad band spectra of AGN and BH transients with disk reflection models \citep[e.g.,][]{jiang19a, jiang19b, 2020MNRAS.492.1947J}. At a higher disk density, the surface temperature of the disk is higher due to stronger free-free absorption \citep[][]{ross07,garcia16} and as a consequence of Eq.~(\ref{xi}) for a constant $\xi$ usually assumed, thus the reflection spectrum shows quasi-thermal emission in the soft X-ray band. Fig.~\ref{f-s-density} shows the impact of higher electron densities on the temperature profile of an illuminated slab and on the reflection spectrum of a disk. Higher disk density models are not only able to explain the broad band spectra of some AGN \citep[e.g.,][]{mallick18} but they can also significantly decrease the super-solar iron abundance inferred in previous reflection-based analyses \citep[e.g.,][]{tomsick18, jiang19b}. The standard thin disk model \citep[e.g.][]{Shakura73,svensson94} predicts an anti-correlation between the disk density parameter and the BH mass/accretion rate, and \citet{jiang19a} and \citet{jiang19b} find evidence for such an anti-correlation from their analysis of stellar-mass and supermassive BH spectra.

\begin{figure}
    \centering
    \includegraphics[width=9cm]{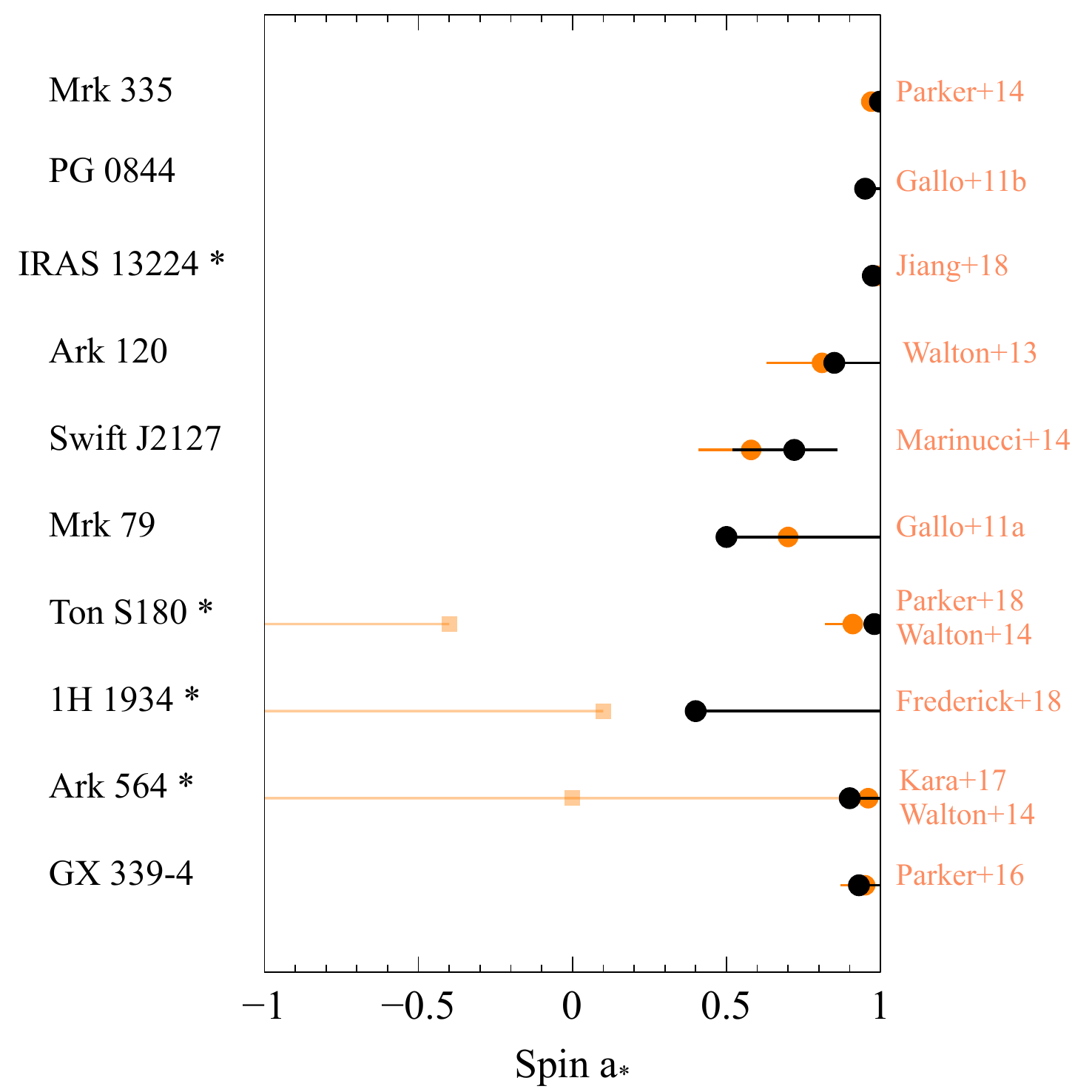}
    \caption{A comparison of spin measurements using high density disk reflection spectroscopy (black circles) with previous reflection based analyses (orange circles). Sources marked with stars are the cases where an additional component, e.g., a blackbody model or a soft power-law model, was used in previous analyses, based on which the spin measurements are shown by the orange squares. From \citet{jiang19b}.}
    \label{pic_high_ne_spin}
\end{figure}

Fig.~\ref{pic_high_ne_spin}~compares the spin measurements using high density disk reflection spectroscopy with previous analyses. Most spin measurements obtained by the high density model \citep[black circles in Fig.~\ref{pic_high_ne_spin}]{jiang19b} are consistent with previous results (orange circles). In a few cases, the results are not consistent. \citet{parker18} (Ton~S180), \citet{frederick18} (1H~1934--063), and \citet{kara17} (Ark~564) employed an additional model to fit the soft excess, e.g., a blackbody or a soft power-law, and found a lower or unconstrained BH spin (orange squares). \citet{walton13} analyzed the \textsl{Suzaku} spectra of Ton~S180 and Ark~564 using only a reflection model and found a consistent spin with the values given by the high density disk model. If the soft excess emission is not dominated by disk reflection, the spin measurements inferred from the full band spectra \citep{walton13,jiang19b} may indeed disagree with those obtained from the iron line alone \citep{kara17,parker18,frederick18}. For example, \citet{2020MNRAS.491.3553B} has suggested a hybrid model combining a warm corona with disk reflection to explain the soft excess and \citet{jiang19b} have shown that the warm corona and the high density disk reflection solutions provide equally good fits for the data of Ton~S180, so a spectral analysis alone does not seem to be able to select the correct model. The combined analysis of other data, like rms \citep[e.g.][]{parker20}, reverberation \citep[e.g.,][]{al20}, and polarization \citep[e.g.,][]{matt93} would help to distinguish the two scenarios.

From the energy conservation, the angle-averaged local reflection/reprocessing component has to contain the same flux as the incident spectrum. It contains two main components. One is due to the Compton backscattering, for which the albedo, $a$, depends on the ionization and the form of the incident spectrum, and is usually in the 0.3--0.7 range (e.g., \citealt{Zdziarski20}). The other contains
$(1-a)$ of the incident/irradiating flux, $F_X$, which is absorbed and re-emitted. The average energy of this part of the re-emitted spectrum is related to the effective irradiation temperature, $\sigma T_{\rm eff}^4=(1-a)F_X$. Its spectral form resembles that of a disk blackbody.

The irradiating flux can be in turn estimated from the source luminosity and geometry, and roughly $F_X\sim L/d^2$, where $d$ is the characteristic dimension of the reflector, e.g., the disk truncation radius. This yields reprocessed spectra peaking at an energy only slightly lower than the peak energy of the viscous blackbody, $E_{\rm peak}$ of equation (\ref{Epeak}). In the lamppost geometry, this component can be calculated using the {\sc xspec} routine {\tt reflkerr\_lpbb}, developed and presented in \citet{Zdziarski20b}. In luminous hard states of BH binaries, such features would clearly be expected if the disk extended close to the ISCO, but are not seen (e.g., in the accreting BH binary XTE J1752--223; \citealt{Zdziarski20b}). The reason that this does not appear as a discrepancy in most of published spectral fitting of BH binaries is basically numerical. Namely, the publicly available reflection codes assume low reflector densities and allow us to fit the ionization parameter $\xi$, defined in Eq.~(\ref{xi}). For typically obtained $\xi\sim 10^3$~erg~cm~s$^{-1}$ from fitting and $n_{\rm e}=10^{15}$~cm$^{-3}$ (assumed in {\tt xillver}), this yields the irradiating flux several orders of magnitude below that expected in BH binaries with $R_{\rm in}$ close to the ISCO, and even $n_{\rm e}=10^{19}$~cm$^{-3}$ available in {\tt relxillD} \citep{garcia16} is not sufficient, as shown in Fig.~1b of \citet{Zdziarski20}.

\subsection{Disk structure \label{ss-disk}}

As pointed out in Section~\ref{s-ref}, the available relativistic reflection models usually assume that the accretion flow is described by an infinitesimally thin Novikov-Thorne disk, that the inner edge of the disk is $R_{\rm in} \ge R_{\rm ISCO}$, and there is no emission of radiation in the plunging region $r < R_{\rm in}$. In reality, the accretion disk has a finite thickness, which increases as the mass accretion rate increases. The angular velocity of the gas may deviate from perfect Keplerian motion, for example because of the presence of magnetic fields. The inner edge of the accretion disk may be inside the ISCO for high values of the mass accretion rate. The plunging region can also contribute, at some level, to the total relativistic reflection spectrum.

The Novikov-Thorne model with $R_{\rm in} = R_{\rm ISCO}$ is normally thought to describe well the accretion disks of sources in the soft state and with an Eddington-scaled disk luminosity between $\sim$5\% to $\sim$30\%~\citep{Steiner10,Penna:2010hu,McClintock:2013vwa}\footnote{At lower luminosities, the disk can still be described by the Novikov-Thorne model, but it may be truncated ($R_{\rm in} > R_{\rm ISCO}$).}. However, relativistic reflection features are usually strong when a source is in the hard or intermediate states, not in the soft state. It is often difficult to obtain accurate measurements of the Eddington-scaled disk luminosity of a source, mainly because of large uncertainties in the measurements of the BH mass and distance. In practice, it is common to employ relativistic reflection models that assume infinitesimally thin Novikov-Thorne disks to fit any BH spectrum showing relativistic reflection features.

\citet{Taylor:2017jep} have recently proposed a simple framework to include the disk thickness in a Novikov-Thorne disk with $R_{\rm in} = R_{\rm ISCO}$, and the model has been further explored in \citet{2020arXiv200309663A} and \citet{2021arXiv210204695T}. The thickness of the disk alters the relativistic effects, as the emission is at a different point of the spacetime off the equatorial plane. Moreover, it can obscure a fraction of the inner part of the accretion disk, and the effect is larger for higher values of the mass accretion rate and of the viewing angle. The actual impact of the disk thickness on the best-fit values of the model parameters mainly depends on the BH spin, mass accretion rate, viewing angle, and disk emissivity profile in a non-trivial way. \citet{Taylor:2017jep} find that the values of the BH spin and of the coronal height are underestimated in simulations of a BH with $a_* = 0.9$ illuminated by a lamppost corona when the theoretical model employs an infinitesimally thin disk. On the contrary, \citet{2020arXiv200309663A} and \citet{2021arXiv210204695T} do not find much difference in the best-fit values of high-quality data of a few sources (GRS~1915+105, EXO~1846--031, and MCG--6--30--15) between the infinitesimally thin disk model and the model with disk of finite thickness, assuming a broken power-law for the disk's emissivity profile.

\begin{figure}[t]
\vspace{0.3cm}
\begin{center}
\includegraphics[trim={3.0cm 13.0cm 1.0cm 3.0cm},clip,width=1\columnwidth]{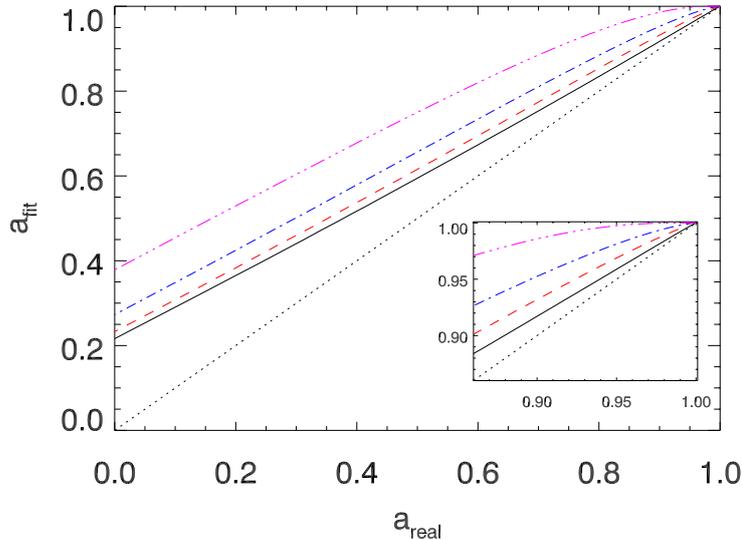}
\end{center}
\vspace{0.0cm}
\caption{Inferred BH spin parameter $a_{\rm fit}$ as a function of the actual BH spin parameter $a_{\rm real}$ from the study in \citet{Reynolds:2007rx} for a viewing angle $i = 30^\circ$ and different values of the vertical scale $h$ outside of the ISCO (from $h = 0.01$~$r_{\rm g}$, solid black line, to $h = r_{\rm g}$, dot-dot-dot-dashed magenta line). See \citet{Reynolds:2007rx} for more details. \copyright AAS. Reproduced with permission. \label{f-RF08}}
\end{figure}

\citet{Reynolds:2007rx} simulated geometrically thin accretion disks in a pseudo-Newtonian potential and estimated the impact of the disk thickness and of the radiation emitted from the plunging region on the BH spin measurement when the reflection spectrum is analyzed with an infinitesimally thin disk model. Within their set-up, they found that the final spin measurement is overestimated, which follows from the fact that the inner edge of the disk is not exactly at the ISCO radius in their simulations but it extends to slightly smaller radii, mimicking a lower value of the ISCO radius. Their findings are summarized in Fig.~\ref{f-RF08}, where we see that the systematic error on the BH spin measurement decreases as the actual BH spin parameter increases.

\begin{figure}[t]
\vspace{0.3cm}
\begin{center}
\includegraphics[width=0.70\columnwidth]{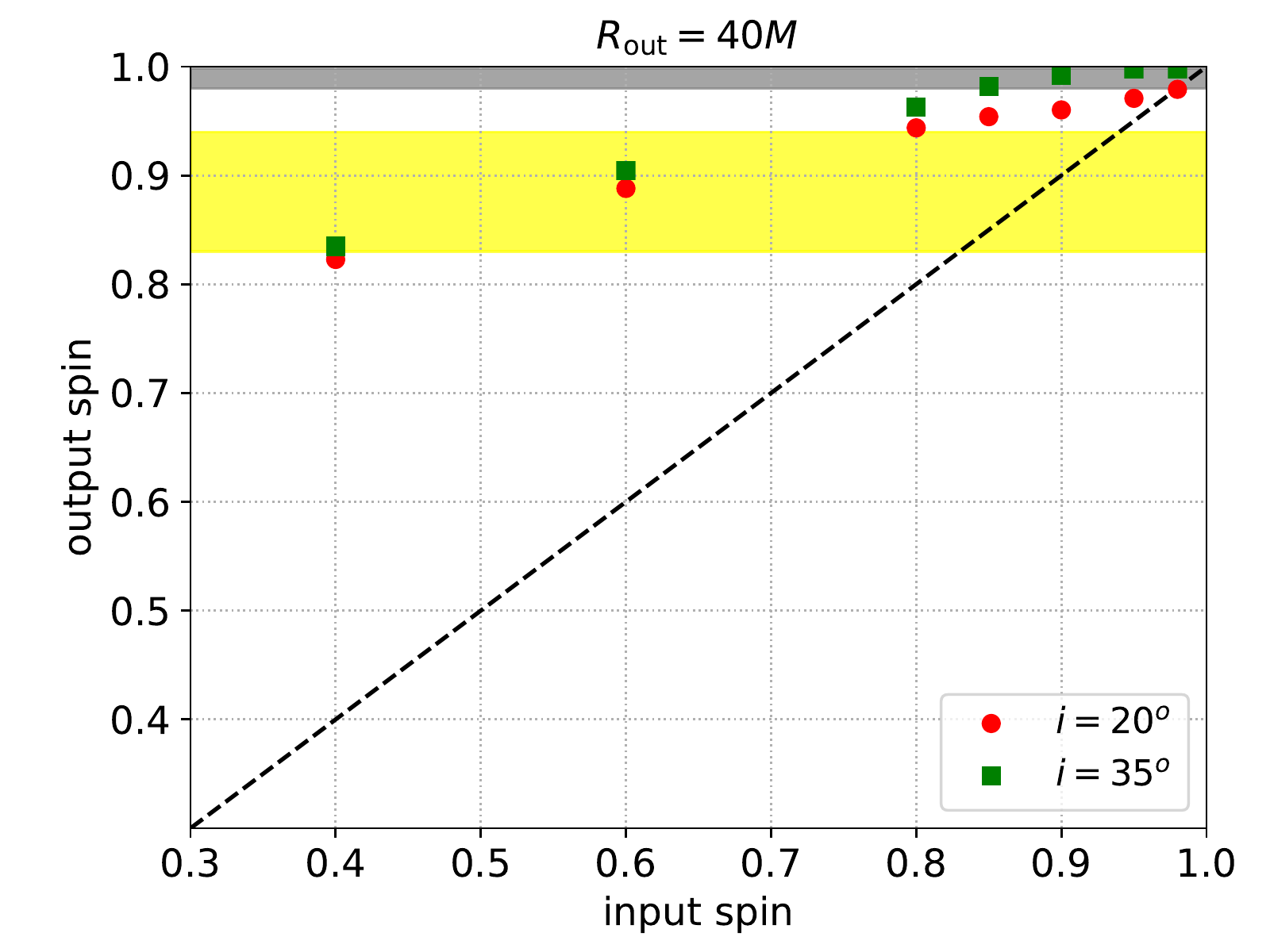} \\ \vspace{0.4cm}
\includegraphics[width=0.70\columnwidth]{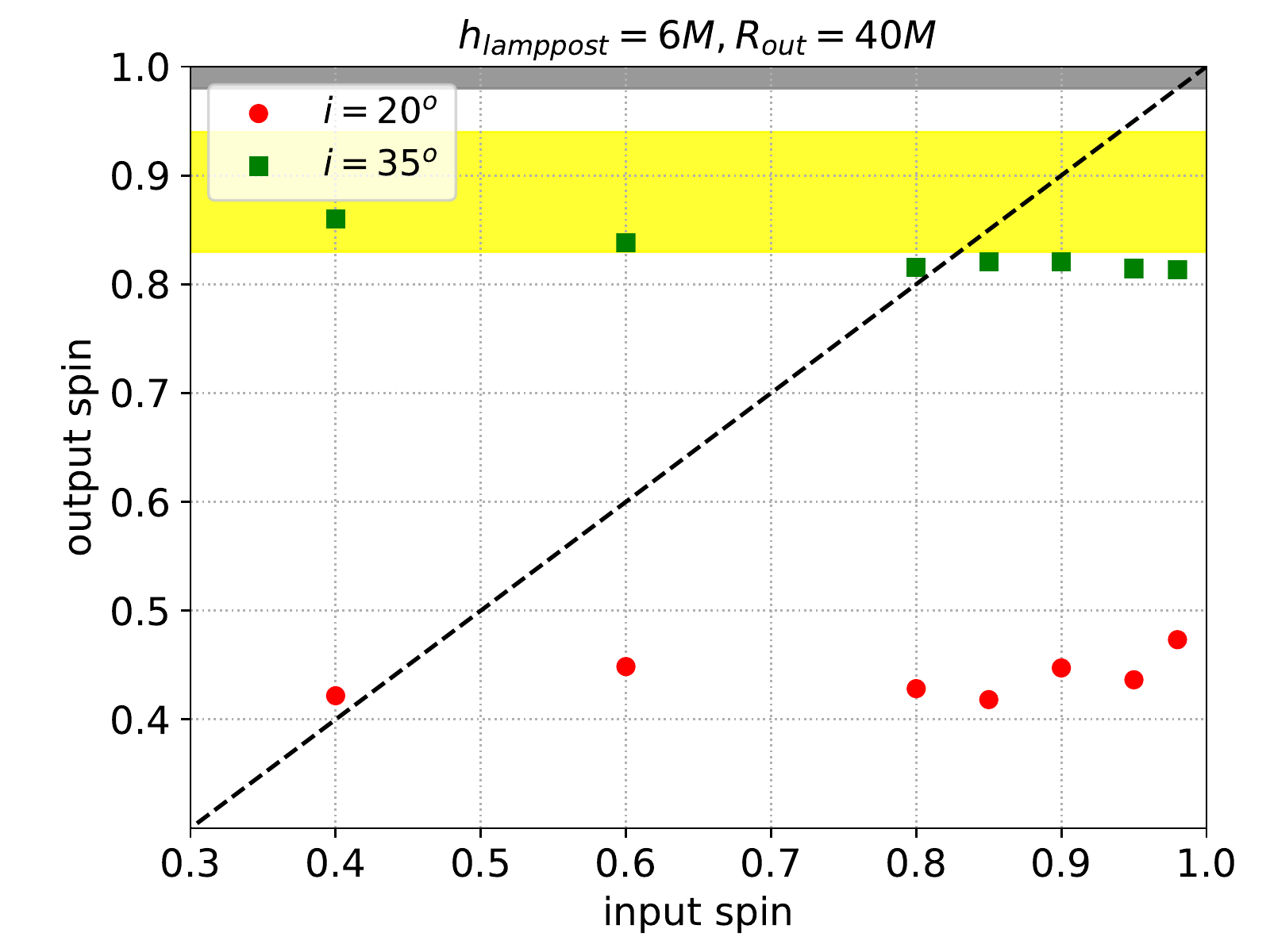}
\end{center}
\vspace{0.0cm}
\caption{Input spin parameter of the simulations vs. best-fit spin parameter from the study in \citet{Riaz:2019kat}. In the simulations, the accretion disk is described by the Polish donut model, the viewing angle is $i = 20^\circ$ (red circles) and $35^\circ$ (green squares), and the emissivity profile is either described by a power-law with emissivity index $q = 9$ (top panel) or that calculated for a lamppost corona with height $h = 6$~$r_{\rm g}$ (bottom panel). The simulated data are fitted with {\tt relxill}, in which the disk is approximated as infinitesimally thin. Statistical uncertainties in the spin parameter are too small to be reported. From~\citet{Riaz:2019kat}. \copyright AAS. Reproduced with permission. \label{f-shafqat}}
\end{figure}

\citet{Riaz:2019bkv} and \citet{Riaz:2019kat} simulated reflection spectra from thick accretion disks, which are expected in sources accreting around, or even above, their Eddington limit. This can be the case in at least some supermassive BHs in narrow-line Seyfert~1 galaxies~\citep{Boroson:1992cf,Grupe:2010me,Gliozzi:2010em,kara17}. They simulated observations with \textsl{NICER} and \textsl{NuSTAR}. While the theoretical model with an infinitesimally thin disk can fit the data well (in the sense that the reduced $\chi^2$ can be close to 1 and the data to best-fit model ratio does not show unresolved features), the final measurements can be incorrect. Fig.~\ref{f-shafqat} shows the discrepancy between the input value of the BH spin parameter and the inferred value by fitting the simulations with an infinitesimally thin disk model for two different emissivity profiles of the disk (power-law with emissivity index $q=9$ in the top panel and lamppost corona with height $h = 6$~$r_{\rm g}$ in the bottom panel) and two different viewing angles ($i=20^\circ$ for the red dots and $i=35^\circ$ for the green squares). Fig.~\ref{f-shafqat} shows that some configurations of the source may lead to precise but inaccurate spin measurements.

The impact of the radiation from the plunging region on the measurements of the model parameters has been recently investigated by \citet{Cardenas-Avendano:2020xtw}. Note that for a steady-state, axisymmetric, and geometrically thin disk, we can evaluate the optical depth in the plunging region as (in units $G_{\rm N} = c = 1$)~\citep{Reynolds:1997ek}
\be\label{eq-tau-pl}
\tau_{\rm e} = \frac{2}{\eta \left| u^r \right|} \left(\frac{r_{\rm g}}{r}\right)  
\left(\frac{L_{\rm d}}{L_{\rm E}}\right) \, ,
\ee
where $\eta$ is the radiative efficiency defined from the disk luminosity $L_{\rm d}$ and the mass accretion rate $\dot{M}$ via $L_{\rm d} = \eta \dot{M}$ and $u^r$ is the radial component of the 4-velocity of the gas in the plunging region. Eq.~(\ref{eq-tau-pl}) shows that the plunging region is optically thick except for very low values of the mass accretion rate: $\tau_{\rm e} > 1$ for $L_{\rm d}/L_{\rm E} > 0.05$ in the Schwarzschild spacetime ($\eta = 0.06$) and for $L_{\rm d}/L_{\rm E} > 0.01$ in the Kerr spacetime with $a_* = 0.998$ ($\eta = 0.3$). The plunging region can thus generate a reflection spectrum. However, the gas in the plunging region is highly ionized~\citep{Reynolds:1997ek,Wilkins:2020pgu} and therefore its spectrum looks like a power-law component: even if we employ a relativistic reflection model that ignores the reflection from the plunging region, the impact on the measurements of the model parameters is often modest~\citep{Cardenas-Avendano:2020xtw}.

In the case of sources with low mass accretion rates, the plunging region is optically thin, as we can see from Eq.~(\ref{eq-tau-pl}). In such a case, the total reflection spectrum of the accretion disk would receive contributions from radiation crossing the equatorial plane in the plunging region and generated either on the other side of the disk or circling the BH one or more time (higher order disk images)~\citep{Zhou:2019dfw}. The contribution of higher order image increases as the size of the plunging region and the viewing angle increase. If the inner edge of the disk is at the ISCO, the effect is maximum for fast-rotating BHs and counterrotating disks and minimum for fast-rotating BHs and corotating disks. For spin measurements with present and near future X-ray missions, the effect is likely always negligible~\citep{Zhou:2019dfw}.

The impact of magnetic fields on the disk structure and X-ray reflection spectroscopy measurements is still controversial and not much studied. \citet{2010ApJ...711..959N} and \citet{2012MNRAS.420..684P} have studied the deviations of a thin accretion disk from the Novikov-Thorne model in the presence of magnetic fields with numerical simulations, finding different results based on employing different magnetic field configurations. \citet{2010ApJ...711..959N} consider a model with a highly magnetized corona and find significant deviations from the Novikov-Thorne model. On the contrary, \citet{2012MNRAS.420..684P} find a negligible effect from magnetic fields on the disk structure within their set-up.

In Eq.~(\ref{xi}), $F_X$, and $n_{\rm e}$ should depend on the radial coordinate $r$, and therefore even the ionization parameter $\xi$ can be expected, in general, to be a function of the radial coordinate $r$. However, $\xi$ is often assumed to be constant over all radii because for most data the fit does not require any ionization gradient. This might occur when the spin of the source is very high, the inner edge of the disk close to the BH, and the corona is compact and low: in such cases, the incident radiation would be highly focused on a small portion of the disk, which can be approximated well by a one-ionization-region. {\tt relxill} and {\tt reltrans} provide the possibility of employing some non-trivial ionization parameter profile $\xi(r)$. \citet{Svoboda:2012cy} and \citet{2019MNRAS.485..239K} show that models with constant $\xi$ may lead to incorrectly estimate very high inner emissivity indices, while they do not find any significant bias in the estimate of the BH spin. 
\citet{2020MNRAS.492..405S}, on the other hand, analyzed a \textsl{NuSTAR} observation of GRS~1915+105 with {\tt reltrans} and found that the model version with a self-consistent ionization gradient returned a significantly smaller disk inner radius (and therefore higher BH spin) than the constant ionization version.
We also note that some studies fitted the data of some sources with a model with two relativistic reflection components \citep[e.g.,][]{2009Natur.459..540F,2015MNRAS.446..759C}, which can be seen as a simple way to approximate a non-trivial ionization gradient in the disk.

\subsection{Coronal geometry \label{ss-cgeo}}

As discussed in Subsection~\ref{ss-corona3}, the corona is some hot plasma near the BH, but its exact morphology is not yet well understood. Since the corona illuminates the disk to produce the reflection spectrum, its geometry is important because eventually determines the emissivity profile of the disk.

At least for some sources, we have clear evidence that the coronal geometry evolves and is different in different emission/spectral states of BH systems. It can thus be useful to consider an outburst of a transient BH binary to illustrate our current understanding of the disk-corona system.
At the beginning of the outburst, the mass accretion rate is low, the disk is truncated, and the low density ADAF (see Subsection~\ref{ss-ad-3}) between the inner edge of the disk and the BH may act as a corona~\citep{1997ApJ...489..865E,1998ApJ...505..854E,2001ApJS..132..377H,2007A&ARv..15....1D}. The outer cold disk/inner hot flow model has been used to explain the measured relatively long reverberation lags of soft X-rays responding to variable hard X-rays \citep{demarco15,DeMarco17, Mahmoud19} and the type C QPOs as precession of the inner hot disk (e.g., \citealt{Ingram16}). However, this explanation of the type C QPOs requires the corona to be geometrically thick compared with its effective Shakura \& Sunyaev viscosity parameter \citep[e.g.,][]{1983MNRAS.202.1181P,2019NewAR..8501524I}, which is not the case for all models of the corona \citep{2021ApJ...906..106M}.

Significant truncation in BH binaries is also implied by considering the re-emission of the X-rays absorbed by the disk \citep{Zdziarski20}, in agreement with the disk truncation found by reflection spectroscopy in some studies (e.g., \citealt{Plant15,Basak16,Dzielak19}). For a truncated disk, the relativistic distortion of reflection from the surrounding cold disk is weak and applications of reflection models allow to estimate geometrical parameters of the disk, like the truncation radius or the inclination angle, but not the BH spin.

However, while there is a common consensus on the fact the disk is truncated at the beginning of the outburst when the mass accretion rate is low, the truncated disk picture is more controversial at high mass accretion rates. Numerous spectral fits in luminous hard states found that the disk inner radius can be very close to the ISCO (e.g., \citealt{2008ApJ...679L.113M,Tomsick08, Reis:2008ja, Reis10,Miller:2013rca,Miller:2014sla,Fuerst:2015ska, Garcia15, Garcia18, Garcia19}).
\citet{kara19} study hard state \textsl{NICER} data of the BH binary MAXI~J1820+070 and conclude that the inner edge of the accretion disk is close to the BH and does not change appreciably over a period of weeks, suggesting that the disk is not truncated \citep[see, however,][for a different interpretation of the same data]{2021arXiv210207811D}. It is possible that, as the mass accretion rate increases, the inner edge of the disk moves to the ISCO radius and the size of the corona decreases~\citep{1997ApJ...489..865E,1998ApJ...505..854E}. If at high mass accretion rates there is no low density advection dominated accretion flow to serve as a corona, the base of jets, which are common either in the hard and intermediate states, may act as a new corona with a lamppost geometry~\citep{markoff05}.

Current spin measurements in the hard state are thus accurate only if the inner edge of the disk is at the ISCO radius. Note, however, that for several sources we have measurements of the BH spin very close to the maximum spin value (see Section~\ref{s-spin}), and in such cases we can get the same estimate of the BH spin even if we do not assume the inner edge at the ISCO and we leave it free in the fit.

If we know the coronal geometry, we can determine the corresponding emissivity profile \citep[see, for instane,][]{2012MNRAS.424.1284W,2017MNRAS.472.1932G}, while the contrary is not so straightforward: if we measure the emissivity profile it may be difficult to infer the coronal geometry because very similar emissivity profiles can be generated by quite different coronal geometries \citep{2017MNRAS.472.1932G}. However, this also implies that, if we do not know the coronal geometry, we can approximate the emissivity profile with some phenomenological model. 
As described in Section \ref{s-ref}, as an empirical approximation for the emissivity profile of coronae of uncertain geometry it is common to employ a broken power-law emissivity in the fitted models. Typically, very steep emissivities are found for the innermost accretion disk~\citep[e.g.,][]{Blum:2009ez,El-Batal:2016wmk,2020arXiv200704324D}.

The lamppost geometry was first considered by \citet{1996MNRAS.282L..53M} for the explanation of an enhanced irradiation of the innermost accretion disk. The lamppost geometry models the corona as an X-ray source at a height $h$ above the rotational axis of the BH. While this is a very simplified description, it can easily produce very steep emissivities, which are often inferred by spectral fitting, when the corona is compact and its height $h$ is very low, at the level of some gravitational radii.
The basic version of the lamppost model assumes that the source is point-like, exactly along the BH spin axis, and has a perfectly constant and isotropic emission of radiation. Depending on the actual coronal properties and quality of the available X-ray data, in some cases these assumptions may oversimplify the actual astrophysical system and therefore the estimate of some model parameters should be treated with care. For example, coronae very close to the BH -- as it is required in spin measurements, see Section \ref{s-spin} -- would be strongly affected by the runaway $e^\pm$ pair production and the BH photon trapping~\citep{2016ApJ...821L...1N}, which motivates the development of more physically consistent models.

For the lamppost geometry, we can calculate the expected emissivity profile by solving photon trajectories from the corona to the disk and the spectrum of the incident radiation at every disk point can be properly calculated. For example, in the simple case of a lamppost corona with spectrum described by a power-law component (photon index $\Gamma$) with exponential cut-off (high energy cut-off $E_{\rm cut}$), we can exactly calculate the redshift factor between the corona and every point on the accretion disk. Such a redshift renders every point on the accretion disk effectively illuminated by an incident spectrum with the same $\Gamma$ but high energy cut-off $E_{\rm cut}' = g E_{\rm cut}$, where $g$ is the redshift factor between the corona and the disk point, $g = E_{\rm disk}/E_{\rm lamppost}$.

The calculation of the redshift between a photon energy in the corona's rest-frame and the corresponding photon energy in the observer's rest-frame is also straightforward and is (in units $G_{\rm N} = c = 1$)
\be\label{eq-g-c-o}
\frac{E_{\rm o}}{E_{\rm lamppost}} = \sqrt{ 1 - \frac{2 M h}{h^2 + a_*^2 M^2} } \, ,
\ee
where $h$ is still the coronal height. We note that the parameters $E_{\rm cut}$/$T_{\rm e}$ in {\tt relxilllp}/{\tt relxilllpCp} and in {\tt reltrans} still refer to the redshifted values in the observer's rest-frame, which are then blueshifted according to Eq.~(\ref{eq-g-c-o}) to infer their original values at the emission point in the corona.
On the contrary, in {\tt reflkerr\_lp} the parameter $T_{\rm e}$ refers to its value in the corona's rest-frame. In {\tt kynxillver}, the {\tt kyn} model using {\tt xillver} tables, the parameter $E_{\rm cut}$ refers to the observed value if positive and to the value at the source if negative\footnote{In the case of a power-law spectrum with an exponential cut-off, different definitions of $E_{\rm cut}$ (i.e. in either the observer's or the source frame) give essentially the same physical results (if all redshift corrections are self-consistently taken into account) because $E_{\rm cut}$ may be simply rescaled by the redshift factor. On the other hand, in the case of a thermal Comptonization continuum, using the temperature defined in the observer's frame may lead to invalid results because the redshifted Comptonized spectrum is not, in general, correctly reproduced by a simple scaling of electron temperature.}.

In the models employing a phenomenological radial emissivity, there is no assumption about the coronal geometry, and therefore we cannot accurately calculate the gravitational redshift between the corona and the observer/disk. {\tt relxill} ({\tt relxillCp}) assumes a constant high energy cut-off $E_{\rm cut}$ of the continuum (coronal temperature $T_{\rm e}$) at each point of the corona (of unknown geometry) and the redshift of the direct radiation from the corona to the observer is neglected. {\tt reflkerr} also assumes a constant coronal temperature $T_{\rm e}$ at each point of the corona (of unknown geometry), but the redshift to the observer is taken into account assuming that the coronal fractional scale height is low.

A major advantage of the lamppost model is that it allows to estimate the distance of the X-ray source, meaning  at which the light-bending and other GR effects may account for the enhanced irradiation of the inner disk, needed to explain some extremely distorted reflection spectra \citep{Dauser:2013xv}.

Additionally, using the assumption in the geometry it is now possible to predict strength of the reflected and the directly emitted and observed radiation. As shown in \citet{2014MNRAS.444L.100D}, using this prediction can yield much more precise constraints (but, of course, inaccurate if the assumed geometry is wrong) on BH parameters like the spin. Typically, the models allow to scale this predicted \emph{reflection fraction} to be able to partly compensate for the assumption of an intrinsically isotropic corona.
In \texttt{relxill}, we have the parameter \texttt{refl\_frac}, which corresponds to the reflection fraction itself. In the other models, we have a parameter quantifying the deviation of the reflection fraction expected in the case of a stationary, isotropic, on-axis lamppost source: it is called \texttt{rel\_refl} in \texttt{reflkerr}, \texttt{Np:Nr} in \texttt{kyn}, and {\tt boost} in {\tt reltrans}. The exact lamppost model is recovered when these parameters are equal to 1 and there are deviations from the lamppost predictions if their value is larger or smaller than 1. For example, an out-flowing corona would correspond to a fitted reflection fraction less than the value expected in the lamppost model (i.e., \texttt{rel\_refl}, \texttt{Np:Nr}, and {\tt boost} $<1$). There are many other physical reasons for the reflection fraction to deviate from that expected from the lamppost model; for example isotropic emission is not expected for Compton scattering in a non-spherical medium, and the reflection fraction for an extended corona will be different from that of a lamppost source. It is important to note that any deviations from an isotropic, stationary lamppost corona will not be fully captured by a free reflection fraction parameter, since they will also change the radial irradiation profile. Therefore, if the fitted reflection scaling strongly deviates from that expected from the lamppost model, the spectral model is not self-consistent, which possibly indicates that the emission is not isotropic or some additional physical effect must be taken into account.

A number of extended versions of the lamppost geometry are available in different models. For example, within the {\tt relxill} package it is possible to consider the case of a point-like or elongated source moving along the BH spin axis, as we could expect if the corona is the material of the jet~\citep{Dauser:2013xv}. In {\tt reflkerr\_elp}, an extended corona corotates with the accretion disk at the ISCO~\citep{2020A&A...641A..89S}. The corona has also been modeled with rings or disks above the BH accretion disk, and the intensity profiles of such coronal geometries seem to explain well at least some data~\citep{Miniutti:2003yd,2006A&A...453..773S,2012MNRAS.424.1284W}.

Phenomenological emissivity profiles (typically a power-law or a broken power-law), which are commonly applied in reflection models, are able to crudely approximate some geometrical configurations. For example, a twice broken power-law can approximate the emissivity of a ring-like or an extended, uniform corona~\citep{Miniutti:2003yd,2006A&A...453..773S,2012MNRAS.424.1284W,2020A&A...641A..89S}. However, wrong conclusions about the corona geometry can be inferred if the emissivity is treated as a free parameter (neglecting constraints imposed by the source geometry) and the fitted empirical parameters are incorrectly related to the geometrical parameters \citep[see discussion in][]{2020A&A...641A..89S}.

Some observations clearly prefer the lamppost scenario over the phenomenological broken power-law intensity profile~\citep[e.g.,][]{Parker:2016ltr}, while in other cases it is the latter to provide a better fit~\citep[e.g.,][]{El-Batal:2016wmk}. In some observations, the two models provide a comparable fit, but the measurement of some parameters may be different, which means that the quality of the data is unsuitable to distinguish the two cases~\citep[e.g.,][]{Miller:2014sla,Xu:2017yrm}. 
In the case of high-quality data, the quality of the fit should tell us which emissivity profile model is more suitable for the description of our astrophysical system, and the fits with other emissivity profiles may return us incorrect measurements of some parameters of the model~\citep[see, e.g., the discussion in][]{Zhang:2019ldz,2019PhRvD..99l3007L}.

Good constraints on the corona can be obtained by combining the spectral and the frequency domain information, which allowed to estimate the evolution of a compact X-ray corona in a BH transient \citep{kara19}. This approach returns very small heights of the X-ray source above the disk ($\simeq 2$--5~$r_{\rm g}$)  in several AGN \citep{2015MNRAS.452..333C,cg18,al20}, and disfavor the geometry with magnetic flares corotating with the Keplerian disk \citep{2005MNRAS.359..308Z}.

Two coronae may coexist at the same time, which inevitably increases the complexity of the problem. Models with a hot and a warm corona have been proposed~\citep[see, e.g.,][]{Petrucci:2017niz,2020MNRAS.491.3553B} and the possibility of the presence of two coronae, with two different photon indices, has been advocated even as a solution of the well known problem of inferred high iron abundance common to several sources~\citep{Fuerst:2015ska}.

\subsection{Returning radiation \label{ss-return}}

The {\it returning radiation} is the radiation emitted by the disk and returning to the disk because of the strong light bending in the vicinity of the BH. This radiation illuminates the disk again and produces a ``secondary'' disk (thermal or reflection) spectrum. All current relativistic reflection models used in data analysis ignore such a radiation in the calculation of synthetic reflection spectra.

\begin{figure}
    \centering
    \includegraphics[width=6.5cm]{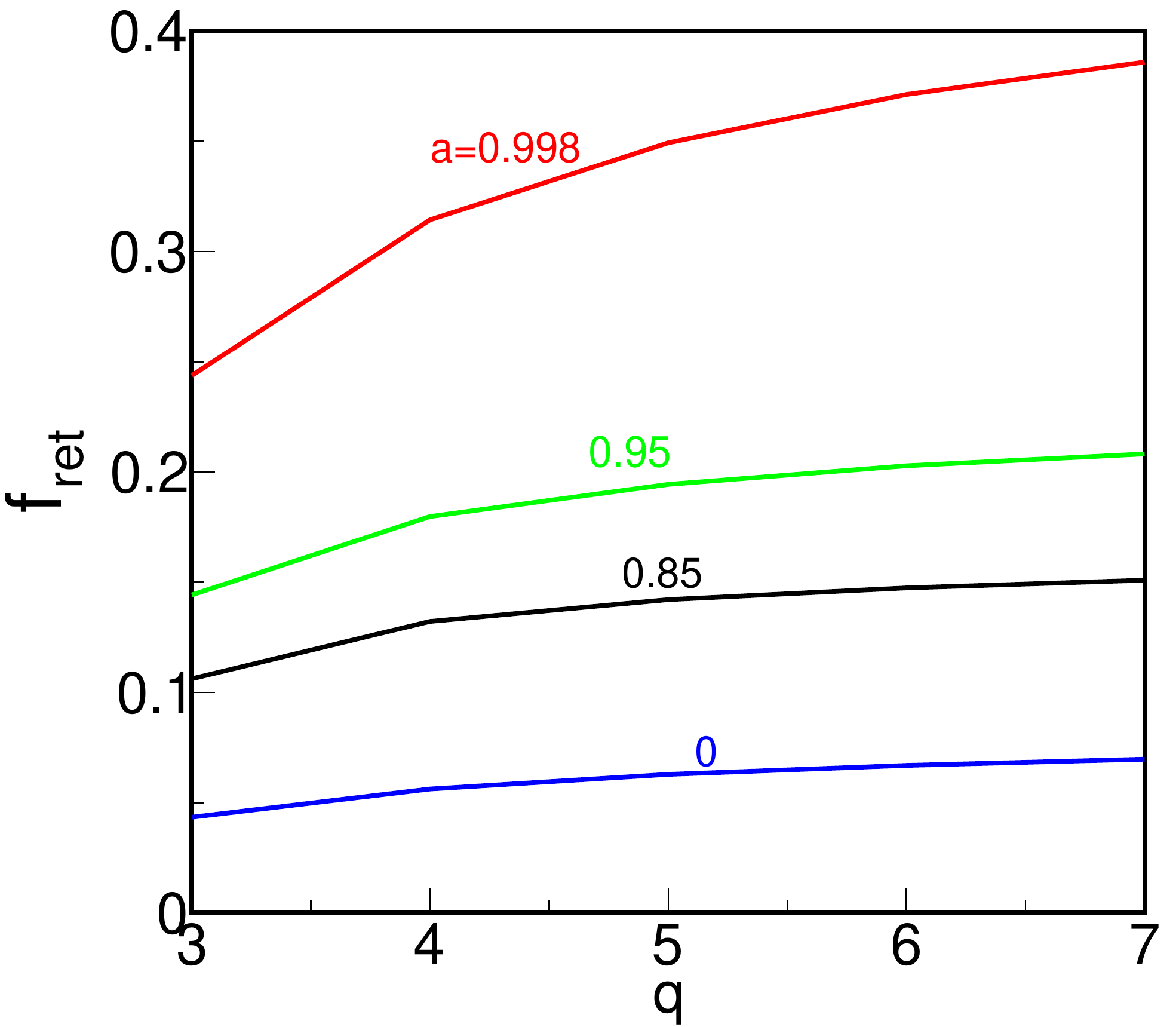}
    \caption{Fraction of radiation returning to the disk as a function of the emissivity index $q$ of the power-law irradiation profile for various values of the BH spin parameter $a_*$. From~\citet{2020arXiv200615838R}. \copyright AAS. Reproduced with permission.}
    \label{f-ret0}
\end{figure}

The light bending is strong enough to return a significant fraction of radiation only in the inner few gravitational radii~\citep{1976ApJ...208..534C,2000ApJ...528..161A,2020arXiv200615838R}. Therefore, the returning radiation is more important for rapidly rotating BHs and for parameters yielding an enhanced emission from the innermost disk, i.e., in the cases when the BH spin measurements are possible, see Section~\ref{s-spin}. Fig.~\ref{f-ret0} shows the fraction of radiation returning to the disk as a function of the emissivity index of the power-law irradiation profile for various values of the BH spin parameters (assuming that the inner edge of the disk is at the ISCO). Note, however, that \citet{2020ApJ...892...47C} interpret the spectrum of XTE~J1550--564, which is thought to be a slow-rotating BH \citep[$a_* \approx 0.5$,][]{2011MNRAS.416..941S}, with a reflection component affected by returning radiation. The returning radiation should be important for a small height \citep[at most a few $r_{\rm g}$;][]{2018MNRAS.477.4269N} of the X-ray source in the lamppost geometry  and for a steep radial emissivity in phenomenological models \citep{2020arXiv200615838R}. For the radial thermal emissivity of a standard accretion disk (i.e., described by the Novikov-Thorne model) the contribution from the inner disk is relatively weak and, therefore, the effect of the returning radiation gives only a small correction, but it can be strongly increased if the non-zero stress inner boundary condition is applied \citep{1997MNRAS.288L..11D,2000ApJ...528..161A}.

The returning radiation effect was first pointed out by \citet{1976ApJ...208..534C}, who included it into the calculation of the thermal emission spectrum of an accretion disk, and found that it can distort the observed spectrum only for rapidly rotating BHs. However, this distortion of the thermal spectrum can be reabsorbed in a higher mass accretion rate in BH spin measurements using the continuum-fitting method~\citep{2005ApJS..157..335L}. In future, the effect of returning radiation may be important in polarization measurements of the thermal component of the disk~\citep{2009ApJ...701.1175S}.

\begin{figure}
    \centering
    \includegraphics[width=11cm]{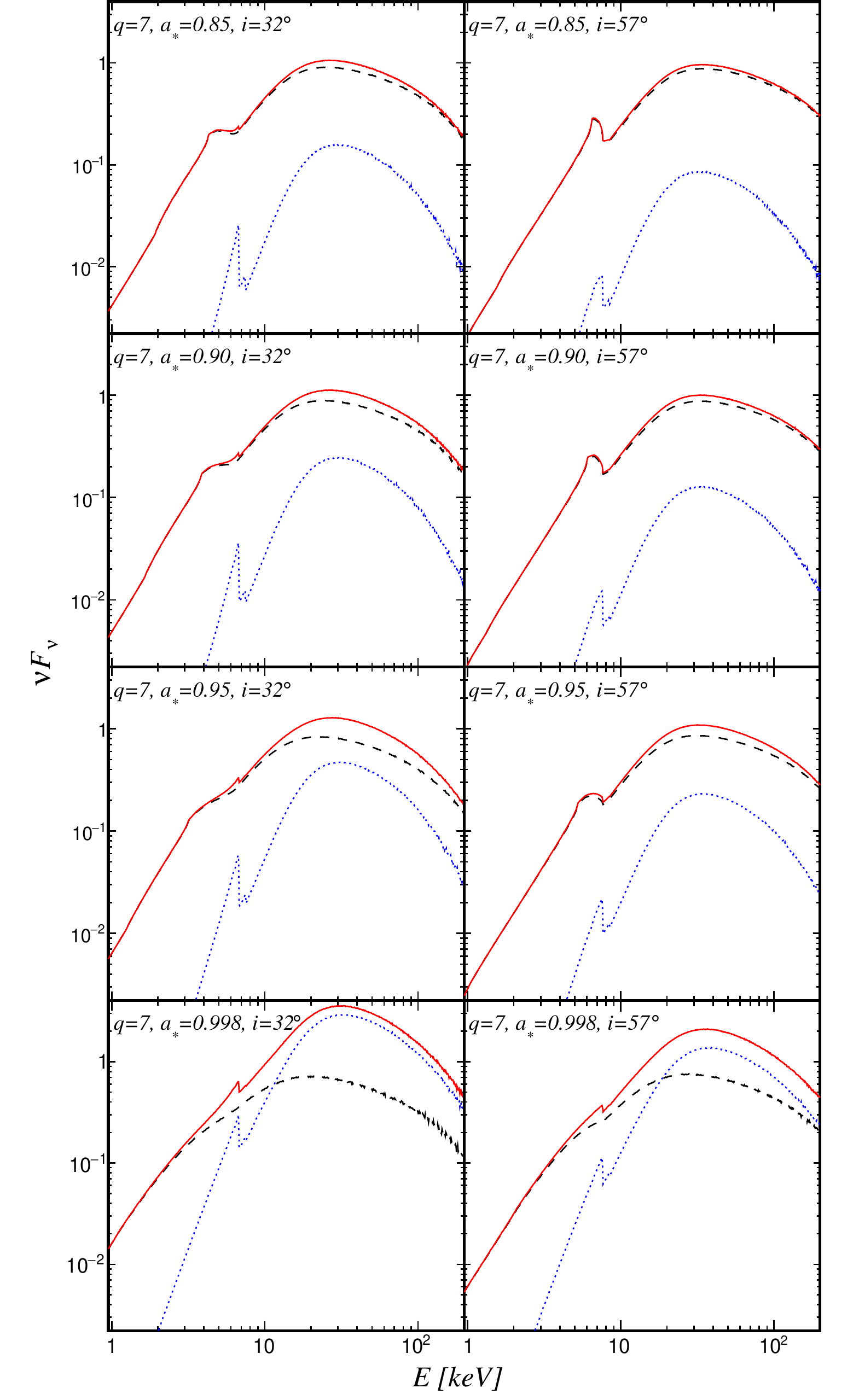}
    \caption{Synthetic reflection spectra for non-ionized accretion disks. The BH spin parameter is $a_* = 0.85$, 0.9, 0.95, and 0.998 (from top to bottom), the inclination angle of the disk is $i = 32^{\circ}$ (left panels) and $i = 57^{\circ}$ (right panels), and we assume that the emissivity profile of the disk is described by a power-law with emissivity index $q = 7$ and that the spectrum of the radiation illuminating the disk is a power-law with photon index $\Gamma=2$. Red solid curves show the spectra calculated including the returning radiation and black dashed curves show the spectra calculated without including the returning radiation. Blue dotted curves represent the differences between the red solid and black dashed curves. From~\citet{2020arXiv200615838R}. \copyright AAS. Reproduced with permission.}
    \label{f-ret1}
\end{figure}

\begin{figure}
    \centering
    \includegraphics[width=11cm]{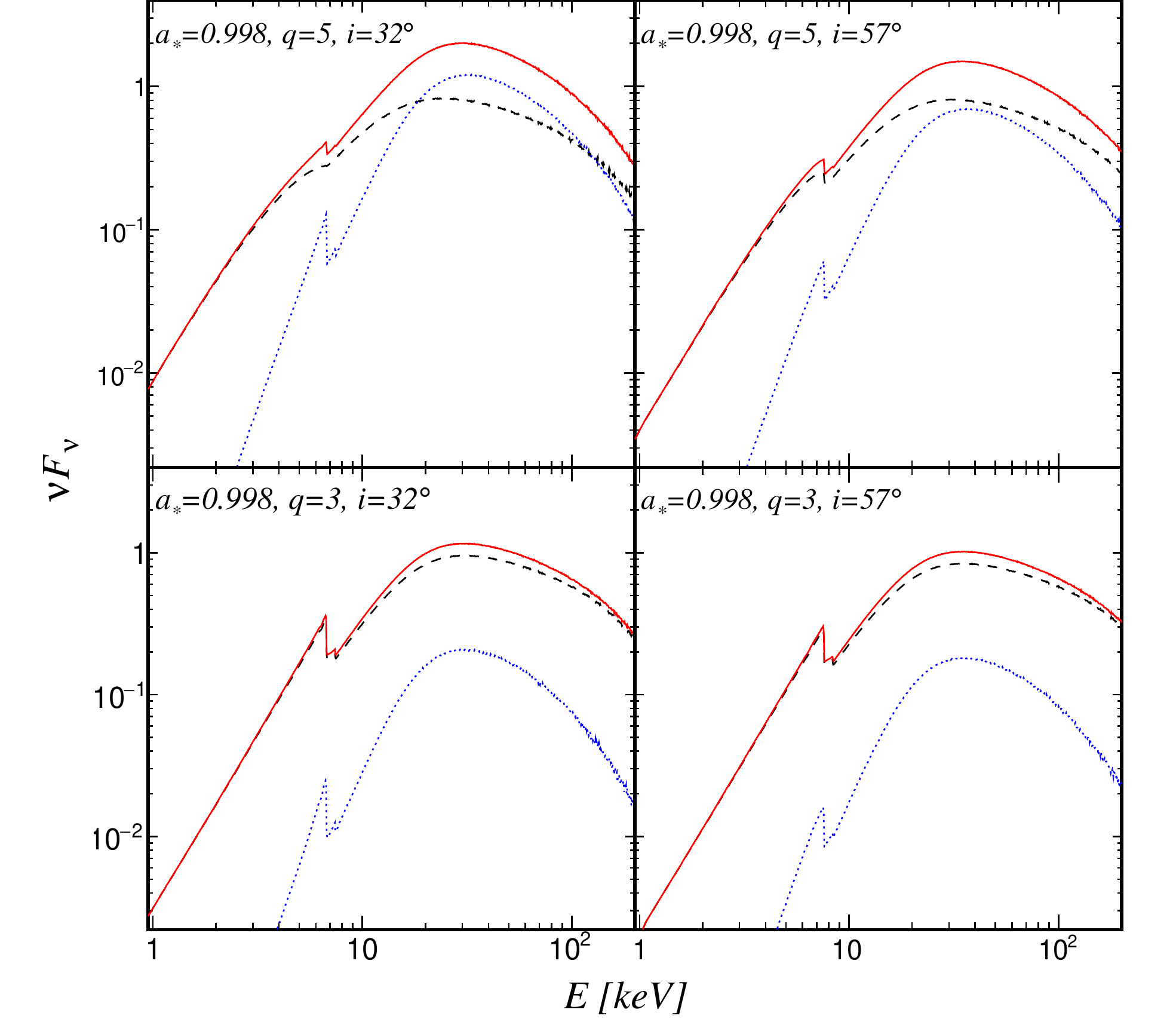}
    \caption{As in Fig.~\ref{f-ret1} for the BH spin parameter $a_* = 0.998$ and assuming an emissivity index $q=5$ (top panels) and $q=3$ (bottom panels). From~\citet{2020arXiv200615838R}. \copyright AAS. Reproduced with permission.}
    \label{f-ret2}
\end{figure}

The impact of the returning radiation on the estimate of the parameters in the fit of the reflection spectrum is much less understood. The effect results in the radial redistribution of reflection, yet it is clear that it cannot be simply reabsorbed e.g.\ in the emissivity indices, because the incident radiation is not an approximate power-law, like the radiation from the corona, but has a reflection spectrum. Precise computations of this effect were presented by \citet{2006A&A...453..773S} and \citet{2016ApJ...821L...1N} who found that the strongest effect of the second order reflection occurs above $\sim 10$~keV. This results, however, from their assumption that the reflection is neutral, in which case the reflected radiation is strongly attenuated by photoabsorption at lower energies. Fig.~\ref{f-ret1} shows the impact of the returning radiation on reflection spectra for non-ionized accretion disks for different values of BH spins and inclination angles assuming that the emissivity profile of the disk is described by a power-law with emissivity index $q = 7$. For lower values of the emissivity index, the contribution of the returning radiation on the spectrum is weaker, as illustrated in Fig.~\ref{f-ret2}.

Calculations for an ionized reflection were presented in \citet{2002MNRAS.336..315R}, but without considering the general-relativistic transfer of radiation, so their results are very approximate. Basing on some qualitative properties, \citet{2002MNRAS.336..315R} conclude that the overall effect of multiple reflection is to make the spectrum resemble a single reflection spectrum with a slightly higher ionization and significantly higher iron abundance. The latter is particularly interesting, because it could explain the significantly super-solar abundance of iron found in many fitting results~\citep{2018ASPC..515..282G}.

The recent work of \citet{2020arXiv200615838R} does not support this possibility, but their study is limited to non-ionized disks. They find that increasing the iron abundance and the increasing contribution of returning radiation produce very different spectral distortions of the low energy part of the Compton hump (which is reduced by the former and enhanced by the latter effect). To compensate for this difference, low values of iron abundance and high reflection fractions are fitted when the standard, i.e., without returning radiation, reflection model is applied to data including this radiation. They note that the data quality around 10~keV is crucial for distinguishing between the effects of returning radiation and super-solar abundance. \citet{2020arXiv200615838R} also find that in the most extreme cases (in particular, for maximally rotating BHs and for very centrally concentrated emissivities) the returning radiation systematically affects the fitted parameters, e.g.\ the BH spin. In such cases, it also produces residuals (at the level of a few per cent) that cannot be compensated by adjusting the parameters of models neglecting the returning radiation, which may be an important issue for interpretation of data from future X-ray missions. Simulations for ionized disks are definitively needed for a complete quantitative assessment of these effects.

In another recent work, \citet{2020MNRAS.tmp.2481W} show the effect of returning radiation in reverberation observations. Within their set-up, they find an enhancement of the iron K$\alpha$ line and the Compton hump relative to the continuum, and the additional light travel time between primary and secondary reflections increases the reverberation time lag measured in the iron K band and in the Compton hump. Their study can be important for future observational facilities, not for the current X-ray observatories.

\begin{figure}[t]
\vspace{0.3cm}
\begin{center}
\includegraphics[type=pdf,ext=.pdf,read=.pdf,width=8cm]{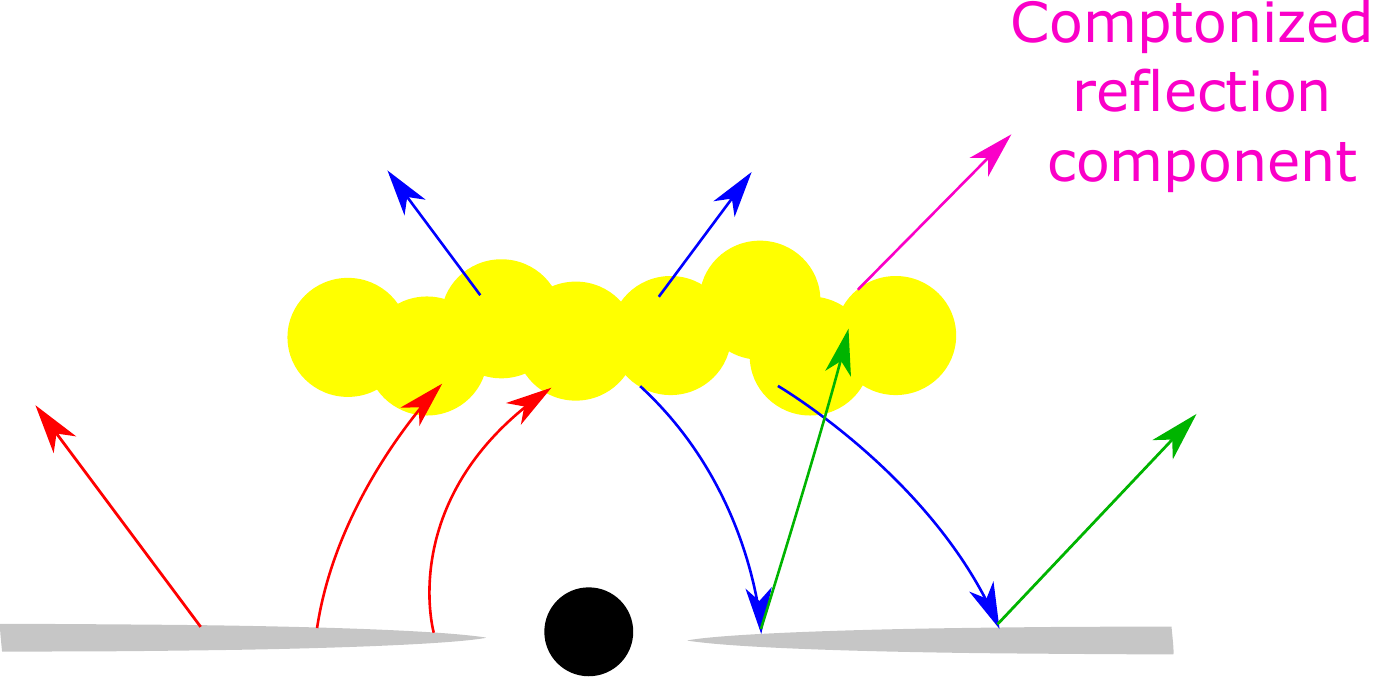}
\end{center}
\vspace{0.3cm}
\caption{Cartoon to illustrate the reflection Comptonization. 
As in Fig.~\ref{f-diskcorona1}, the red arrows indicate thermal photons from the disk, the blue arrows are for the continuum component, and the green arrows represent the reflection component. A fraction of the reflection radiation inverse Compton scatter off free electrons in the corona, producing the Comptonized reflection component (magenta arrow). 
\label{f-diskcorona2}}
\end{figure}

\subsection{Reflection Comptonization}

The hot corona is widely thought to Compton up-scatter cool thermal disk emission.  Frequently, the Compton depth is of order unity, with a result that a substantial fraction of the underlying emission will have scattered before reaching the observer.  By extension, it is argued that the relativistic reflection component -- emitted by the same inner accretion disk in which thermal disk emission is predominantly produced -- should also be Compton scattered by the coronal electrons, as shown in Fig.~\ref{f-diskcorona2}. However, this effect is widely overlooked in the reflection literature.

\begin{figure}[]
\begin{center}
\includegraphics[trim={0cm 0cm 0
  -0.75cm},clip,width=0.85\columnwidth]{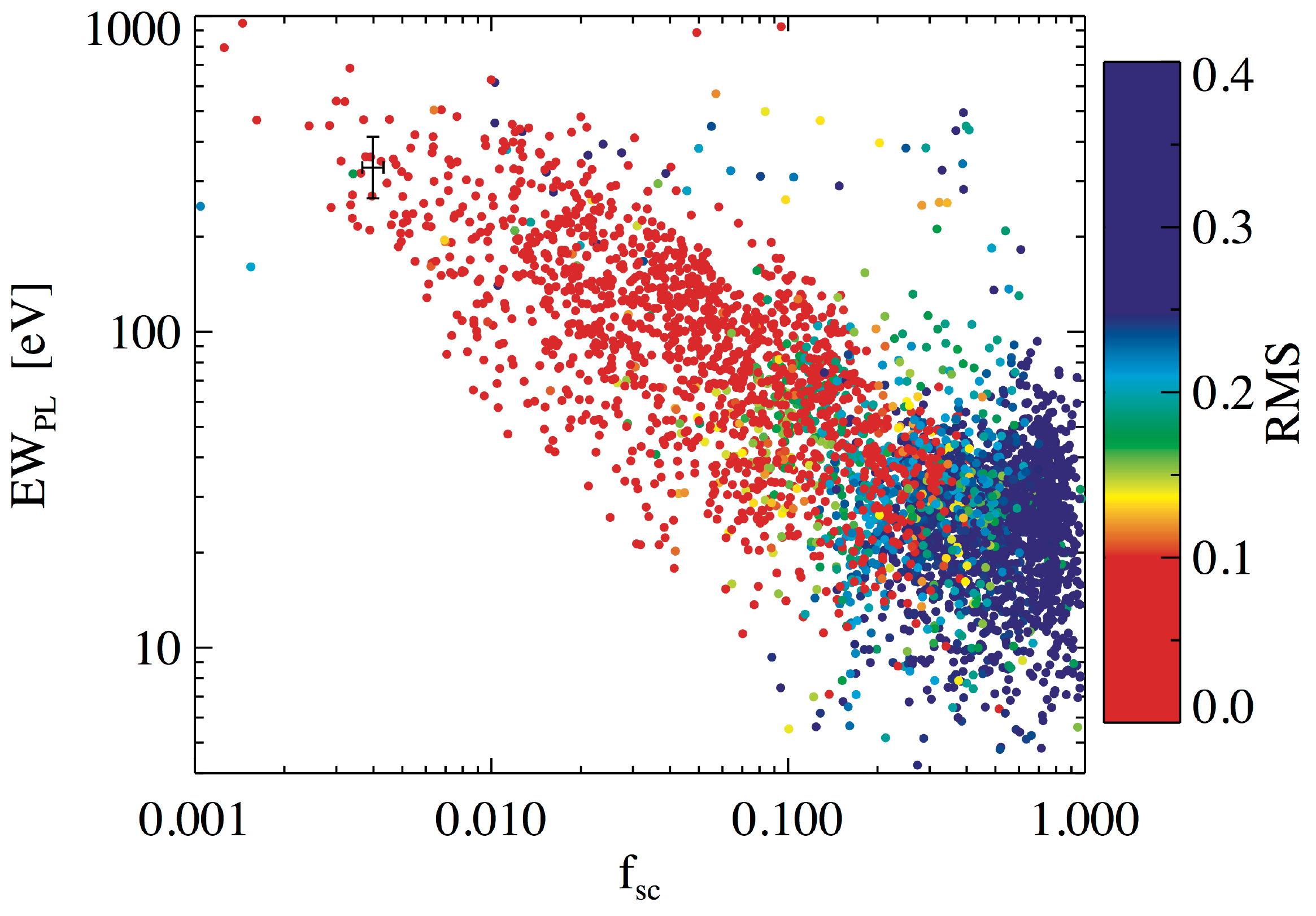}
\caption{The strong anti-correlation between increased scattering-depth in the corona ($f_{\rm sc}$) and the equivalent-width of the fluorescent iron line ($EW$) {\em with respect to the power-law continuum} is naturally explained by the dilution of the narrow line into continuum by the scattering process.  Here, $EW$ is a proxy for the observed reflection amplitude.  The color of the data point is matched to its rms flicker-noise (i.e., the integrated PDS).  Soft states are reddest with lowest rms and hard states are bluest with highest rms. From~\citet{Steiner16}. \copyright AAS. Reproduced with permission.}\label{fig:S16}
\end{center}
\end{figure}

\begin{figure}[]
        \begin{center}
          \includegraphics[width=0.85\columnwidth]{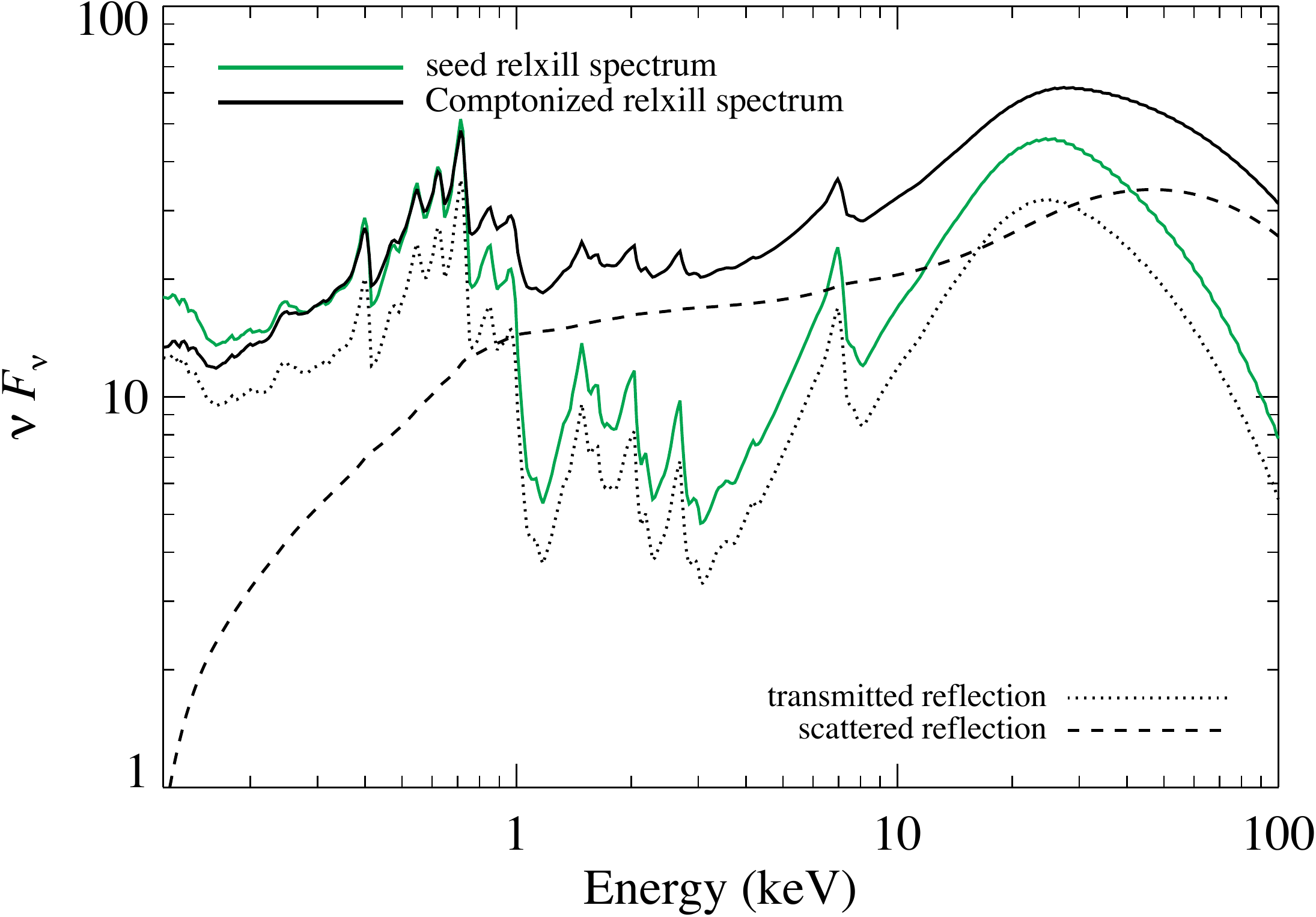}
          \vspace{0.1cm}
          \caption{Impact of Compton scattering on a reflection component ({\tt relxill}) is illustrated for a fiducial corona with $f_{\rm sc}=0.3$ (equivalent to a uniform corona with $\tau=0.35$), $\Gamma=2$, $E_{\rm cut}=100$~keV, and $\log\xi =2$.  The intrinsic reflection Component is shown as a green solid line, and the black solid line shows the Comptonized output.  The total number of photons is the same in both solid curves (the higher photon count of the green curve at low energies compensates the higher photon count of the black curve above 1~keV). Dashed and dotted black lines show the portions which have respectively scattered or been transmitted through the corona. From~\citet{Steiner17}. \copyright AAS. Reproduced with permission.}\label{fig:S17}
        \end{center}
\end{figure}

\cite{Steiner16} and \cite{Steiner17} discuss this effect in two ways.  \citet{Steiner16} demonstrates that the Compton scattering of reflection emission accounts for the apparent dilution of Fe-line strength in hard (and hard-intermediate) states, see Fig.~\ref{fig:S16}. \citet{Steiner17} demonstrates how this effect manifests as a significant underestimate of the reflection fraction when coronal Compton scattering is omitted\footnote{On the other hand, the high reflection fraction observed in some AGN suggests that the reflection scattering is weak for those sources and, in turn, that the corona is not very extended and does not cover well the inner part of the accretion disk (like in the lamppost model).} . With ramifications for measuring spin or Kerr deformation parameters, the line-profile is also subject to deformation by the Compton scattering kernel.  Harder spectra with lower $\Gamma$ produce significant down-scattering tails to intrinsically narrow line features.  The result is that narrower Fe-lines and reflection features undergoing scattering can somewhat mimic a mixture of broad and narrow reflection components (see Fig.~\ref{fig:S17}). It is proposed that reflection analyses should incorporate the Comptonization in a scattering kernel in data modeling.

\citet{2015MNRAS.448..703W} study reflection Comptonization in scenarios with extended coronae. In such a case, the corona should be patchy. They find that the reflection fraction is systematically underestimated if the effect is not taken into account and the covering fraction is high. The detection of reflection-dominated spectra necessarily requires low covering fractions. \citet{2015MNRAS.448..703W} also note that high covering fractions may make the disk appear to be truncated, which can lead to underestimate the BH spin, while the measurements of other parameters, like the viewing angle, iron abundance, and ionization, are not significantly affected.

The degree of reflection Compton-scattering in the corona will scale, of course, with the coronal emission strength, but the precise relationship will depend on disk-coronal geometry. Centrally compact coronae are likely to be less efficient at intercepting disk and disk-reflection emission compared to other disk-hugging planar geometries such as the surface (``warm'') corona described by \citet{2020MNRAS.491.3553B}.

A new tool in {\sc xspec}, providing an accurate convolution routine for thermal Comptonization of arbitrary seed spectra is {\tt thcomp}~\citep{Zdziarski20a}. It is significantly more accurate, especially at $kT_{\rm e} \gtrsim 50$~keV, than {\tt simplcut}~\citep{Steiner17}, which is based on an older, and less accurate, code {\tt nthcomp}~\citep{1996MNRAS.283..193Z}.

We note that different reflection models should be used in slightly different ways with the Componization models {\tt simplcut} and {\tt thcomp}. In {\tt simplcut} and {\tt thcomp}, $T_{\rm e}$ is the coronal temperature, which is supposed to be constant at every point of the corona. However, the coronal temperature $T_{\rm e}$ is the redshifted value to the observer's rest-frame in {\tt relxillpCp} and {\tt reltrans} and the actual coronal temperature in {\tt reflkerr\_lp}. If we use {\tt relxillpCp} or {\tt reltrans}, we should blueshift $T_{\rm e}$ according to Eq.~(\ref{eq-g-c-o}) before connecting its value to the coronal temperature of the Comptonization model. If we use {\tt reflkerr\_lp}, we can directly link its coronal temperature to that of the Comptonization model. In {\tt kyn}, where only the power-law with exponential cut-off is implemented, $E_{\rm cut}$ refers to the value of the cut-off energy at the detection point if positive and at the source if negative, so in the latter case we can directly link $E_{\rm cut}$ to the coronal temperature in {\tt simplcut} or {\tt thcomp}, e.g. $E_{\rm cut} = - 2 T_{\rm e}$. However, a lamppost corona is probably too small to appreciably Comptonize the reflection spectrum from the disk (Wang et al., in preparation); see, however, \citet{Garcia18}, where to achieve a good fit it is required that 83\% of the reflection component is Comptonized in the lamppost corona.

\section{Instrumental uncertainties \label{s-instr}}

An understanding of instrumental and foreground effects is crucial for measurements of BH properties, otherwise we risk obtaining results that may be precise, but are not accurate. An in-depth discussion of instrumental effects (often subsumed under the term ``calibration uncertainties'') is beyond the scope of this review, mainly because these effects are strongly dependent on the properties of a given instrument. We therefore provide a brief overview of the general effects of calibration uncertainties in the context of X-ray reflection spectroscopy, along with an example of how these uncertainties manifest in the form of systematic errors on model parameters such as BH spin.

In general, the total measured X-ray counts per spectral channel $C(i)$, with $i$ denoting the $i$-th detector spectral channel, can be expressed as 
\begin{equation}
    C(i) = t_\mathrm{exp} \cdot \int \dd E \, R(i,E) A(E) f_E(E) + B(i)
\end{equation}
\citep[cf.][]{Davis_2001a} with $f_E(E) = \dd N_\mathrm{ph} / \dd t \, \dd A \, \dd E$ denoting the photon source spectrum, $R(i,E)$ representing the redistribution matrix function (RMF; the map of detector channels to energy space), $A(E)$ the ancillary response function (ARF; the effective area of the detector), $t_\mathrm{exp}$ the exposure time of the observation, and $B(i)$ the instrumental background contribution. The source spectrum itself, $f_E(E)$, consists of a contribution from the BH system, including the continuum and reflection components; from foreground effects such as absorption, both interstellar and local to the system; and from scattering. $C(i)$ cannot be directly converted into the photon flux spectrum $f_E(E)$; to compare with observations, models for $f_E(E)$ are folded through the instrumental response and the comparison happens in the instrumental count space. Though different X-ray analysis packages may differ in details of the implementation \citep[e.g.,][]{Nowak_2005a}, the underlying formalism remains the same and is inherent to the detection set-up.

Calibration uncertainties may include (but are not limited to): uncertainty in the total instrument effective area, pile up effects \citep[e.g.,][]{Miller_2010a,2010MNRAS.407.2287D}, rate-dependent charge transfer inefficiency \citep[e.g.,][]{Kolehmainen_2011a}, contamination buildup on the detector, gain shift effects \citep{Duro_2016a}, as well as less specific uncertainties in both the RMF and ARF. Additionally, it is worth emphasizing that the above effects are dynamic and evolve as detectors and instruments age \citep[e.g.,][]{Plucisky_2018}, but may also depend on detector temperature or presence of nearby bright (optical) sources. Resources on calibration are provided by individual instrumental teams and, e.g., by the International Astrophysical Consortium for High Energy Calibration (IACHEC)\footnote{\url{https://iachec.org}}.

In practice, these calibration uncertainties are often addressed through the addition of blanket systematic errors to the data across the entire energy range of an instrument, usually on the order of $\lesssim$1--5\%. Given the precision required in reflection spectroscopy studies, such an approach is often unsatisfactory, especially considering that statistical errors (i.e., measurement uncertainties on the fitted model parameters) have typically been equal or greater than systematic errors.

Generally speaking, the goal of all of spin measurements using current X-ray satellites is to get $<10$\% precision in order to use spin as a metric to differentiate between accretion vs. mergers as the dominant growth mechanism for supermassive BHs \citep[e.g.,][]{2008ApJ...684..822B}.  In practice, this level of precision has historically referred to only statistical errors.  Since 2012, \textsl{NuSTAR} has yielded high signal-to-noise (S/N) data out to 79~keV, capturing the high energy end of the reflection spectrum and enabling significantly improved statistical precisions and accuracies of our spin measurements
\citep[see, e.g.,][]{Risaliti:2013cga,Marinucci:2014ita,Parker:2016ltr,2020arXiv200704324D}. 
\textsl{NuSTAR} has been especially effective when paired with observatories that provide higher spectral resolution at lower energies \citep[e.g., \textsl{XMM-Newton}; see][for an example of this pairing to study BH spin]{2014MNRAS.440.2347M}.  Over the next two decades, with improving data quality and state-of-the-art reflection models, spin measurements are poised to become systematics-limited.

The \textsl{Athena} X-ray observatory is currently being designed and fabricated in an effort led by ESA.  As described later in Section~\ref{s-c}, the X-IFU microcalorimeter instrument will yield spectra of unprecedented sensitivity.  In order to verify that the proposed calibration requirements for the X-IFU allow the mission to meet its stated science objectives, a team of scientists is examining how the uncertainty on the knowledge of the X-IFU effective area can affect measurements of BH spin.  A vetted approach to this question is to create a distribution of ARFs perturbed around the nominal ARF of the instrument and bounded by the proposed acceptable uncertainty on the effective area \citep[e.g.,][]{2006SPIE.6270E..1ID}.  The team then uses a Monte Carlo algorithm to draw a randomized array of, e.g., 1000 simulated ARFs from this distribution.  Using these perturbed ARFs and the nominal RMF and instrument background, the team simulates 1000 BH system spectra using a fiducial input model including a {\tt relxill} component for the inner disk reflection as well as a power-law continuum and intrinsic absorption, if desired. They then perform spectral fits to each of the 1000 simulated spectra using the nominal ARF, RMF, background and input model, leaving all relevant parameters free to vary in the fit.  The best-fit BH spin is extracted with its 1-$\sigma$ statistical error for each model fit. 

Using this approach, the team is able to evaluate the effect that adopting a given calibration uncertainty on the X-IFU effective area has on the final measured BH spin.  For the distribution of 1000 simulated spectra that sample the proposed ARF calibration uncertainty, the team computes the average measured BH spin and corresponding average 1-$\sigma$ statistical error and compares these values to the input spin in the model.  The standard deviation of the distribution of measured spins can be used as a proxy for the systematic error introduced by the calibration uncertainty on the effective area.  If the net result is that the combination of statistical and systematic error on BH spin exceeds the desired 10\%, the team can then repeat this exercise, iteratively revising the calibration uncertainty until the resulting spin error satisfies the desired science requirement.  This result can then inform the mission’s ground and in-flight calibration planning.

The reader is referred to \citet{2019A&A...628A...5B} for a detailed description of a similar exercise conducted to evaluate \textsl{Athena}’s ability to constrain supermassive BH spins across a wide range in spectral model parameter space.


\section{Spin measurements \label{s-spin}}

X-ray reflection spectroscopy can measure BH spins in both BH binaries and AGN and is currently the only technique capable of precisely measuring the spin of supermassive BHs, while the spins of stellar-mass BHs can be measured also by analyzing the thermal spectrum of the disk~\citep{Zhang:1997dy,2001MNRAS.325.1253G,McClintock:2011zq,McClintock:2013vwa} or the gravitational wave signal of a coalescing binary system~\citep{Abbott:2016blz}.

Spin measurements are crucial to understand how BHs form, evolve, and interact with the host environment~\citep{2020arXiv201108948R}. In the case of stellar-mass BHs in X-ray binaries, the value of the BH spin parameter is normally determined by the formation mechanism, because the amount of mass that can be transferred from the companion star to the BH is modest and cannot appreciably change the BH mass and spin angular momentum~\citep[see, e.g.,][]{1999MNRAS.305..654K}. On the contrary, for supermassive BHs in AGN, the value of their spin parameter is determined by the mass transferred from the host environment to the compact object and spin measurements can test different scenarios of galaxy evolution and merger~\citep[see, e.g.,][]{2012MNRAS.423.2533B,2014ApJ...794..104S}.

The reflection spectrum in Eq.~(\ref{eq-flux}) depends on the metric of the spacetime, which enters the calculations of the photon trajectories connecting the emission and the detection points (eventually encoded in the Jacobian $|\partial(X,Y)/\partial(g^*,r_{\rm e})|$ and the gravitational redshift between the emission and the detection points) and the calculations of the disk structure (in particular, the inner edge of the disk if we impose $R_{\rm in} = R_{\rm ISCO}$ and the velocity of the gas in the disk with the associated Doppler boosting). However, it turns out that the impact of the BH spin on the reflection spectrum is often weak except for the location of the inner edge of the disk. As a result, with current detectors it is challenging to measure the BH spin if $a_*$ and $R_{\rm in}$ are independent and both free in the fit, while it is possible to estimate the BH spin if we assume $R_{\rm in} = R_{\rm ISCO}$. An exception is when $R_{\rm in}$ is very close to the value of the ISCO radius of a Kerr BH with $a_* = 0.998$ (the maximum spin value in most relativistic reflection models) and therefore even if we relax the assumption $R_{\rm in} = R_{\rm ISCO}$ the fit requires that the BH spin is close to $a_* = 0.998$.

\begin{figure}[t]
\vspace{0.3cm}
\begin{center}
\includegraphics[trim={1.0cm 0.0cm 0cm 0cm},clip,width=0.8\columnwidth]{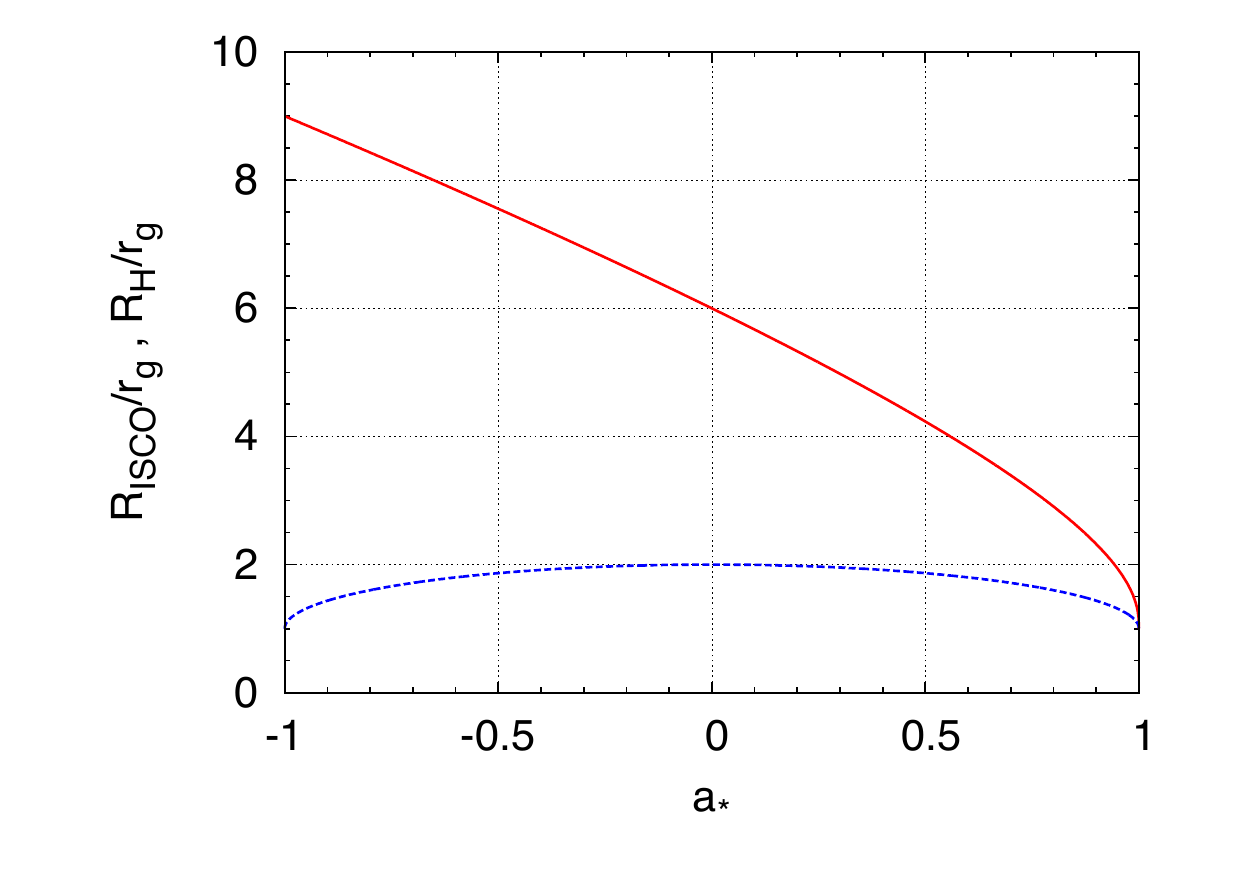}
\end{center}
\vspace{-0.3cm}
\caption{Radial coordinates of the ISCO radius (red solid curve) and of the event horizon (dashed blue curve) in the Kerr spacetime in Boyer-Lindquist coordinates. $a_* > 0$ ($< 0$) means the orbital angular momentum is parallel (anti-parallel) to the BH spin. $r_{\rm g} = G_{\rm N}M/c^2$ is the gravitational radius. \label{f-isco}}
\end{figure}

In the Kerr spacetime, there is indeed a one-to-one correspondence between the BH spin parameter $a_*$ and the ISCO radius (when measured in units of the gravitational radius $r_{\rm g}$). The analytic formula is~\citep{Bardeen:1972fi}
\be
R_{\rm ISCO} &=& r_{\rm g} \left[ 3 + Z_2 \mp 
\sqrt{\left(3 - Z_1\right)\left(3 + Z_1 + 2 Z_2\right)} \right] \, , \nonumber\\
Z_1 &=& 1 + \left( 1 - a_*^2\right)^{1/3} 
\left[ \left( 1 + a_*\right)^{1/3} + \left( 1 - a_*\right)^{1/3}\right] \, , \nonumber\\
Z_2 &=& \sqrt{3 a_*^2 + Z_1^2} \, , 
\ee
and the plot $a_*$ vs. $R_{\rm ISCO}$ is shown in Fig.~\ref{f-isco}. The validity of the assumption $R_{\rm in} = R_{\rm ISCO}$ is still controversial, as already discussed in Subsection~\ref{ss-cgeo}, and, especially in the hard state of BH binaries, there are situations in which the accretion disk is clearly truncated and $R_{\rm in} > R_{\rm ISCO}$ \citep[see, e.g.,][and reference therein]{2018ApJ...855...61W}.

\citet{Dauser:2013xv} analyzed synthetic relativistically broadened iron lines assuming a lamppost coronal geometry and concluded that reliable spin measurements are only possible when the BH spin is high and the corona is compact and close to the object. In such a case, we can measure the BH spin from the broad iron line shape. In the case of a slow-rotating BH or an extended/distant corona, the iron line is narrow and we cannot distinguish the case of a slow-rotating BH from that of an extended/distant corona. A similar result is found in \citet{2014MNRAS.439.2307F}, where the conclusion is, again, that accurate measurements require that the inner part of the accretion disk is illuminated well. Within their set-up, \citet{2014MNRAS.439.2307F} find that it is possible to have both robust measurement of the BH spin and of the inner edge of the accretion disk if the corona is quasi-static and at a height less than 10~gravitational radii. A more detailed study, considering simulated observations of the full reflection spectrum and including other typical spectral components, is reported in~\citet{Kammoun:2018ltv}, but the conclusions are still quite similar. Generally speaking, reliable measurements of BH spins are only possible in the case of fast-rotating BHs and in the presence of a corona close to the compact object and illuminating well the inner part of the accretion disk. In such a case, relativistic effects of the spin are strong enough to permit us to measure this parameter and break its degeneracy with other parameters. On the contrary, in the case of BHs with a low or moderate value of the spin parameter, the impact of the background metric is not so strong on the reflection spectrum and it is very challenging to break the parameter degeneracy.

Accurate spin measurements also require observations with a good energy resolution near the iron line (which is often the most informative feature to determine the BH spin) and data covering a broad energy band (which is extremely helpful to both select the correct astrophysical model and get accurate measurements of other parameters, such as the photon index $\Gamma$, the high energy cut-off $E_{\rm cut}$ and the ionization parameter $\xi$, with benefits to the estimate of $a_*$ as well). In this regard, the launch of \textsl{NuSTAR}~\citep{2013ApJ...770..103H} in 2012 has significantly contributed to progress in the field and has proved to be fundamental to disentangle the contribution from the accretion disk and from much more distant material to the total reflection fraction. Further advances are expected from instruments such as \textsl{XRISM}~\citep{2020SPIE11444E..22T} and \textsl{Athena}~\citep{2013arXiv1306.2307N} with excellent energy resolutions around the Fe~K region, which, if combined with a broad bandpass, will help disentangle reflection and absorption signatures.

\begin{figure}[t]
\vspace{0.3cm}
\begin{center}
\includegraphics[width=0.9\columnwidth]{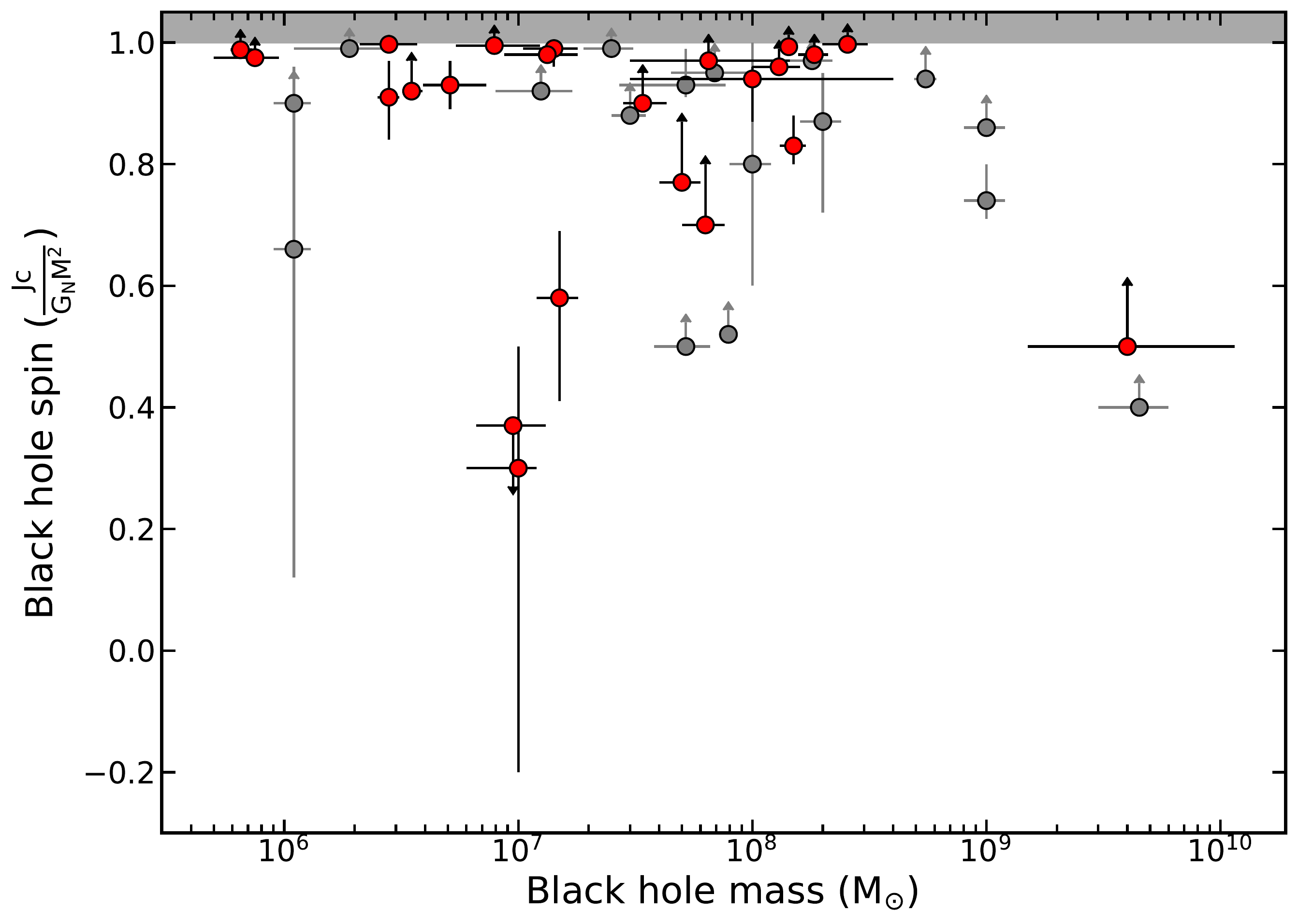}
\end{center}
\caption{The BH spin $a_*$ is plotted against the BH mass, for our sample of 40~AGN. red (gray) dots are for BH spin measurements that include (do not include) \textsl{NuSTAR} data. \label{AGN_spin} }
\end{figure}

Tab.~\ref{t-bhb} and Tab.~\ref{t-agn} show the spin measurements of, respectively, BH binaries and AGN. When multiple measurements are present in the literature, we reported the most recent one (thus obtained with a more advanced reflection model) and using the broadest energy band. In the AGN table, we assumed a 20\% uncertainty on BH mass estimates with no associated error bars in the literature. In Tab.~\ref{t-agn}, we have 40~objects and Fig.~\ref{AGN_spin} shows the corresponding plot BH spin vs. BH mass, where the 23~objects for which the BH spin is inferred with \textsl{NuSTAR} data are indicated by red dots and the 17~objects for which the BH spin measurement does not include \textsl{NuSTAR} data are gray dots.
A list with older spin measurements can be found in \citet{Reynolds:2013qqa}, \citet{2013mams.book.....B}, \citet{Reynolds:2019uxi}, and \citet{2020arXiv201108948R}.

The analysis of relativistic reflection features in a number of BH binaries and AGN find a high, or even very high, iron abundance. The origin of such high iron abundances is currently unknown. It is often thought that it is the result of some deficiency in current reflection models or in our understanding of these systems. Depending on the actual reason for these super-solar iron abundances, this may lead to undesirable large systematic uncertainties in the estimate of the BH spin parameter too~\citep{Reynolds:2013qqa}.
In the previous sections we have mentioned some of the explanations proposed in the literature:   
\begin{enumerate}
\item The value of the disk electron density is higher than that used in the calculations (Subsection~\ref{ss-disk}).
\item There are two coronae, so that the value of the photon index of the radiation illuminating the disk and responsible for the reflection spectrum can be different from the value of the photon index of the continuum (Subsection~\ref{ss-cgeo}).
\item Returning radiation is not included in the calculations (Subsection~\ref{ss-return}).
\end{enumerate}
The estimate of the iron abundance is thus thought to be highly uncertain. Not only modeling bias, but even real physical mechanisms have been proposed, like radiative levitation of metal ions in the inner part of the accretion disk~\citep{2012ApJ...755...88R}.

\begin{table}
    \centering
    \scalebox{0.8}{
    \renewcommand\arraystretch{1.5}
    \begin{tabular}{ccccc}
    \hline\hline
    Source & Spin ($a_*$) & Instrument(s) & State$^\star$ & Reference \\
    \hline\hline
    4U~1543--475 & $0.67_{-0.08}^{+0.15}$ & \textsl{RXTE} & Steep Power Law & \citet{2020MNRAS.493.4409D} \\
    \hline 
    4U~1630--472 & $0.985_{-0.014}^{+0.005}$ & \textsl{NuSTAR} & Intermediate & \citet{King:2014sja} \\
    \hline    
    Cyg~X-1& $>0.83$ & \textsl{Suzaku}+\textsl{NuSTAR} & Soft & \citet{Tomsick:2013nua} \\
    & $>0.97$ & \textsl{Suzaku}+\textsl{NuSTAR} & Hard & \citet{Parker:2015fja} \\
    & $0.93 \sim 0.96$ & \textsl{NuSTAR} & Soft & \citet{Walton:2016hvd} \\
    \hline
    EXO 1846--031 & $0.997_{-0.002}^{+0.001}$ & \textsl{NuSTAR} & Hard Intermediate & \citet{2020arXiv200704324D} \\
    \hline    
    GRO~J1655--40 & $> 0.9$ & \textsl{XMM-Newton} & High/Soft & \citet{2009MNRAS.395.1257R} \\    
    \hline
    GRS~1716--249 & $>0.92$ & \textsl{Swift}+\textsl{NuSTAR} & Hard Intermediate & \citet{2019ApJ...887..184T} \\
    \hline
    GRS~1739--278 & $0.8 \pm 0.2$ & \textsl{NuSTAR} & Low/Hard & \citet{Miller:2014sla} \\
    \hline
    GRS~1915+105 & $0.98 \pm 0.01$ & \textsl{NuSTAR} & Low/Hard & \citet{Miller:2013rca} \\
    \hline    
    GS~1354--645 & $>0.98$ & \textsl{NuSTAR} & Hard & \citet{El-Batal:2016wmk} \\
    \hline
    GX~339--4 & $0.95_{-0.08}^{+0.02}$ & \textsl{Swift}+\textsl{NuSTAR} & Very High & \citet{Parker:2016ltr} \\
    \hline
    LMC~X-1 & $0.97_{-0.25}^{+0.02}$ 
    & \textsl{RXTE}+\textsl{Suzaku} & Soft & \citet{2012MNRAS.427.2552S} \\
    \hline
    MAXI~J1535--571 & $>0.84$ & \textsl{NuSTAR} & Hard & \citet{Xu:2017yrm} \\
    \hline
    MAXI~J1631--479 & $>0.94$ & \textsl{NuSTAR} & Soft & \citet{Xu:2020vil} \\
    \hline
    MAXI~J1836--194 & $0.88 \pm 0.03$ & \textsl{Suzaku} & Hard-Intermediate & \citet{2012ApJ...751...34R} \\    
    \hline
    SAX~J1711.6--3808 & $0.6_{-0.4}^{+0.2}$ & \textsl{BeppoSAX} & Low/Hard & \citet{2009ApJ...697..900M} \\    
    \hline    
    Swift~J1658.2--4242 & $>0.96$ & \textsl{Swift}+\textsl{NuSTAR} & Hard & \citet{Xu:2018lfo} \\
    \hline
    Swift~J174540.2--290037 & $0.92_{-0.07}^{+0.05}$ & \textsl{Chandra}+\textsl{NuSTAR} & Hard & \citet{2019ApJ...885..142M} \\    
    \hline
    Swift~J174540.7--290015 & $0.94_{-0.10}^{+0.03}$ & \textsl{Chandra}+\textsl{NuSTAR} & Soft & \citet{2019ApJ...885..142M} \\
    \hline
    Swift~J1753.5--0127 & $0.76_{-0.15}^{+0.11}$ & \textsl{XMM-Newton} & Low/Hard & \citet{2009MNRAS.395.1257R} \\    
    \hline
    Swift~J1910.2--0546 & $< -0.32$ & \textsl{XMM-Newton} & Intermediate & \citet{2013ApJ...778..155R} \\ 
    \hline
    V404~Cyg & $>0.92$ & \textsl{NuSTAR} & Hard$^\star$ & \citet{Walton:2016fso}\\
    \hline
    XTE~J1550--564 & $0.76 \pm 0.01$ & \textsl{ASCA} & Very High & \citet{2009ApJ...697..900M} \\
    \hline
    XTE~J1650--500 & $0.79 \pm 0.01$ & \textsl{XMM-Newton} & Low/Hard & \citet{2009ApJ...697..900M} \\    
    \hline  
    XTE~J1652--453 & $0.45 \pm 0.02$ & \textsl{RXTE}+\textsl{XMM-Newton} & Hard-Intermediate & \citet{2011MNRAS.411..137H} \\    
    \hline
    XTE~J1752--223 & $0.92 \pm 0.06$ & \textsl{RXTE} & Hard & \citet{Garcia18} \\    
    \hline
    XTE~J1908+094 & $0.75 \pm 0.09$ & \textsl{BeppoSAX} & Low/Hard & \citet{2009ApJ...697..900M} \\ 
    \hline\hline
\end{tabular}}
\vspace{0.2cm}
\caption{Spin measurements of stellar-mass BHs using X-ray reflection spectroscopy. $^\star$ State of the source during the observation used for the spin measurement. All states listed in the table are explicitly mentioned in reference papers except for V404~Cyg. The spectra of V404 Cyg in~\citet{Walton:2016fso} exhibit a very hard power-law ($\Gamma \approx 1.4$) and a weak disk blackbody component, so they can be classified in the hard state. \label{t-bhb}}
\end{table}

\begin{table}
\centering
\scalebox{0.8}{
{\renewcommand{\arraystretch}{1.2}
\begin{tabular}{ccccc}
\hline\hline
    Source & Spin ($a_*$) &  Mass (M$_{\odot}$) & Instrument(s) & References \\
    \hline\hline
     1H0419--577 & $>0.96$ & $(1.3\pm0.3)\times10^8$& \textsl{XMM-Newton}+\textsl{NuSTAR}& \citet{Jiang:2018aln} \\
     & & & & \citet{Grupe:2010me} \\
    \hline
    1H0707--495    & $>0.988$ &$(6.5_{-0.6}^{+0.8})\times10^5$&\textsl{XMM-Newton}+\textsl{NuSTAR}  & \citet{Kara:2015kda}\\
    & & & & \citet{ponti12} \\
    \hline
    1H0323+342 & $> 0.9$ &$(3.4_{-0.6}^{+0.9})\times10^7$& \textsl{NuSTAR}+\textsl{Suzaku}+\textsl{Swift} & \citet{2018MNRAS.479.2464G}\\
    & & & & \citet{2016ApJ...824..149W} \\
    \hline
    3C 120   & $> 0.95$ &$(6.9_{-2.4}^{+3.1})\times10^7$& \textsl{Suzaku}+\textsl{XMM-Newton} & \citet{2016MNRAS.458.2012V}\\
    & & & & \citet{2017ApJ...849..146G} \\
    \hline
    4C 74.26   & $>0.5$ &$(4.0_{-2.5}^{+7.5})\times10^9$& \textsl{Swift}+\textsl{NuSTAR}  & \citet{lfb17}\\
    & & & & \citet{wh02} \\
    \hline
    Ark~120 & $0.83^{+0.05}_{-0.03}$ &$(1.50\pm0.19)\times10^8$& \textsl{XMM-Newton}+\textsl{NuSTAR}  & \citet{pdr19} \\
    & & & & \citet{pet04} \\
    \hline
    Ark 564   & $> 0.9$ &$(1.1 \pm 0.2)\times10^6$& \textsl{XMM-Newton} & \citet{jiang19b}\\
    \hline
    ESO 362--G18   & $> 0.92$ &$(1.25 \pm 0.45)\times10^7$& \textsl{Suzaku}+\textsl{XMM-Newton} & \citet{2016MNRAS.458.2012V}\\
    \hline
    Fairall~9 & $>0.997$ & $(2.55\pm0.56)\times10^8$&\textsl{XMM-Newton}+\textsl{NuSTAR} & \citet{Lohfink:2016gxc} \\
     & & & & \citet{pet04} \\
     \hline
    Fairall 51   & $0.8 \pm 0.2$ & $(1.0 \pm 0.2)\times10^8$ & \textsl{Suzaku} & \citet{svoboda2015}\\
    & & & & \citet{2006AA...459...55B} \\
    \hline
    H1821+643   & $> 0.4$ &$(4.5 \pm 1.5)\times10^9$& \textsl{Suzaku} & \citet{2016MNRAS.458.2012V}\\
    \hline 
    HE~1136--2304    & $>0.995$ &$(7.9_{-2.5}^{+4.5})\times10^6$& \textsl{XMM-Newton}+\textsl{NuSTAR}  & \citet{pkk16}\\
    \hline
    IRAS~00521--7054 & $>0.77$ & $(5.0\pm1.0)\times10^7$&\textsl{XMM-Newton}+\textsl{NuSTAR}&  \citet{Walton:2019utn} \\
    \hline
    IRAS~09149--6206 & $0.94^{+0.02}_{-0.07}$ & $(1.0_{-0.7}^{+3.0})\times10^8$&\textsl{Swift}+\textsl{XMM-Newton}+\textsl{NuSTAR}&  \citet{2020arXiv200910734W} \\
    \hline    
    IRAS 13197--1627 & $\geq 0.7$ & $(6.3\pm1.3)\times10^7$&\textsl{XMM-Newton}+\textsl{NuSTAR}&  \citet{wal18} \\
        & & & & \citet{vas10} \\
    \hline    
    IRAS~13224--3809 & $>0.975$ & $(7.5_{-2.5}^{+2.0})\times10^5$&\textsl{XMM-Newton}+\textsl{NuSTAR}&  \citet{jj18} \\
    & & & & \citet{ponti12} \\
    \hline
    IRAS 13349+2438 & $0.3^{+0.2}_{-0.5}$ &$(1.0_{-0.4}^{+0.2})\times10^7$& \textsl{XMM-Newton}+\textsl{NuSTAR}&  \citet{pma20} \\
    & & & & \citet{ponti12} \\
    \hline
    MCG--6--30--15& $0.91_{-0.07}^{+0.06}$ &$(2.8\pm0.3)\times10^6$& \textsl{XMM-Newton}+\textsl{NuSTAR}  & \citet{Marinucci:2014ita} \\
    & & & & \citet{ponti12} \\
    \hline
    Mrk 79   & $> 0.5$ &$(5.2 \pm 1.4)\times10^7$& \textsl{XMM-Newton} & \citet{jiang19b}\\
    \hline
    Mrk 110   & $> 0.99$ &$(2.5 \pm 0.6)\times10^7$& \textsl{XMM-Newton} & \citet{jiang19b}\\
    \hline
    Mrk~335 & $0.99_{-0.03}^{+0.01}$ &$(1.42\pm0.37)\times10^7$& \textsl{Swift}+\textsl{NuSTAR} & \citet{Parker:2014kna} \\
    & & & & \citet{pet04} \\
    \hline
    Mrk 359   & $0.66_{-0.54}^{+0.30}$ &$(1.1 \pm 0.2)\times10^6$& \textsl{Suzaku} & \citet{2016MNRAS.458.2012V}\\
    \hline
    Mrk~509 & $>0.993$ & $(1.43\pm0.12)\times10^8$&\textsl{Swift}+\textsl{Suzaku}+\textsl{NuSTAR} & \citet{Garcia:2018ckp}  \\
    & & & & \citet{pet04} \\
    \hline
    Mrk~766 & $>0.92$ & $(3.5_{-0.3}^{+0.4})\times10^6$&\textsl{Swift}+\textsl{XMM-Newton}+\textsl{NuSTAR} & \citet{bpk18}  \\
    & & & & \citet{ponti12} \\
    \hline
    Mrk 841   & $> 0.52$ &$(7.9 \pm 0.2)\times10^7$& \textsl{Suzaku} & \citet{2016MNRAS.458.2012V}\\
    \hline
    Mrk~1044 & $0.997_{-0.001}^{+0.016}$ &$(2.8_{-0.7}^{+0.9})\times10^6$& \textsl{Swift}+\textsl{XMM-Newton}+\textsl{NuSTAR} & \citet{mallick18}  \\
    & & & & \citet{du15} \\
    \hline
    Mrk~1501 & $\ge 0.98$ & $(1.84 \pm 0.27)\times10^8$ & \textsl{XMM-Newton}+\textsl{NuSTAR} & \citet{Chamani20} \\
    & & & & \citet{Grier12} \\
    \hline
    NGC~1365 & $>0.97$ &$(6.5_{-3.5}^{+8.0})\times10^7$ &\textsl{XMM-Newton}+\textsl{NuSTAR} & \citet{Walton:2014pca}  \\
    & & & & \citet{risa09} \\
    \hline
    NGC 3783   & $> 0.88$ &$(3.0 \pm 0.5)\times10^7$& \textsl{Suzaku} & \citet{2016MNRAS.458.2012V}\\
    \hline
    NGC 4051  & $> 0.99$ &$(1.9 \pm 0.8)\times10^6$& \textsl{Suzaku} & \citet{2016MNRAS.458.2012V}\\
    \hline
    NGC~4151 & $0.98 \pm 0.01$  &$(1.33\pm0.46)\times10^7$ &\textsl{Suzaku}+\textsl{NuSTAR} & \citet{Keck:2015iqa} \\ 
    & & & & \citet{pet04} \\
    \hline 
    NGC~5506 & $0.93 \pm 0.04$  &$(5.1_{-1.2}^{+2.2})\times10^6$ & \textsl{Suzaku}+\textsl{XMM-Newton}+\textsl{NuSTAR} & \citet{sun18} \\ 
    & & & & \citet{niko09} \\
    \hline
    PG 0804+761   & $> 0.94$ &$(5.5 \pm 0.6)\times10^8$& \textsl{XMM-Newton} & \citet{jiang19b}\\
    \hline
    PG 1229+204   & $0.93_{-0.02}^{+0.06}$ &$(5.7 \pm 2.5)\times10^7$& \textsl{XMM-Newton} & \citet{jiang19b}\\
    \hline
    PG 2112+059   & $> 0.86$ &$(1.0 \pm 0.2)\times10^9$& \textsl{XMM-Newton} & \citet{Schartel2010}\\
    & & & & \citet{2006ApJ...641..689V} \\
    \hline
    Q2237+305   & $0.74_{-0.03}^{+0.06}$ &$(1.0 \pm 0.2)\times10^9$& \textsl{Chandra} & \citet{2014ApJ...792L..19R}\\
    & & & & \citet{2011ApJ...742...93A} \\
    \hline
    RBS 1124   & $> 0.97$ &$(1.8 \pm 0.4)\times10^8$& \textsl{Suzaku} & \citet{walton13}\\
    & & & & \citet{2010MNRAS.401.1315M} \\
    \hline
    RXS J1131--1231   & $0.87_{-0.15}^{+0.08}$ &$(2.0 \pm 0.4)\times10^8$& \textsl{Chandra}+\textsl{XMM-Newton} & \citet{2014Natur.507..207R}\\
    & & & & \citet{sluse2012} \\
    \hline
    Swift~J2127.4+5654 & $0.58_{-0.17}^{+0.11}$ &$(1.5\pm0.3)\times10^7$ & \textsl{XMM-Newton}+\textsl{NuSTAR} & \citet{2014MNRAS.440.2347M}  \\
    & & & & \citet{malizia08} \\
    \hline
    Ton S180 & $<0.37$ & $(9.5_{-2.9}^{+3.6})\times10^6$&\textsl{XMM-Newton}+\textsl{NuSTAR} & \citet{mnp20}  \\
\hline\hline
\end{tabular}}}
\vspace{0.2cm}
\caption{Spin measurements of supermassive BHs. Note that \citet{2016MNRAS.458.2012V} only present a summary of spin measurements and refers to the original papers for the spectral analysis. 
\label{t-agn}}
\end{table}


\section{Testing Einstein's gravity in the strong field regime \label{s-gr}}

In the calculations of the relativistic reflection spectrum of an accretion disk, we normally assume Einstein's theory of GR, and thus: $i)$ the background metric is described by the Kerr solution, $ii)$ massive and massless particles follow the geodesics of the spacetime (Weak Equivalence Principle), and $iii)$ atomic physics in the accretion disk is the same as that we can study in our laboratories on Earth (Local Lorentz Invariance and Local Position Invariance)\footnote{Details on the Weak Equivalence Principle, the Local Lorentz Invariance, and the Local Position Invariance, as well as the classification of gravity theories in which these principles are preserved or violated, can be found in \citet{Will:2014kxa}.}. Relaxing one of these assumptions, we can use X-ray reflection spectroscopy to test Einstein's theory of GR in the strong field regime~\citep{Johannsen:2016uoh,Bambi:2015kza,Krawczynski:2018fnw}.

We note that X-ray reflection spectroscopy and other electromagnetic techniques can test different sectors of the theory from tests using gravitational waves and are thus complementary to the latter~\citep{Barausse:2008xv,Bambi:2013mha,Li:2019lsm,2021arXiv210308978R}. For example, the existence of a new coupling between the gravity and the electromagnetic sectors may lead photons to violate the Weak Equivalence Principle or may alter the value of the electromagnetic constant $\alpha$ in the strong gravity region around the BH, violating the Local Position Invariance. Both phenomena can likely have an impact on the electromagnetic spectrum of the accretion disk without affecting the gravitational wave signal of coalescing objects.

Violations of assumptions $i)$ or $ii)$ would affect the transfer function $f$ in Eq.~(\ref{eq-flux}). Violations of assumption $iii)$ would modify the reflection spectrum at the emission point $I_{\rm e}$ in Eq.~(\ref{eq-flux}). We note that most studies presented in the literature explore the possibility of a violation of assumption $i)$, calculating the reflection spectrum in some non-Kerr background and assuming geodesic motion and standard microphysics in the rest-frame of the gas in the disk, while very little has been done to study the impact of violations of assumptions $ii)$ and $iii)$~\citep{Bambi:2013mha,2021arXiv210308978R}.

Early studies focused on the impact of the background metric on the shape of a broadened iron line~\citep{Lu:2002vm,Schee:2008fc,Bambi:2012at,Johannsen:2012ng}. An important step forward in this research line is represented by the reflection model {\tt relxill\_nk}\footnote{\url{http://www.physics.fudan.edu.cn/tps/people/bambi/Site/RELXILL_NK.html}}~\citep{Bambi:2016sac,Abdikamalov:2019yrr}, which is an extension of the {\tt relxill} package to non-Kerr spacetimes and permits to test GR from the analysis of the whole reflection spectrum. {\tt relxill\_nk} works with a spacetime metric more general than the Kerr solution and that includes the Kerr solution as a special case. Deformations from the Kerr geometry are quantified by a number of {\it deformation parameters}, which are additional parameters of the new model. By analyzing reflection-dominated spectra of accreting BHs with {\tt relxill\_nk}, it is possible to estimate the values of these deformation parameters and, like in a null experiment, check {\it a posteriori} whether the data are consistent with the Kerr BH hypothesis or require deviations from the Kerr background~\citep{Cao:2017kdq}. {\tt relxill\_nk} can also be used to test specific non-GR theories of gravity in the case the rotating BH solution is known in analytic form; this was done for some conformal theories of gravity~\citep{2018PhRvD..98b4007Z,2019EL....12530002Z}, Kaluza-Klein theories~\citep{2020arXiv200500184Z}, Einstein-Maxwell dilaton-axion gravity~\citep{2021arXiv210307593T}, and asymptotically safe gravity~\citep{2020arXiv200512958Z}.

\begin{figure}
    \centering
    \includegraphics[width=5.5cm]{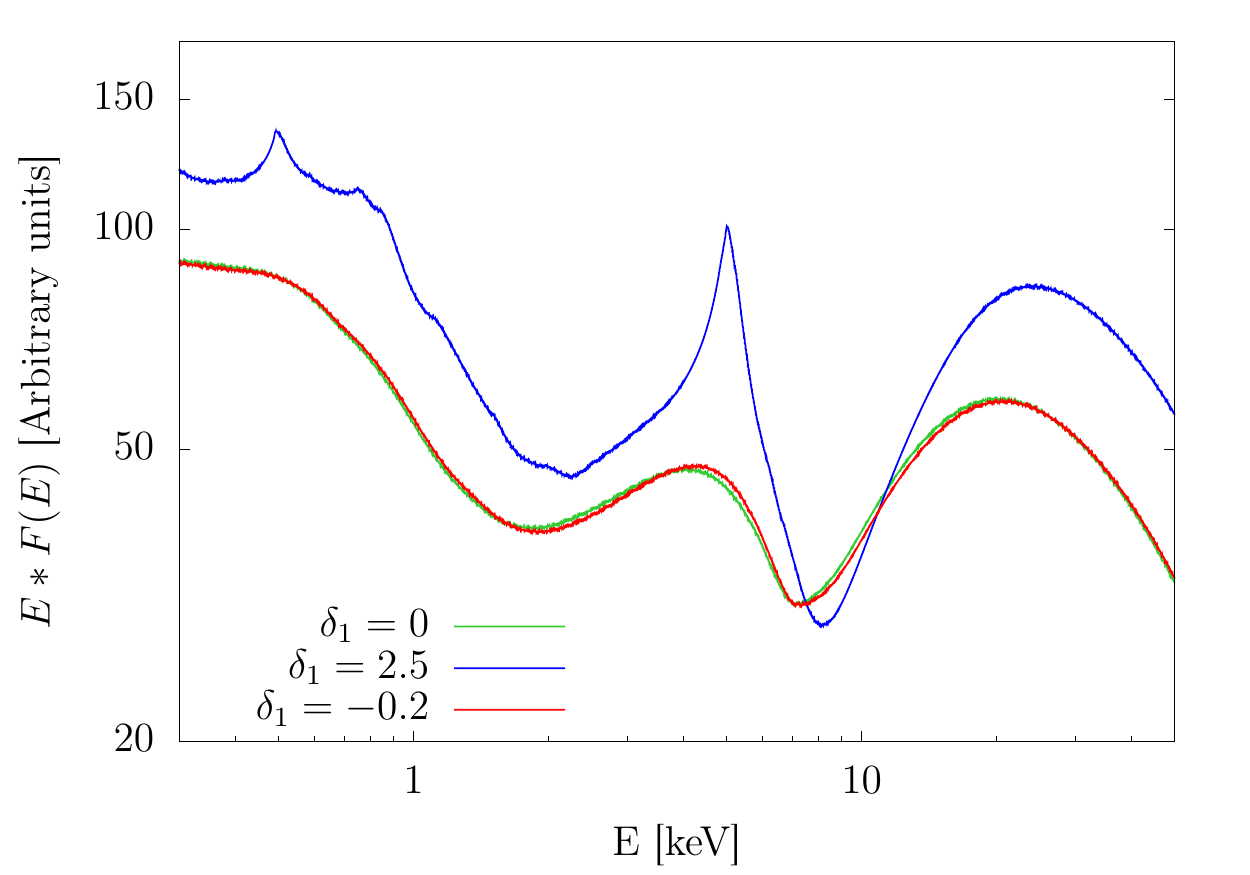}
    \includegraphics[width=5.5cm]{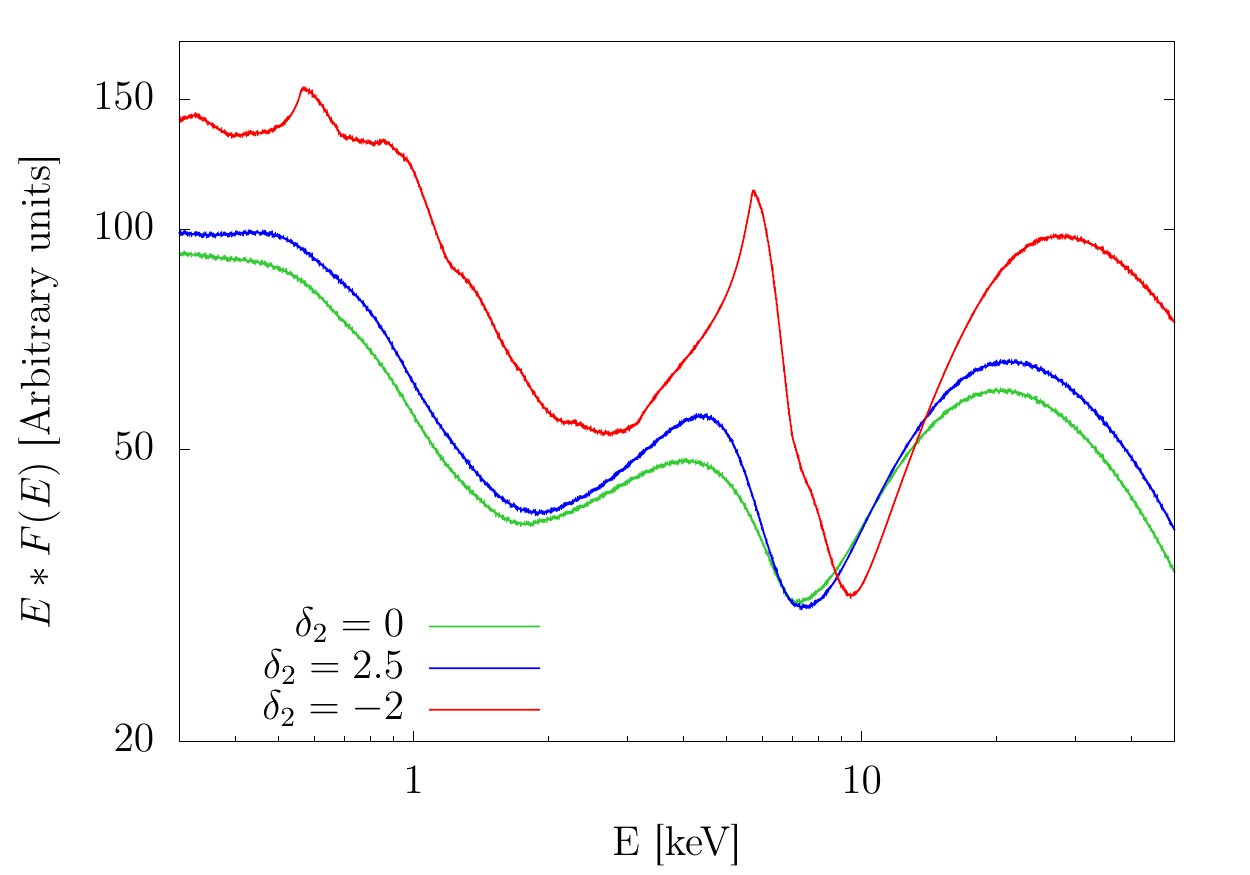} \\ \vspace{0.5cm}
    \includegraphics[width=5.5cm]{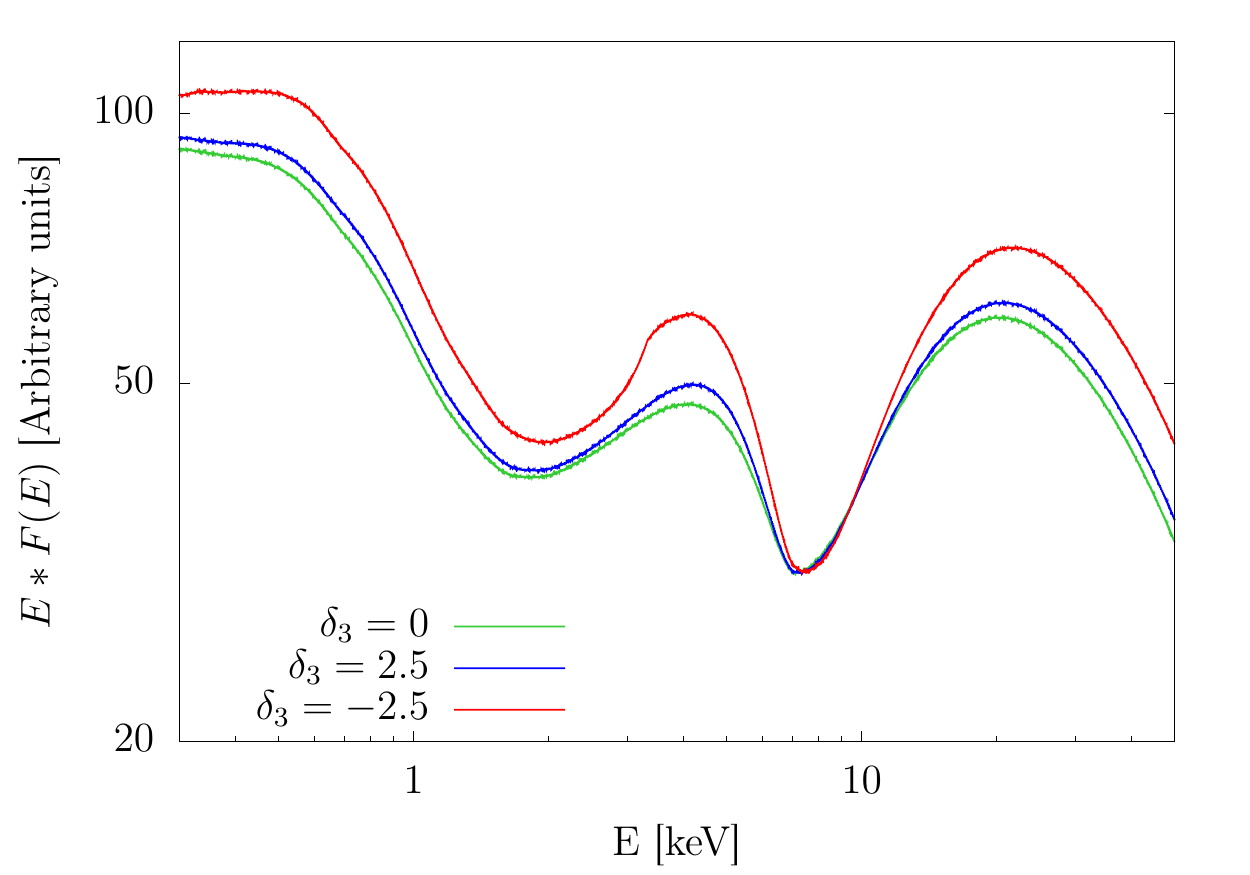}
    \includegraphics[width=5.5cm]{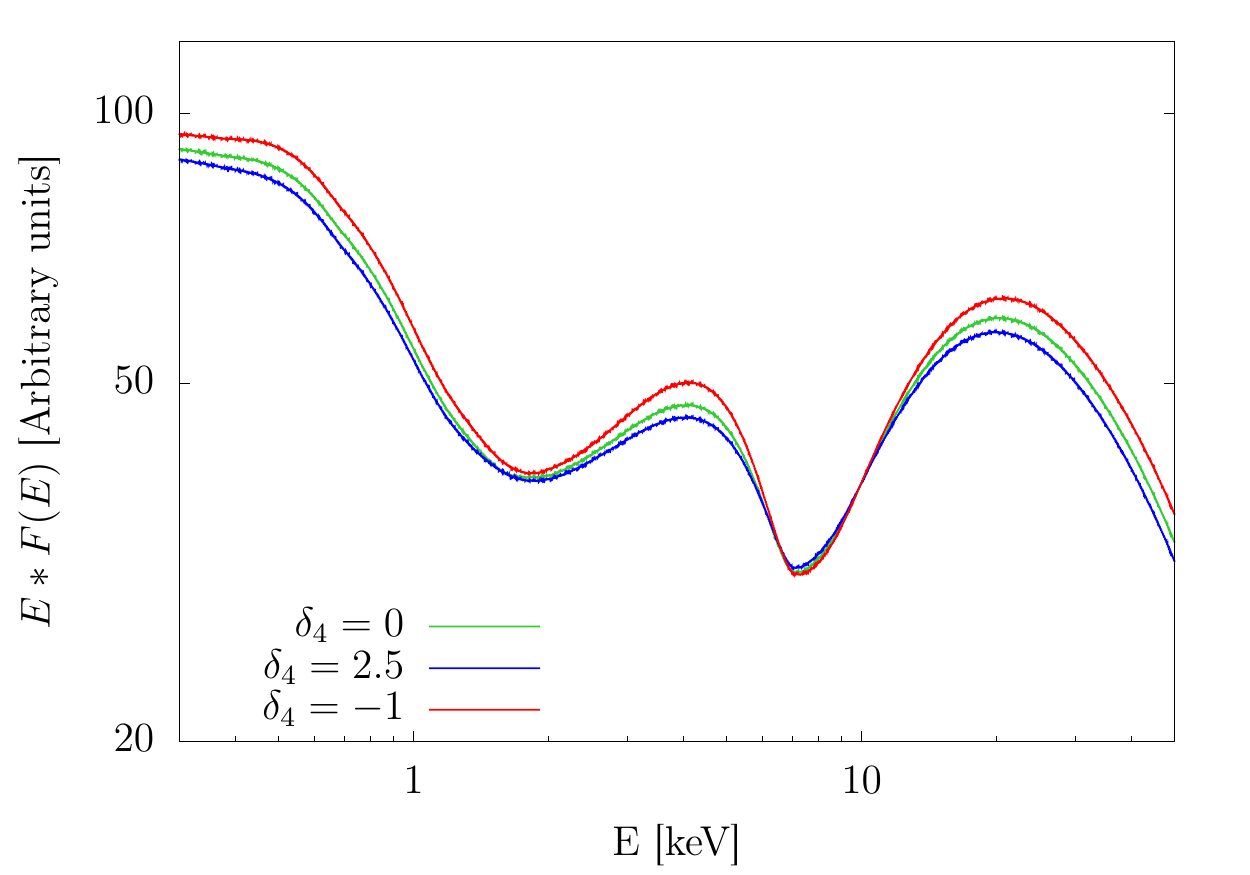} \\ \vspace{0.5cm}
    \includegraphics[width=5.5cm]{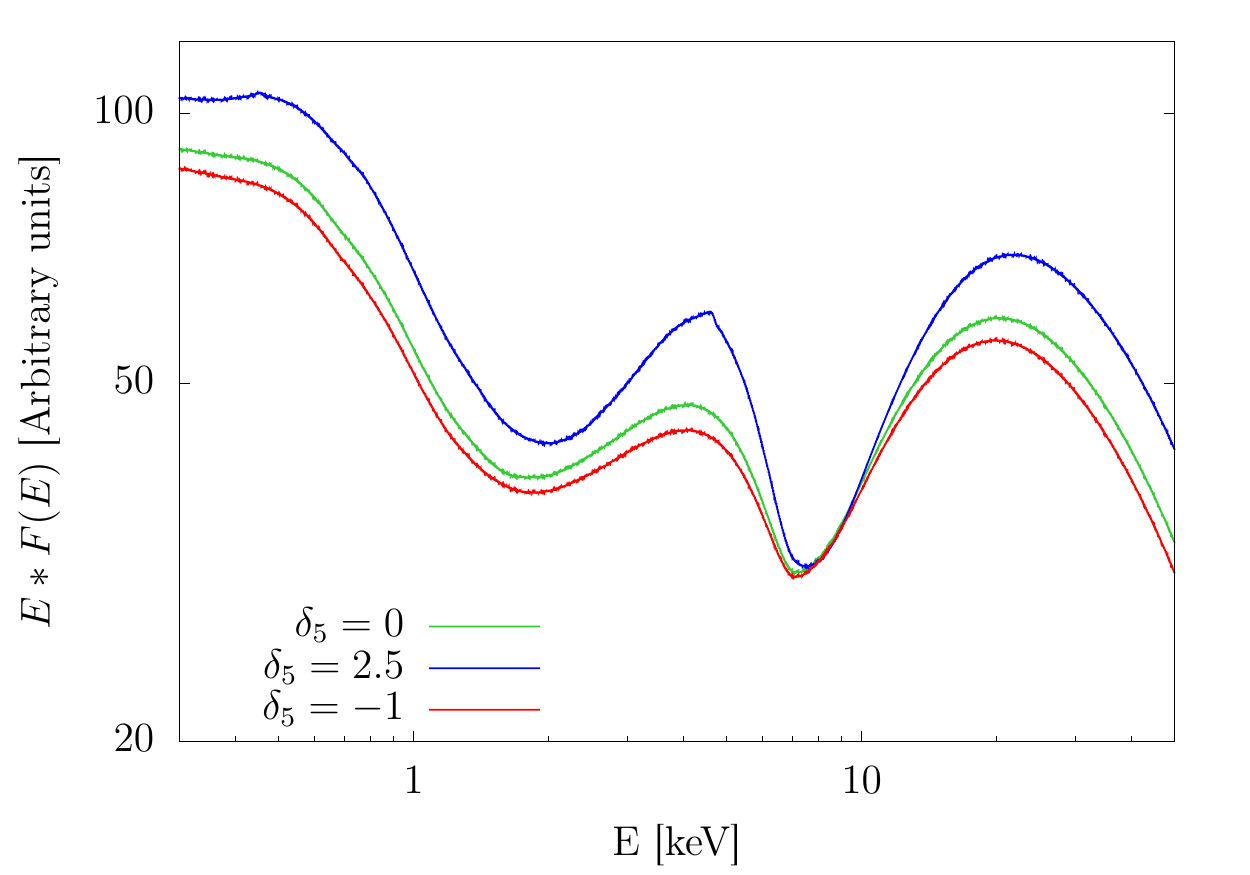}
    \includegraphics[width=5.5cm]{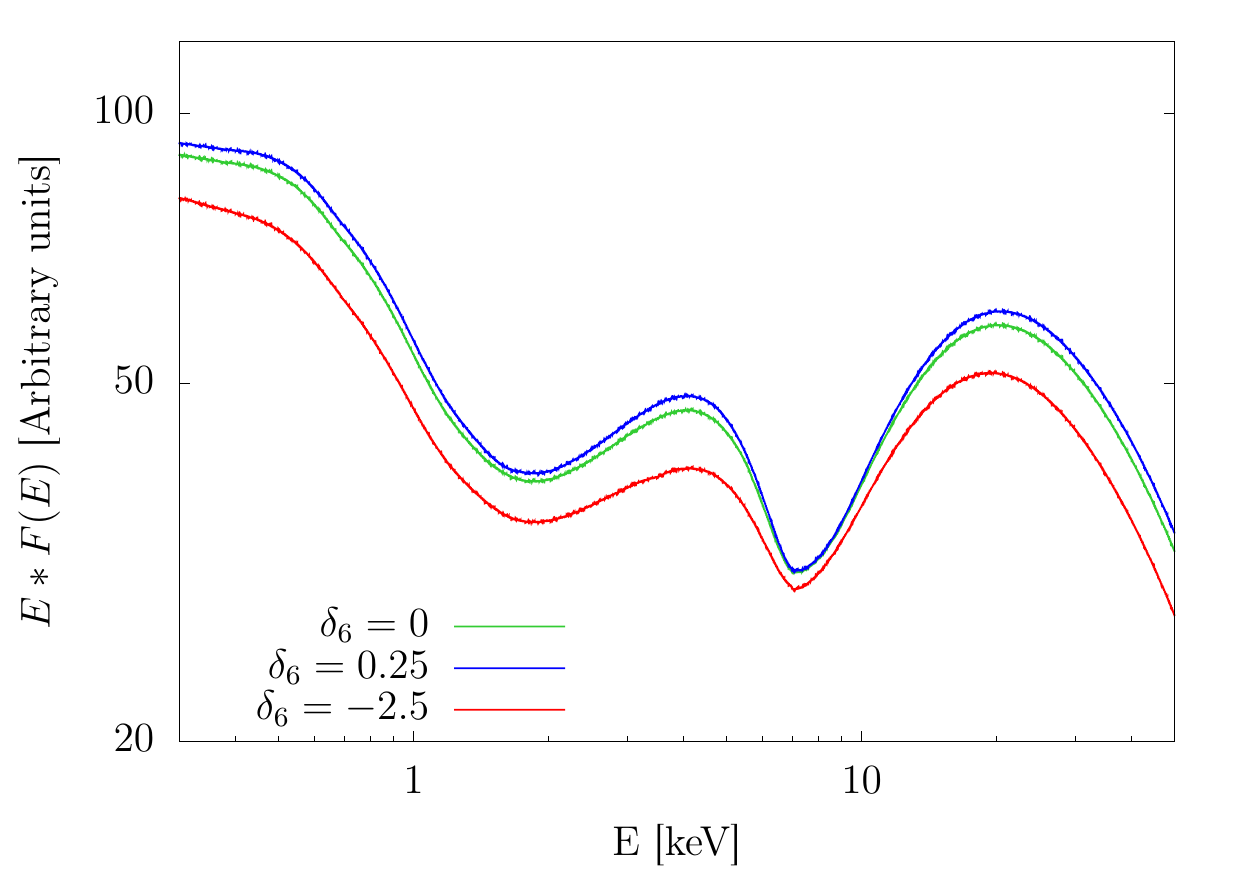} 
    \vspace{0.1cm}
    \caption{Impact of the deformation parameters $\delta_i$ ($i=1$, 2, ... 6) defined in \citet{2019arXiv190312119N} for the parametric BH spacetime proposed in \citet{2016PhRvD..93f4015K}. The $\delta_i = 0$ case corresponds to the Kerr solution and in every panel only one of the deformation parameters is allowed to be non-vanishing. All spectra are calculated assuming the following values of the model parameters: $a_* = 0.99$, $i = 30^\circ$, $\log\xi = 3.1$, $A_{\rm Fe} = 5$, $\Gamma = 2$, $E_{\rm cut} = 300$~keV, $q = 6$, and $R_{\rm in} = R_{\rm ISCO}$. Figure readapted from \citet{2019arXiv190312119N}. }
    \label{f-gr}
\end{figure}

Depending on the kind of deformation from the Kerr metric, the impact on the reflection spectrum is different \citep[see, e.g.,][]{2019PhRvD..99h3001T,2021arXiv210404183A}. Generally speaking, if the metric of a stationary and axisymmetric BH spacetime is written in canonical form and the metric deformations involve the metric coefficients $g_{tt}$, $g_{t\phi}$, or $g_{\phi\phi}$, we alter the structure of the accretion disk, including the position of the ISCO radius and the orbital velocity of the gas in the disk: these deformations normally produce quite significant modifications in the reflection spectrum and, modulo some degeneracy with other model parameters, it is relatively easy to constrain these modifications to the Kerr metric. Metric deformations involving the metric coefficients $g_{rr}$ and $g_{\theta\theta}$ are, in general, more elusive, as the structure of an infinitesimally thin accretion disk on the equatorial plane is not affected by these deformations and only the photon motion from the emission point to the detection point changes, normally with a quite weak impact on the shape of the reflection spectrum. Fig.~\ref{f-gr} shows the impact of different deformation parameters in the non-Kerr spacetime proposed in \citet{2016PhRvD..93f4015K} on the reflection spectrum of an accreting BH. The case of vanishing value of the deformation parameter corresponds to the Kerr solution. Note that the value of these deformation parameters is allowed to vary within some certain range that depends on the specific deformation parameter, as for values outside such a range the spacetime may not describe a BH or may present pathological properties (e.g., the existence of closed time-like curves permitting to go backward in time).

Generally speaking, GR tests are as difficult as accurate spin measurements. If it is possible to get a precise and accurate spin measurement of a source from a certain observation, we can also get a precise and accurate constraint on possible deviations from GR from the same observation. Since there is normally one more parameter in the model, there may be new parameter degeneracy, but this depends on the specific deformation from the Kerr solution. On the other hand, it may not be necessary to test GR with all sources, and one may only analyze those in which the systematic uncertainties are small and under control. This point may be different from BH spin measurements. For the spin, it is important to have a robust technique that can be applied to as many sources as possible, because the measurement of the spin parameter of a particular BH has a limited interest while we want to know the spin distribution over the BH population to learn about the BH formation, evolution, and its interaction with the host environment. In the case of modifications of GR, we may have three different scenarios. $a)$ The spacetime metric around astrophysical BHs is not described by the Kerr solution but is the same for all objects: in such a case, it is enough to test the Kerr metric with a source/observation that are well-understood and for which we can have precise and accurate measurements using X-ray reflection spectroscopy. $b)$ Astrophysical BHs have some new ``hairs'': they are not fully characterized by their mass and spin angular momentum, but they have other features, and the value of the associated physical quantity can be different for every object. In such a context, getting very stringent constraints on possible deviations from GR from a specific source does not imply that the same result holds for all BHs. $c)$ Astrophysical BHs have the so-called ``secondary hairs'': their non-GR property depends on the mass and/or the spin, so it is not a new independent feature. This is the case, for example, in Einstein-dilaton-Gauss-Bonnet gravity, where BHs have a scalar charge, but its value depends on the BH mass~\citep{1996PhRvD..54.5049K}: an accurate test for an object of a specific class (e.g., a stellar-mass BH in an X-ray binary) holds for all objects of that class.

As of now, all studies reported in the literature are consistent with the hypothesis that the spacetime metric around astrophysical BHs is described by the Kerr solution and present constraints on the value of certain deformation parameters~\citep[see, e.g.,][]{Xu:2018lom,Tripathi:2018lhx,Zhang:2019ldz,2020arXiv200309663A,2020arXiv201013474T,2020arXiv201210669T}. For stellar-mass BHs, the currently most precise and accurate GR tests are obtained from \textsl{NuSTAR} data of EXO~1846--031 \citep{2020arXiv201210669T} and GX~339--4 \citep{2020arXiv201013474T}. In the case of supermassive BHs, the currently most precise and accurate GR tests are reported from the analysis of simultaneous \textsl{NuSTAR} and \textsl{XMM-Newton} observations of MCG--06--30--15 \citep{Tripathi:2018lhx}. In these specific sources and observations, the systematic uncertainties seems to be well under control. \citet{2020arXiv201013474T} \citep[and see also][]{2020arXiv201210669T} estimated the systematic uncertainties related to the issues discussed in the previous sections of this review article and representing the current limitations for more precise tests with our present knowledge of these systems and with the available theoretical models. Selecting sources and observations in which the inner edge of the accretion disk is very close to the BH, we can often break, or at least limit, the degeneracy between the deformation parameter and the other model parameters~\citep{2020arXiv201210669T}.

\begin{figure}
    \centering
    \includegraphics[width=11cm]{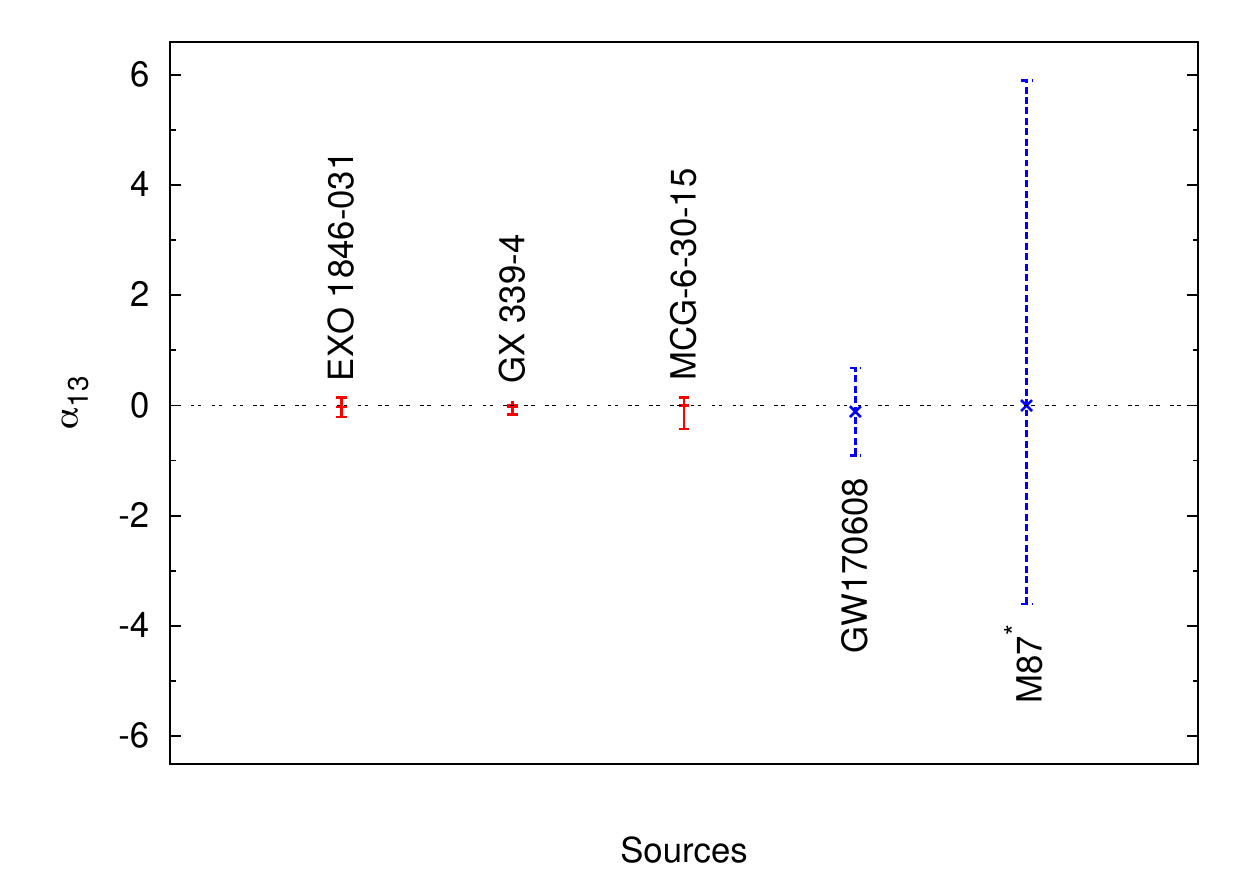}
    \vspace{0.1cm}
    \caption{3-$\sigma$ measurements of the Johannsen deformation parameter $\alpha_{13}$ from the analysis of the reflection features of the stellar-mass BHs in EXO~1846--031 \citep{2020arXiv201210669T} and GX~339--4 \citep{2020arXiv201013474T} and of the supermassive BH in MCG--06--30--15 \citep{Tripathi:2018lhx}. We also show the 3-$\sigma$ constraint from GW170608 \citep{2020CQGra..37m5008C} and the 1-$\sigma$ constraint from M87* \citep{2020PhRvL.125n1104P}. See the text for more details. Figure readapted from \citet{2020arXiv201210669T}. }
    \label{f-gr2}
\end{figure}

The 3-$\sigma$ constraints on the Johannsen deformation parameter $\alpha_{13}$~\citep{2013PhRvD..88d4002J} from the analysis of the reflection features of EXO~1846--031, GX~339--4, MCG--06--30--15 are reported in Fig.~\ref{f-gr2}, where the horizontal black dotted line at $\alpha_{13} = 0$ marks the Kerr solution. These constraints obtained from X-ray reflection spectroscopy can be compared with the constraints obtained from other techniques. From mm VLBI observations of the BH in the galaxy M87, \citet{2020PhRvL.125n1104P} obtained a much weaker constraint on $\alpha_{13}$ and in Fig.~\ref{f-gr2} we report their 1-$\sigma$ limit. If we assume that the emission of gravitational waves in a coalescing binary system can be described well by the Einstein Equations, it is possible to constrain $\alpha_{13}$ from the LIGO/Virgo data. \citet{2020CQGra..37m5008C} tested the Kerr metric with the events reported in the LIGO-Virgo Catalog GWTC-1 and obtained the strongest constraint from GW170608. As shown in Fig.~\ref{f-gr2}, the 3-$\sigma$ constraints on $\alpha_{13}$ from GW170608 is weaker than the most stringent constraints obtained from X-ray reflection spectroscopy. We note, however, that gravitational wave tests can normally get strong constraints from the dynamical aspects of the theory (e.g., dipolar radiation, ringdown, inspiral-merger, consistency tests...), which are not possible when only the spacetime metric is specified.

As in the case of BH spin measurements, even GR tests can benefit from future X-ray missions like \textsl{Athena}. However, to fully exploit the higher quality data of these missions, it is mandatory the development of more sophisticated theoretical models capable of addressing the simplifications described in the previous sections of this review.

\section{Concluding remarks \label{s-c}}

X-ray reflection spectroscopy can be a powerful technique for studying accreting BHs. The reflection process mainly occurs in the very inner part of the accretion disk. The analysis of the relativistic reflection features of a source can thus permit to study the accretion process in the strong gravity region and the physical properties of BHs.

The last decade has seen significant efforts in the development of X-ray reflection spectroscopy, with important progress in the comprehension of the physics and astrophysics of BHs. Today we have successful spin measurements of about 70~sources among stellar-mass and supermassive BHs with this technique. Work is in progress to use X-ray reflection spectroscopy for testing GR in the strong field regime.

In this review article, we have presented the state-of-the-art in reflection modeling, listed the available packages to analyze reflection features in X-ray observations of BHs, and pointed out the differences among different models. Despite the significant advancements in the last decade, all the available reflection models present a number of simplifications that can, at different levels, have an impact on precision measurements of BHs. In Section~\ref{s-ref}, we have discussed simplifications and assumptions in the calculations of synthetic reflection spectra in the rest-frame of the gas, structure of the accretion disk, and coronal modeling. We have discussed our current understanding of the impact of returning radiation and reflection Comptonizations, two effects that are often neglected but may bias final measurements. Systematic uncertainties are not only in our current theoretical models, and attention should be taken for instrumental uncertainties as well. For specific sources and observations, accurate and precise measurements are already possible and we have, for example, reliable estimates of some BH spins or accurate tests of the Kerr metric~\citep[see, for instance, the discussion on the systematic uncertainties in][]{2020arXiv201013474T,2020arXiv201210669T}.

Over the course of the next two decades, advances in space-based X-ray telescopes and detectors will usher in a new era of precision spectral and timing studies that will significantly improve our ability to identify and characterize reflection signatures in BH systems.  Missions such as \textsl{XRISM} \citep[projected launch in $\sim 2022$][]{2020SPIE11444E..22T} and \textsl{Athena} \citep[projected launch in $\sim 2034$][]{2013arXiv1306.2307N} will combine the power of microcalorimeter spectral resolution and increases in collecting area to enhance the diagnostic power of the data, particularly when analyzed with the ever-improving, latest generation of reflection models discussed herein.

\begin{figure}
    \centering
    \includegraphics[width=11cm]{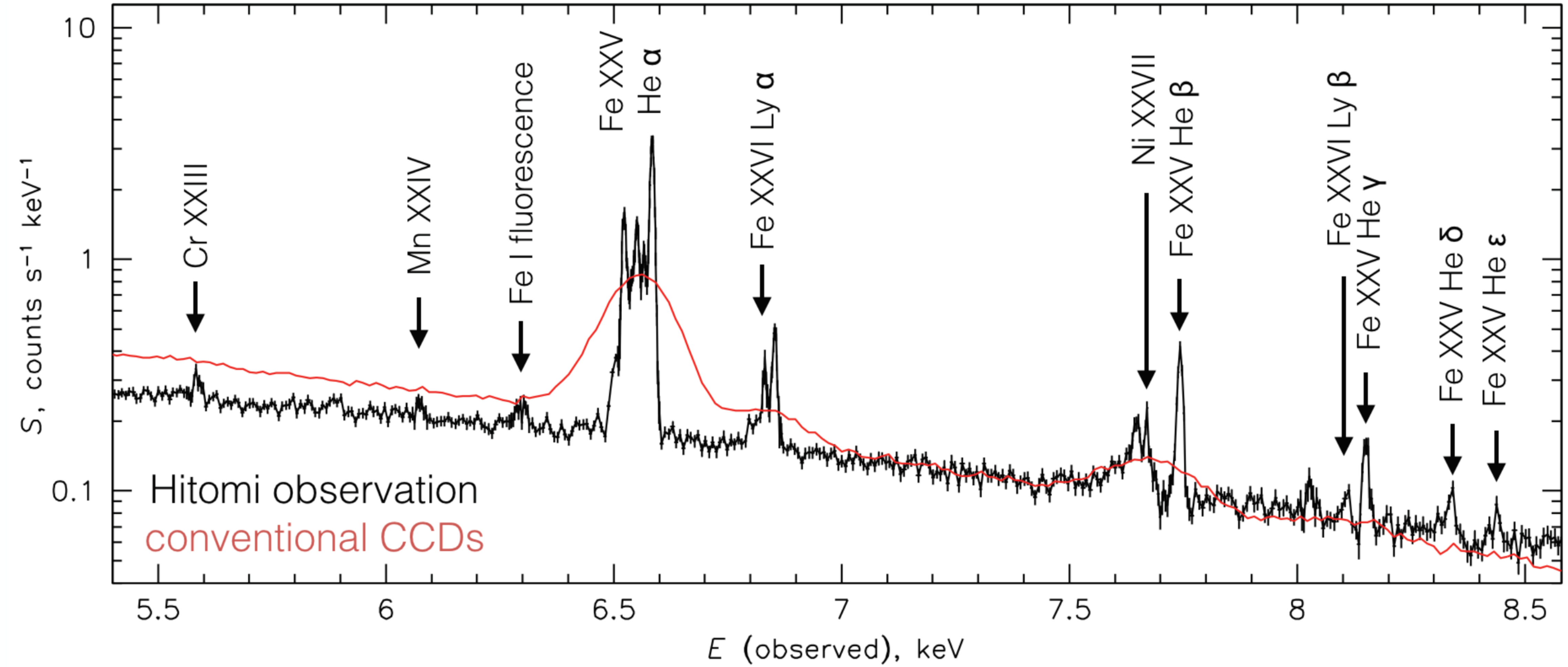}
    \caption{Spectrum of the Perseus cluster core observed with \textsl{Hitomi}'s microcalorimeter (black) and \textsl{Suzaku}'s CCD imaging spectrometer (red). The \textsl{XRISM}/Resolve microcalorimeter will provide high-resolution spectra in the 0.3--12~keV band for extended sources similar to \textsl{Hitomi}. Figure from~\citet{2020arXiv200304962X}.} 
    \label{f-xrism}
\end{figure}

The \textsl{XRISM}/Resolve microcalorimeter \citep{2017xru..conf..219T} will achieve a spectral resolution of $\sim 5$~eV across a 0.3--12~keV energy range, with up to $\sim 400$~cm$^2$ of collecting area over that bandpass, see Fig.~\ref{f-xrism}. This instrument is essentially a copy of the \textsl{Hitomi}/SXS, which provided revolutionary insight into gas microphysics in a handful of astrophysical objects before the mission’s untimely demise in 2016 \citep[e.g.,][]{2018PASJ...70....9H}.  Resolve’s spectral resolution will enable measurements of Fe~K region line widths down to $\sim 250$~km/s in velocity space and centroids to $\sim 100$~km/s, which will significantly improve our ability to characterize narrow and broad spectral features in both emission (e.g., reflection contributions from distant material) and absorption (e.g., outflowing winds along our line of sight to the BH).   In order to accurately and precisely measure BH spin, we must first isolate the signatures of relativistic reflection from the inner accretion disk; microcalorimeters will allow us to do with unprecedented confidence.

\textsl{XRISM} will observe multiple AGN and BH binaries during the performance verification phase over the first 9~months of the mission.  As an example of the spectral power of the Resolve instrument, a 100~ks observation of the AGN MCG--6--30--15 will yield independent, $<10$\% constraints on the BH spin; a goal frequently cited by observers in order to meaningfully compare data with numerical simulations and ascertain whether accretion or mergers has been the primary BH growth mechanism in recent epochs \citep[e.g., ][]{2008ApJ...684..822B}.  With this exposure time, Resolve will be able to separate out narrow emission and absorption features at the 10-$\sigma$ level for Fe~K$\alpha$ lines and at the 3-$\sigma$ level for Fe~K$\beta$ lines.  Resolve’s unique ability to conclusively disentangle broad and narrow spectral features will ensure that our spectral fitting is accurate as well as precise, and it will yield these results in less than a third of the time than present-day instruments such as \textsl{XMM-Newton}.

The \textsl{Athena}/X-IFU microcalorimeter~\citep{2018SPIE10699E..1GB} will further improve data quality, achieving a spectral resolution of $\sim 2.5$~keV from 0.2--12~keV with a projected $>1$~m$^2$ across most of that bandpass.  In addition to providing more spectral sensitivity to narrow line features, this advance in collecting area will be crucial in enabling time-resolved reflection spectroscopy for both X-ray binaries and AGN.  Because we know that certain system parameters cannot change on short timescales (e.g., mass and spin of the BH), we can link these variables between multiple observations of a given system.  This will allow us to better characterize the features of the system that are changing (e.g., outflowing winds, coronal properties), as well as to place tighter constraints on the BH spin. A summary of \textsl{Athena}’s capabilities in measuring supermassive BH spins is given in \citet{2019A&A...628A...5B}: $<10$\% spin measurements will be possible for hundreds of sources compared with the tens we have to date. Further, the data quality will ensure that spin measurements will finally enter an era of being systematics-limited rather than statistics-limited.
Note that \textsl{Athena} will be able to get even spin measurements of sources at higher redshift \citep[for more details, see][where the authors simulate observations of sources up to $z=2.5$]{2019A&A...628A...5B}.

Similar advances in instrumental data quality raise the question of whether X-ray astronomy will move to a regime closer to that of optical spectroscopy, where global models are rarely used, and, if so, whether X-ray reflection spectroscopy will still be a useful tool for studying accreting BHs. While it is plausible that a part of X-ray astronomy will indeed move to a regime closer to that of optical astronomy, this will be unlikely the case for the study of the strong gravity region of BHs, where relativistic effects make narrow features broad.

Future developments in the field are not only expected from data with higher spectral resolution. Measurements of the polarization of reflection spectra will help spectral analyses to constrain better the properties of a source~\cite[see, e.g., ][]{matt93,kmd18}. For example, it is currently difficult to get accurate estimates of the inclination angle of a disk, while the polarization signal is quite sensitive to it. In the more distant future, X-ray interferometry techniques could permit us even to image the very inner part of the accretion disk around BHs, opening new possibilities of studying the reflection features in strong gravity region of these systems~\citep{2019arXiv190803144U}.



\begin{acknowledgements}
We wish to thank Barbara De~Marco and Erin Kara for useful comments and suggestions.
All authors are members of the International Team~458 at the International Space Science Institute (ISSI), Bern, Switzerland. J.G. and A.A.Z.\ are also members of the International Team~486 at the International Space Science Institute (ISSI), Bern, Switzerland. The work of C.B., H.L., and A.T.\ is supported by the Innovation Program of the Shanghai Municipal Education Commission, Grant No.~2019-01-07-00-07-E00035, the National Natural Science Foundation of China (NSFC), Grant No.~11973019, and Fudan University, Grant No.~JIH1512604. V.G.\ is supported through the Margarete von Wrangell fellowship by the ESF and the Ministry of Science, Research and the Arts Baden-W\"urttemberg. J.J.\ is supported by the Tsinghua Shui'Mu Scholar Program and the Tsinghua Astrophysics Outstanding Fellowship. R.M.\ acknowledges the financial support of INAF (Istituto Nazionale di Astrofisica), Osservatorio Astronomico di Roma, ASI (Agenzia Spaziale Italiana) under contract to INAF: ASI~2014-049-R.0 dedicated to SSDC. A.N.\ is supported by the Polish National Science Centre under the grants 2015/18/A/ST9/00746 and 2016/21/B/ST9/02388. A.A.Z.\ is supported by the Polish National Science Centre under the grants 2015/18/A/ST9/00746 and 2019/35/B/ST9/03944. 
 \end{acknowledgements}


\bibliographystyle{mnras}
\bibliography{bbb}

\end{document}